\documentclass[twocolumn]{article}

\usepackage{subfigure}
\usepackage{multirow,tabularx}
\usepackage{algorithm}
\usepackage[math]{easyeqn}
\usepackage{algorithm, algorithmicx, algpseudocode}
\usepackage{graphicx}
\usepackage{xcolor}
\usepackage{authblk}

\newcommand{\mybm}[1]    {\mbox{\boldmath{$#1$}}}

\newcommand{\boldb}    {{\mybm b}}

\newcommand{\boldo}    {{\mybm 0}}

\newcommand{\boldbp}   {{\mybm P}}

\newcommand{\boldu}    {{\mybm u}}

\newcommand{\defor}    {{\mybm \varepsilon}}

\newcommand{\vectu}    {\mathbf{u}}

\newcommand{\vectb}    {\mathbf{b}}

\newcommand{\vectp}    {\mathbf{p}}

\newcommand{\vectr}    {\mathbf{r}}

\newcommand{\vectv}    {\mathbf{v}}

\newcommand{\vectx}    {\mathbf{x}}
\newcommand{\vecty}    {\mathbf{y}}
\newcommand{\vectz}    {\mathbf{z}}

\newcommand{\matra}    {\mathbf{A}}

\newcommand{\matrq}    {\mathbf{Q}}
\newcommand{\matrr}    {\mathbf{R}}

\newcommand{\matrtau}  {\mathbf{\tau}}
\usepackage{hyperref}

\definecolor{col1}{HTML}{1E88E5}
\definecolor{col2}{HTML}{D81B60}
\definecolor{col3}{HTML}{43A047}
\definecolor{col4}{HTML}{F4511E}

\date{}

\usepackage[normalem]{ulem}
\usepackage{xspace}

\begin{document}

\title{MPI+X: task-based parallelization and dynamic load balance \\ of finite element assembly}
\author {Marta Garcia-Gasulla, Guillaume Houzeaux, Roger Ferrer, Antoni Artigues, Victor L\'opez, Jes\'us Labarta and Mariano V\'azquez}
\affil[]{Barcelona Supercomputing Center, c) Jordi Girona 29, 08034 Barcelona, Spain}


\twocolumn[
\begin{@twocolumnfalse}
\maketitle
\begin{abstract}
The main computing tasks of a finite element code(FE) for solving partial differential equations (PDE's)
  are the algebraic system assembly and the iterative solver. This work focuses on the first task, in the context 
  of a hybrid MPI+X paradigm. Although we will describe algorithms in the FE context, a similar strategy 
  can be straightforwardly applied to other discretization methods, like the finite volume method.  
  The matrix assembly consists of a loop over the elements of the MPI partition to compute
  element matrices and right-hand sides and their assemblies in the local system to each MPI partition.
  In a MPI+X hybrid parallelism context, X has consisted traditionally of loop parallelism using OpenMP.
  Several strategies have been proposed in the literature to implement this loop parallelism, like
  coloring or substructuring techniques to circumvent the race condition that appears when assembling the 
  element system into the local system. The main drawback of the first technique is the decrease
  of the IPC due to bad spatial locality. The second technique avoids this issue but requires extensive 
  changes in the implementation, which can be cumbersome when several element loops should be treated.
  We propose an alternative, based on the task parallelism of the element loop using some extensions to the 
  OpenMP programming model. The taskification of the assembly solves both aforementioned
  problems. In addition, dynamic load balance will be applied using the DLB library, especially
  efficient in the presence of hybrid meshes, where the relative costs of the different elements is
  impossible to estimate a priori.
  This paper presents the proposed methodology, its implementation and its validation through the solution
  of large computational mechanics problems up to 16k cores.
\end{abstract}
\end{@twocolumnfalse}
\vspace{10mm}]

%







\maketitle

\section{Introduction}
The two most intensive computing tasks of computational mechanics codes for unstructured meshes
are the algebraic system assembly and the iterative solver to solve it. In this paper we will focus on 
improving the performance and execution of the first task, the algebraic system assembly.

The algebraic system assembly consists of a loop over elements, in the Finite Element (FE) context,
and faces or cells in the Finite Volume (FV) context. Although this work will focus on the first family
of methods, all the strategies described here can be applied to the second one.

For each element, the system assembly consists of two main steps: 
\begin{itemize}
 \item Compute the element matrix and right-hand side.
 \item Assembly the element system into the local algebraic system of each MPI partition.
\end{itemize}
The element loop is local to each MPI partition and does not involve any communication. It is thus well-suited for shared memory parallelism.
In an MPI+X hybrid parallelism context, X has consisted traditionally of loop parallelism using OpenMP.
However, assembling the element system into the local one involves an update of a shared variable which limits drastically
the efficiency of the straightforward use of OpenMP pragmas.

Several strategies have been proposed in the literature to circumvent this weakness, like the
coloring or substructuring techniques to avoid the race condition appearing in the assembly
of the element system into the local system. The main drawback of the first technique is the drop
of the number of instructions per cycle (IPC) due to the bad spatial locality inherent to the coloring.
The second technique solves this issue but requires intensive recoding, which can be cumbersome when
several element loops should be treated. These techniques will be summarized in Section \ref{sec:FE_assembly}.

We propose an alternative, based on the task parallelism of the element loop using an extension to the OpenMP programming
model and implemented in the OmpSs model\cite{ompss}\cite{ompss-web}. The taskification of the assembly that we propose solves both aforementioned problems. The technique will be described in Section \ref{sec:task}.

In addition, in the context of MPI parallelization load imbalance is an issue that can degrade the performance and does not have a straightforward solution. The main issue when load balancing an MPI application comes from the fact that the data is not shared among the different MPI processes. Consequently, application developers put a lot of effort at obtaining a well balanced data partition \cite{partitioning} \cite{Ant-colony_partition} \cite{multilevel_partition}.

Unfortunately a well balanced partition is not always easy to obtain as we will see in Section \ref{sec:dlb}. And, even, if a well balanced partition is achieved it does not imply a well balanced execution. In some cases the load can change during the execution, i.e. particles moving or a dam breaking.
In this case, a runtime solution is necessary. One of the solutions proposed in the literature is
to repartition the mesh during the execution to obtain a better balanced distribution~\cite{openFoam}. This kind of solutions
implies a redistribution of data and cannot be applied each timestep because of the overhead they introduce. Moreover, they cannot
react to punctual load changes or load imbalance introduced by system noise. We will apply a dynamic load balance that does
not require to modify the application neither to redistribute data.

Finally, in Section \ref{sec:results}, the efficiency of the proposed taskifying strategy will be compared to classical
loop parallelism with OpenMP using an element coloring strategy. In this section we will also present the performance evaluation 
of the load balancing library. And we will demonstrate that both mechanisms can be useful to scale a finite element code up to 16386 cores.

\section{Fluid and structure dynamics}
\label{sec:fluid}

In this work we consider two different sets of partial differential equations (PDE's), modeling incompressible flows and large deformations
of structures. We will put more emphasis on the first set of equations, as the numerical
modeling and system solution are more complex. Apart from the sets of equations to be solved, we will introduce
as well the case examples selected to carry out the proposed optimizations. In the case of the Navier-Stokes
equations we will consider the airflow in the respiratory system, while for structure mechanics, we will consider
a fusion reactor.

\subsection{Fluid solver}

The high performance computational mechanics code used in this work is Alya \cite{Vazquez15d}, developed at BSC-CNS,
and part of the Unified European Application Benchmark Suite (UEABS) \cite{ueabs}. This suite
provides a set of scalable, currently relevant and publically available codes and datasets, of a size which can
realistically be run on large systems, and maintained 
into the future. In this section, will briefly describe the CFD module of Alya and its parallelization.

\subsubsection{Physical and Numerical models}
\label{sec:numerical_model}

The equations governing the dynamics of an incompressible fluid are the so-called incompressible Navier-Stokes equations.
They express the Newton's second law for a fluid continuous medium, whose unknowns are the velocity $\boldu$
and the pressure $p$ of the fluid. Two physical properties are involved, namely $\mu$ be the viscosity, and $\rho$ the density. 
At the continuous level, the problem is stated as follows: find the velocity $\boldu$ and pressure $p$ in a domain $\Omega$ such 
that they satisfy in a given time interval 
\begin{eqnarray}
    \displaystyle{\rho \frac{\partial \boldu}{\partial t}}
  + \rho (\boldu \cdot \nabla)\boldu
  - \nabla \cdot[ 2 \mu \defor(\boldu) ]
  + \nabla p = \boldo, \label{eq:momentum}\\
   \qquad\qquad\qquad 
   \nabla \cdot \boldu = 0,
   \label{eq:continuity}
\end{eqnarray}
together with initial and boundary conditions. The velocity strain rate is defined as
$\defor(\boldu) := \frac{1}{2} ( \nabla \boldu + \nabla \boldu^t)$.

The variational multiscale (VMS) method is applied to discretize this set of equations, as
extensively described in \cite{GHouzeaux_JPrincipe08}. In addition, the velocity subgrid 
scale is tracked in convection and time. This means that apart from solving for the previous unknowns
$\boldu$ and $p$, an additional equation is solved to obtain the subgrid scale $\tilde{\boldu}$.
A typical assembly for the grid scale equations consists in a loop over the elements of the mesh,
as shown in Algorithm \ref{alg:sequential_assembly}.
\begin{algorithm}[h!bt]
\begin{algorithmic}[1]  
  \For   {elements $e$}
  \State Compute element matrix and RHS: $\matra^e$, $\vectb^e$ \label{ite:element_a_b}
  \State Assemble matrix $\matra^e$ into $\matra$ \label{ite:assembly_a}
  \State Assemble RHS    $\vectb^e$ into $\vectb$ \label{ite:assembly_b}
  \EndFor
\end{algorithmic}
\caption{Assembly of a generic matrix $\matra$ and vector $\vectb$.}
\label{alg:sequential_assembly} 
\end{algorithm} 

\subsubsection{Algebraic system solution}

After the assembly step, the following monolithic algebraic system for
the grid scale unknowns, velocity $\vectu$ and pressure $\vectp$, is obtained:
\begin{eqnarray} 
   \left[ \begin{array}{ll}
          \matra_{uu} & \matra_{up} \\
          \matra_{pu} & \matra_{pp}
   \end{array} \right]
   \left[ \begin{array}{l}
          \vectu \\
          \vectp
   \end{array} \right]
   =
   \left[ \begin{array}{l}
          \vectb_u \\
          \vectb_p
   \end{array} \right].
   \label{eq:monolithic}
\end{eqnarray}

This system can be solved directly using a Krylov solver and efficient preconditioner \cite{YSaad03}. However, an algebraic-split
approach is used instead in this work. We extract the pressure Schur complement of the pressure unknown $\vectp$
and solve it with the Orthomin(1) method, as detailed in \cite{GHouzeaux_RAubry_MVazquez09}. The resulting algorithm
is shown in Algorithm \ref{alg:mom-orthomin}.
\begin{algorithm}[h!]
\begin{algorithmic}[1]
  \State Solve momentum eqn $\matra_{uu} \vectu^{k+1} = \vectb_u - \matra_{up} \vectp^k$ \label{ite:mom1}
  \State Compute Schur complement residual $\vectr^k = [\vectb_p - \matra_{pu} \vectu^{k+1}] - \matra_{pp} \vectp^k$
  \State Solve continuity eqn $\matrq \vectz = \vectr^k$ \label{ite:con1}
  \State Solve momentum eqn $\matra_{uu} \vectv = \matra_{up} \vectz$ \label{ite:mom2}
  \State Compute $\vectx = \matra_{pp} \vectz - \matra_{pu} \vectv$
  \State Compute $\alpha = (\vectr^k,\vectx)/(\vectx,\vectx)$
  \State Update velocity and pressure
         \begin{eqnarray} \left\{ \begin{array}{rclcl} 
               \vectp^{k+1} & = & \vectp^k     + \alpha \vectz\\
               \vectu^{k+2} & = & \vectu^{k+1} - \alpha \vectv
         \end{array} \right. \label{eq:mom-orthomin} \end{eqnarray}                
\end{algorithmic}
\caption{Algebraic solver: Orthomin(1) method for the pressure Schur complement.}
\label{alg:mom-orthomin} 
\end{algorithm}
In the algorithm, matrix $\matrq$ is the pressure Schur complement preconditioner, computed here as an algebraic
approximation of the Uzawa operator, and as explained \cite{GHouzeaux_RAubry_MVazquez09}. On the one hand,
the momentum equation is solved with the GMRES method and diagonal preconditioning in steps \ref{ite:mom1} and \ref{ite:mom2}
of the algorithm. On the other hand, the continuity equation is solved with the Deflated Conjugate
Gradient (DCG) method \cite{RLohner_FMut10} with linelet preconditioning \cite{Soto} in step \ref{ite:con1} of the algorithm.
The deflation provides a low frequency damping across the domain, especially very efficient for the case study
considered in this work, where the geometry is elongated. The linelet preconditioner consists
of a tridiagonal preconditioner applied in the normal direction to the boundary layer mesh near the walls.

%

At each time step, this system is solved until convergence is achieved.
Convergence is necessary because the original equation is non-linear (the convective term makes matrix $\matra_{uu}$ depend on
$\vectu$ itself). For any information concerning the parallel solution 
system \ref{eq:monolithic} on distributed memory supercomputers, see \cite{GHouzeaux_MVazquez08a,GHouzeaux_RAubry_MVazquez09}.
Only a brief description will be given herein in Section \ref{sec:par_mpi}. 

\subsubsection{Subgrid scale}

Once the velocity $\vectu$ and pressure $\vectp$ are obtained on the nodes of the mesh, the velocity
subgrid scale vector is obtained through a general equation of the form
\begin{eqnarray}
  \tilde{\vectu} = \matrtau^{-1} \left( \matrr \vectu + \vectb_{\tilde{\vectu}} \right),
  \label{eq:utilde}
\end{eqnarray}
where $\matrtau$ is the so-called stabilization diagonal matrix and $\matrr$ is the residual rectangular matrix,
as the subgrid scale is obtained element-wise and not node-wise. Both $\matrtau$ and $\matrr$ may depend on 
$\tilde{\vectu}$ and thus Equation \ref{eq:utilde} can be non-linear.
Note finally that in practice, $\tilde{\vectu}$ is obtained via a simple loop over the elements of the mesh
and the system of the equation does not need to be explicitly formed.

\subsubsection{Solution strategy}

In practice, the iterations of the Orthomin(1) iterative solver to solve for the pressure Schur complement are
coupled to the non-linearity iterations of the Navier-Stokes equations, which include not only the convective term
but also the subgrid scale. The resulting workflow is shown in Algorithm \ref{alg:solution_strategy}.

\begin{algorithm*}
\begin{algorithmic}[1]  
  \For   {time steps}
    \While  {until convergence}
    \State Assemble global matrices and RHS of Equation \ref{eq:monolithic} using Algorithm \ref{alg:sequential_assembly} 
    \State Solve momentum equation with GMRES, step \ref{ite:mom1} of Algorithm \ref{alg:mom-orthomin} 
    \State Solve continuity equation with DCG, step \ref{ite:con1} of Algorithm \ref{alg:mom-orthomin} 
    \State Solve momentum equation with GMRES, step \ref{ite:mom2} of Algorithm \ref{alg:mom-orthomin} 
    \State Compute subgrid scale using Equation \ref{eq:utilde}
    \EndWhile
  \EndFor
\end{algorithmic}
\caption{Solution strategy for solving Navier-Stokes equations.}
\label{alg:solution_strategy} 
\end{algorithm*}

The workflow consists of three main computational kernels. The {\it assembly} which carries out operations on 
the elements of the mesh in order to construct the algebraic system; the {\it algebraic solver}, that is the algorithm 
for the pressure Schur complement, which consists in solving twice the momentum equation and once
the continuity equation; finally, the {\it subgrid scale} calculation which is computed on the elements of the mesh
and thus involves a loop over the elements. 

\subsubsection{Case example: respiratory system}
\label{sec:meshresp}

\begin{figure}[h!tbp]
 \centering
 \includegraphics[width=0.47\textwidth]{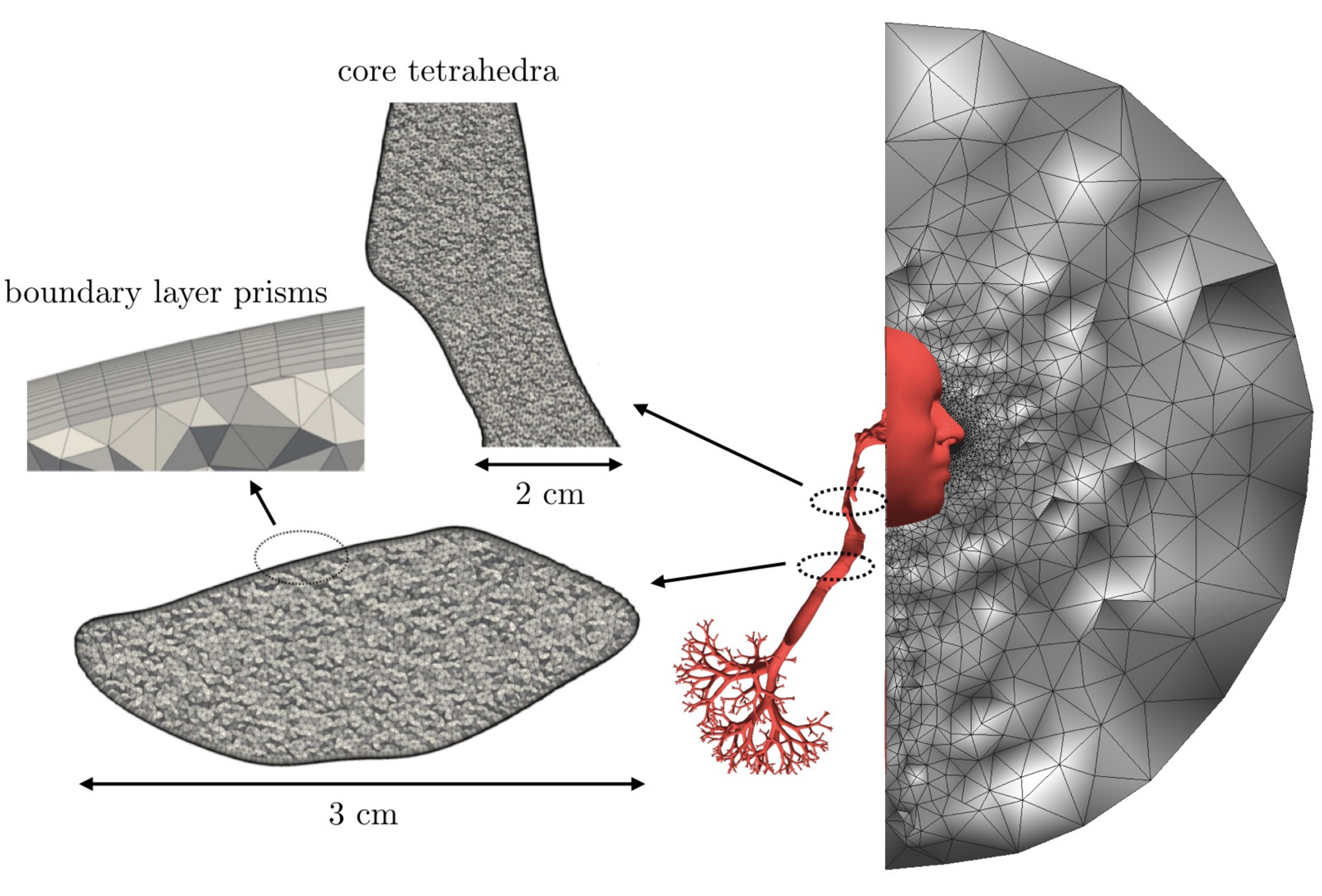}
 \caption{Respiratory system, details of the mesh.}
  \label{fig:resp_mesh}
\end{figure}

For the evaluation of the different techniques described in the following sections we will consider
the case of the respiratory system, similar to that described in \cite{Calmet16b}.
The mesh is hybrid and composed of 17.7 million elements: prisms to resolve accurately the boundary 
layer; tetrahedra in the core flow; pyramids to enable the transition from prism quadrilateral 
faces to tetrahedra. This kind of mesh is quite representative in fluid dynamics, as most of the 
fluid problems of interest involve boundary layers and a core flow.
Figure \ref{fig:resp_mesh} shows some details of the mesh, and in particular the
prisms in the boundary layer. We will see in Section \ref{sec:mpi_lb} how the presence of different
types of elements makes difficult the control of the load balance when using the mesh partitioner METIS. 

\subsection{Structure solver}

The structure mechanics solver is extensively described in \cite{Casoni14}. For the sake of completeness, we will only
briefly describe the set of equations to be solved.

\subsubsection{Physical and Numerical models}

The equation of balance of momentum with respect to the reference configuration can be written as
\begin{eqnarray} 
    \rho_0 \frac{\partial^2 \boldu}{\partial t^2} - \nabla_0 \cdot \boldbp = \boldb_0,
   \label{eq:equationLocalBalanceMomentum} 
\end{eqnarray} 
where $\rho_0$ is the mass density (with respect to the reference volume) and
$\nabla_0 \cdot$ is the divergence operator with respect to the reference configuration.
Tensor $\boldbp$ and vector $\boldb_0$ stand for the first Piola-Kirchhoff stress and the
distributed body force on the undeformed body, respectively.
Equation \ref{eq:equationLocalBalanceMomentum} must be supplied with initial and boundary
conditions.

To discretize this equation, the Galerkin method is used in space and the Newmark method \cite{BelytschkoBook} in time.
A Newton-Raphson method is used to solve the linearized system.
For each time step, and until convergence, one has to assemble the algebraic system using
Algorithm \ref{alg:sequential_assembly} (where $\matra$ is the Jacobian and $\vectb$ is the
residual of the equation) and then solve the corresponding algebraic system for the displacement
unknown. According to the characteristic of this system, the GMRES or the DCG methods are considered.

\subsubsection{Case example: Iter}

\begin{figure}[h!tbp]
 \centering
 \includegraphics[width=0.45\textwidth]{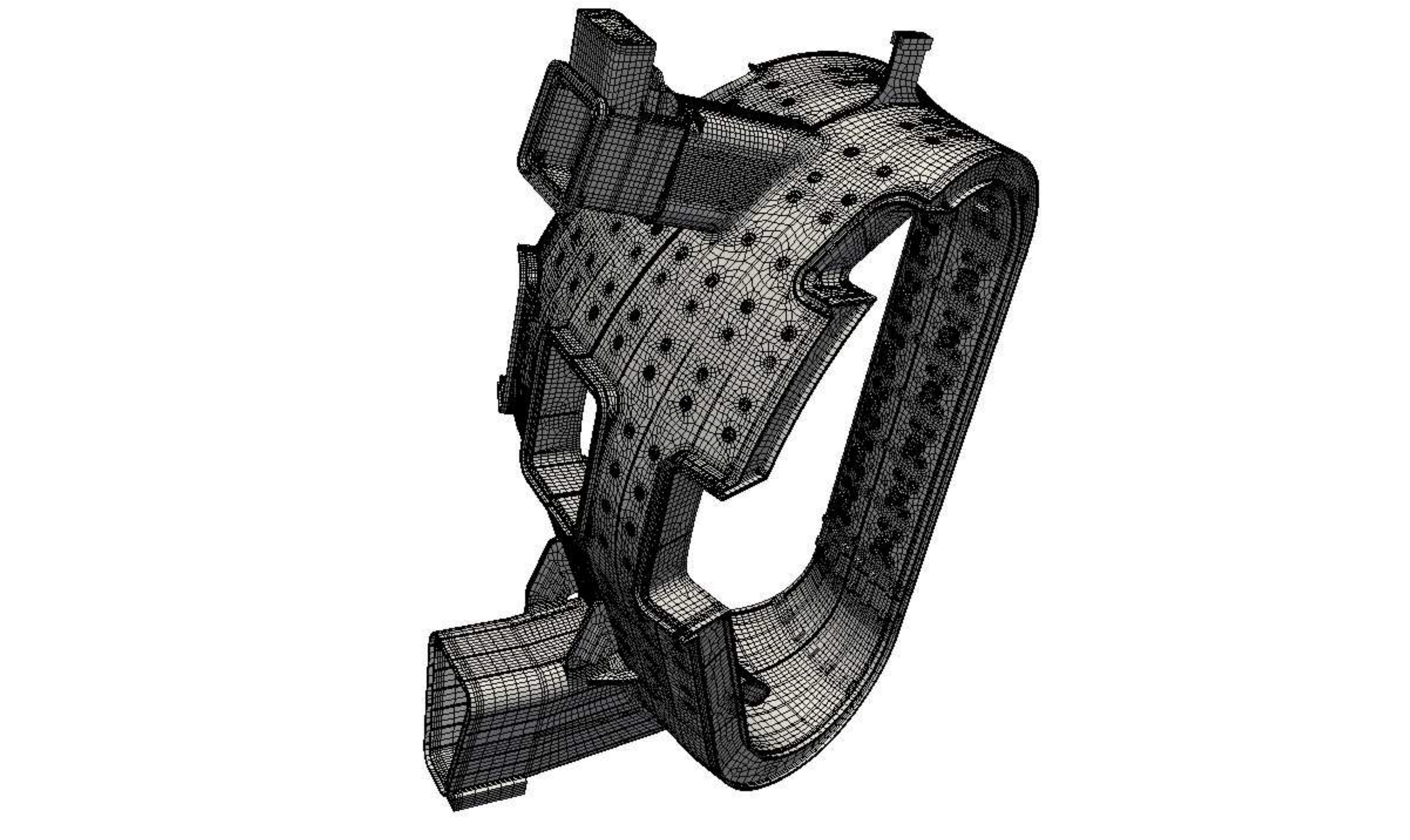}
  \caption{Iter, mesh.}
  \label{fig:Iter_mesh}
\end{figure}

The mesh is a slice of a torus shaped chamber, representing the center part of a nuclear
fusion reactor called the vacuum vessel. In Figure~\ref{fig:Iter_mesh} we
can see the representation of the mesh made of 31.5 million hexahedra, prisms, pyramids and tetrahedra elements.



\section{Parallelization of Algebraic System Assembly}
\label{sec:FE_assembly}

For the sake of completeness, this section describes briefly the classical parallelization techniques of a finite element
assembly in an HPC environment.

\subsection{Distributed Memory Parallelization Using MPI}
\label{sec:par_mpi}

In the finite element context, two options are available when defining a distributed
memory parallelization, which we refer to as partial-row and full-row methods
\cite{GHouzeaux16_book}.
In the finite difference context or in the finite volume context using a cell centered
discretization scheme, the second method is generally considered, as it emerges naturally.

In the finite element community, the partial-row method is quite natural when
partitioning the original element set (mesh) into disjoint subsets of elements (subdomains).
In this case, nodes belonging to several subdomains (interface nodes) are duplicated.
As the matrix coefficients come from element integrations, the coefficients of the edges involving interface nodes
are never fully assembled. The resulting local (to each MPI subdomain) matrices are therefore square matrices.
However, if one considers iterative solvers to solve the resulting algebraic system (which is
the case in this work), the main operation consists of Sparse Matrix-Vector Products (SpMV)
$\vecty = \matra \vectx$. Thus, there is no need for obtaining these
global coefficients, as the result $\vecty$ can be computed locally on each subdomain and
then assembled later on interface nodes using MPI send and receive messages
(by associativity of the multiplication operation).

The full-row approach consists in assigning nodes exclusively to one subdomain. This strategy leads
to local rectangular matrices, and these matrices are fully assembled blocks of rows of the global matrix.
Two options are thus possible. A first option in which the global matrix coefficients of the interface nodes can be obtained by
summing up the contributions coming from neighboring subdomains, thus obtaining full rows. This can be achieved through MPI communications. 

A second option in which the mesh is partitioned into disjoint subsets of nodes. The full rows of the matrix are thus obtained by assigning all the
necessary elements to the subdomains in order to get all the element contributions. In this case,
interface elements must be duplicated (halo elements), leading to duplication of the work during the assembly
process on these halo elements. As far as SpMV is concerned, MPI communications are performed
before the product on the multiplicand $\vectx$.

The advantage of the partial-row approach is that the load balance of the assembly can be
controlled, in principle, when partitioning the mesh. With partitioners like METIS \cite{Metis}, the number of elements per subdomain can in addition
be constrained with the minimization of the interface sizes. The main drawback is that while
balancing the number of elements per subdomain, one loses control on the number of nodes, which
dictates the balance of SpMV. On the other hand, the presence of halo elements in the full-row
approach limits the scalability, due to the duplicated work on them and to the lack of control on the
number of elements per subdomain.

When considering hybrid meshes, an additional difficulty arises, as one has to estimate the
relative weights of the different element types in order to balance the total weight per
subdomain, as we will show in Section \ref{sec:results}.
Thus, no matter if full-row or partial-row is ultimately chosen, load imbalance will occur before starting the simulation. In this work, the
partial-row approach is considered and described in \cite{Vazquez15d}, while load imbalance will
be treated in Section \ref{sec:dlb}.

\subsection{Shared Memory Parallelization Using OpenMP}

\subsubsection{Loop parallelism}

\begin{figure*}[h!tbp]
 \centering
 \includegraphics[width=0.85\textwidth]{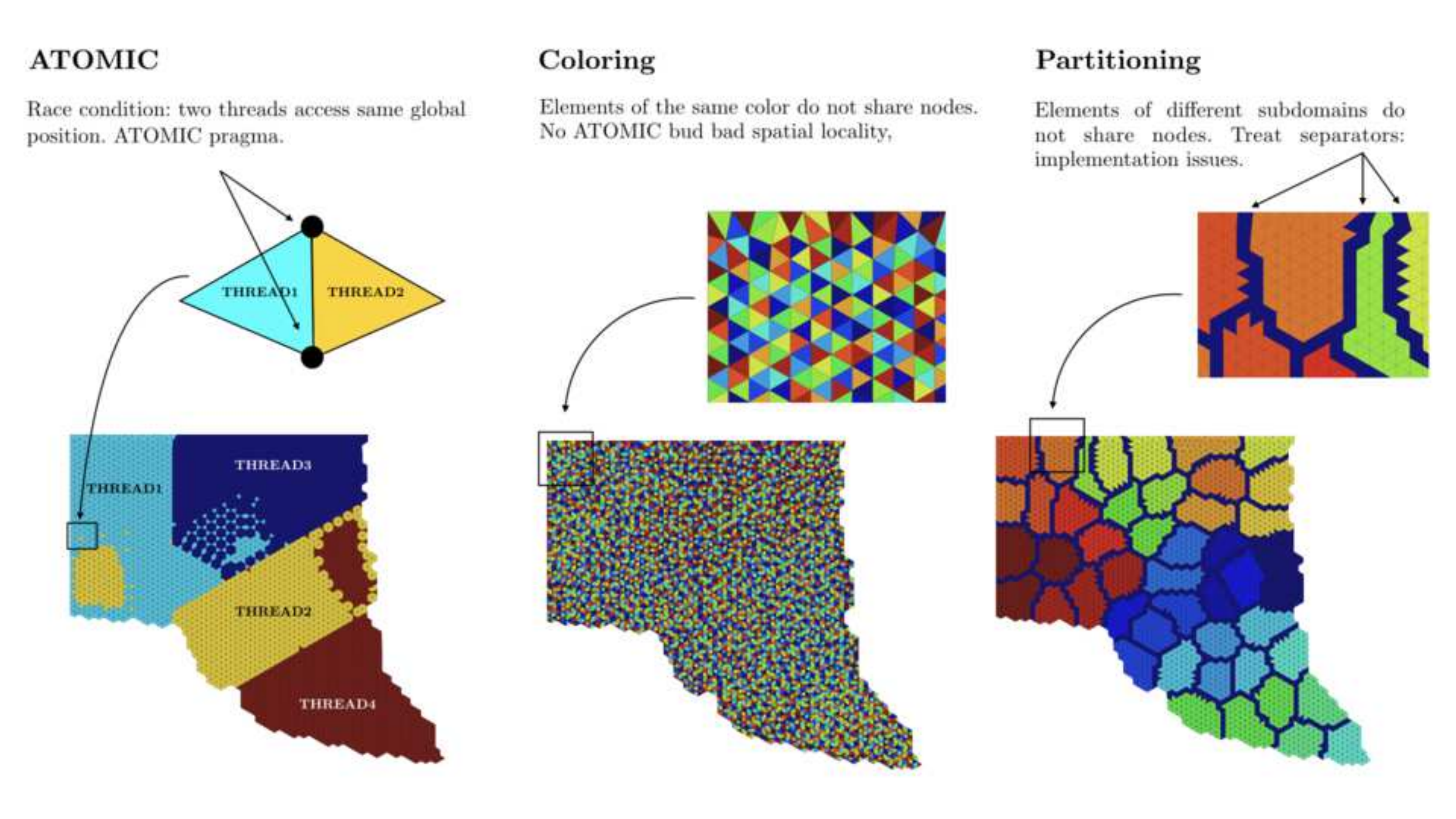}
 \caption{Shared memory parallelism techniques using OpenMP.
   (a) ATOMIC pragma. (b) Element coloring.
   (c) Local partitioning.}
  \label{fig:openmp}
\end{figure*}

During the last decade, the predominance of general purpose clusters have obliged parallel code designers to devise distributed
memory techniques, mainly based on MPI, as briefly described in last subsection. Then, while the number of CPUs has been multiplied,
the number of subdomains has been increasing. The side effect is the increase of communication which limits the strong
scalability, and the increase of number of MPI subdomains, which limits the weak scalability.
Nowadays, supercomputers offer a great variety of architectures, with many
cores on nodes (e.g. Xeon Phi). Thus, shared memory parallelism is gaining more and more attention as a it offers
more flexibility to parallel programming. This parallelism has traditionally been based on OpenMP, a programming model
enabling a straightforward parallelization through simple pragmas. Finite element assembly consists in computing
element matrices and right-hand sides ($\matra^{(e)}$ and $\vectb^{(e)}$) for each element $e$, and assembling them into the local
matrices and RHS of each MPI process, namely $\matra$ and $\vectb$, as shown in Algorithm \ref{alg:sequential_assembly}.
This assembly has been treated using mainly three techniques, as illustrated in Figure \ref{fig:openmp}.

All of these techniques are based on loop parallelism, each of them offering different advantages and drawbacks. The main issue
is the race condition appearing in the assembly scattering the element arrays $\matra^{(e)}$ and $\vectb^{(e)}$ into the local ones,
$\matra$ and $\vectb$.
The first method consists in avoiding the race condition using ATOMIC pragmas to protect these shared variables
(Figure \ref{fig:openmp} (left)). The cost of the ATOMIC limits the scalability of the assembly. This strategy is
shown in Algorithm \ref{alg:no_Coloring}.
\begin{algorithm}[h!tbp]
 \begin{algorithmic}[1]
   \State \verb+!$OMP PARALLEL DO &+
   \State \verb+!$OMP SCHEDULE (DYNAMIC,Chunk_Size) &+
   \State \verb+!$OMP PRIVATE (...) &+
   \State \verb+!$OMP SHARED  (...)  +      
   \For   {elements $e$}
   \State Compute element matrix and RHS: $\matra^e$, $\vectb^e$
   \State \verb+!$OMP ATOMIC+ \label{alg:atomic1}
   \State Assemble matrix $\matra^e$ into $\matra$
   \State \verb+!$OMP ATOMIC+ \label{alg:atomic2}
   \State Assemble RHS    $\vectb^e$ into $\vectb$
   \EndFor
 \end{algorithmic}
\caption{Matrix Assembly without coloring in each MPI partition}
\label{alg:no_Coloring} 
\end{algorithm}

In the context of vectorization, the element coloring technique has been proposed \cite{Farhat89,MISRA1992131}. By coloring elements such that elements with the same color do not share 
nodes, no ATOMIC is required to protect $\matra$ and $\vectb$. The main drawback of this method is that any spatial locality of data is lost which implies a low IPC (instructions per
cycle). In the performance evaluation based in hardware counters included in Section~\ref{sec:results}, we will show that the loss
in IPC is up to 100\% due to the ATOMIC pragma, while it is of 50\% using the coloring technique respect using a pure MPI version.
This strategy is shown in Algorithm \ref{alg:Coloring}. To unify the terminology with other
techniques, we define a subdomain as a set of elements of the same color, and $nsubd$ the total number of subdomains,
that is the total number of colors.
\begin{algorithm}[h!tbp]
  \begin{algorithmic}[1]
    \State Partition local mesh in $nsubd$ subdomains using a coloring strategy 
    \For   {$isubd = 1,nsubd$}
    \State \verb+!$OMP PARALLEL DO &+
    \State \verb+!$OMP SCHEDULE (DYNAMIC,Chunk_Size) &+
    \State \verb+!$OMP PRIVATE (...) &+
    \State \verb+!$OMP SHARED  (...)  +      
    \For   {elements $e$ in $isubd$}
    \State Compute element matrix and RHS: $\matra^e$, $\vectb^e$
    \State Assemble matrix $\matra^e$ into $\matra$
    \State Assemble RHS    $\vectb^e$ into $\vectb$
    \EndFor
    \State \verb+!$OMP END PARALLEL DO +    
    \EndFor
  \end{algorithmic}
 \caption{Matrix assembly with coloring in each MPI partition}
 \label{alg:Coloring} 
 \end{algorithm} 
 
 To circumvent these two inconveniences, local partitioning techniques (in each MPI partition independently)
 have been proposed \cite{NME:NME2973,thebault13}.
 Here, classical partitioners like METIS or Space Filling Curve based partitioners can be used.
Elements are assigned to subdomains, and subdomains are unconnected
through separators (layer of elements) such that elements of neighboring subdomains do not share nodes.
By assigning elements to a subdomain, the loop over elements is substituted by the parallelization of the loop
over subdomains. This techniques guarantees spatial locality and avoids the race condition. However,
this force us to treat the separators differently (e.g. by re-decomposition) and makes its implementation
more complex. The algorithm is shown in Algorithm \ref{alg:subd}.
\begin{algorithm}[h!tbp]
  \begin{algorithmic}[1]
    \State Partition local mesh in $nsubd$ subdomains using METIS 
    \For   {$isubd = 1,nsubd$}
    \State \verb+!$OMP PARALLEL DO &+
    \State \verb+!$OMP SCHEDULE (DYNAMIC,Chunk_Size) &+
    \State \verb+!$OMP PRIVATE (...) &+
    \State \verb+!$OMP SHARED  (...)  +      
    \For   {elements $e$ in $isubd$}
    \State Compute element matrix and RHS: $\matra^e$, $\vectb^e$
    \State Assemble matrix $\matra^e$ into $\matra$
    \State Assemble RHS    $\vectb^e$ into $\vectb$
    \EndFor
    \State \verb+!$OMP END PARALLEL DO +    
    \EndFor
    \State Treat separators   
  \end{algorithmic}
 \caption{Matrix assembly with local partitioning in each MPI partition}
 \label{alg:subd} 
 \end{algorithm} 
 We can observe the similitude between the loops of the coloring and local partitioning techniques. The differences
 are in the way the subdomains are obtained (coloring \textit{vs} METIS) and the existence of separators
 in the local partitioning technique.

\subsubsection{Task parallelism}
\label{sec:task}

It is possible to implement another strategy by forgoing the loop parallelism
approaches shown above and using a task parallelism approach instead.

As of OpenMP 3.0 a new tasking model was introduced which allows the OpenMP
programmer to parallelize a set of problems with irregular parallelism. When a
thread of the OpenMP program encounters a \texttt{TASK} construct it creates a
task which is then run by one of the threads of the parallel region. In
principle the order, i.e. the schedule, in which the tasks are run is not
determined by the creation order. OpenMP 4.0 allows constraining the schedule
by adding the possibility of defining dependences between tasks. This way the
OpenMP programmer can use a data-flow style for irregular parallelism.

While intuitive, the dependency approach based on input and output dependences
is too strict for a problem like the finite element assembly. It forces the
runtime to determine a particular order (necessarily influenced by the task
creation order) that fulfills the dependences when executing the tasks.

The research in the OmpSs programming model \cite{multideps} led to
the proposal of a new kind of dependency between tasks called
\texttt{COMMUTATIVE}. This new dependency kind means that two tasks cannot be
run concurrently if they refer to the same data object but does not impose any
other restriction in the particular order in which such exclusive execution
happens. This kind of dependency is suitable for our problem as, in principle,
we do not really care which subdomain is processed first as long as two
subdomains that share a border are not processed concurrently.

A further complication exists, though, for the current dependency support in
OpenMP 4.0 implies that the number of dependences is statically defined at
compile time. This is inconvenient as each subdomain may have a variable
number of neighbors. To address this we use the multidependences extensions in which
each task may have a variable number of dependences \cite{multideps}.

\begin{figure}[h!tbp]
 \centering
 \includegraphics[width=0.47\textwidth]{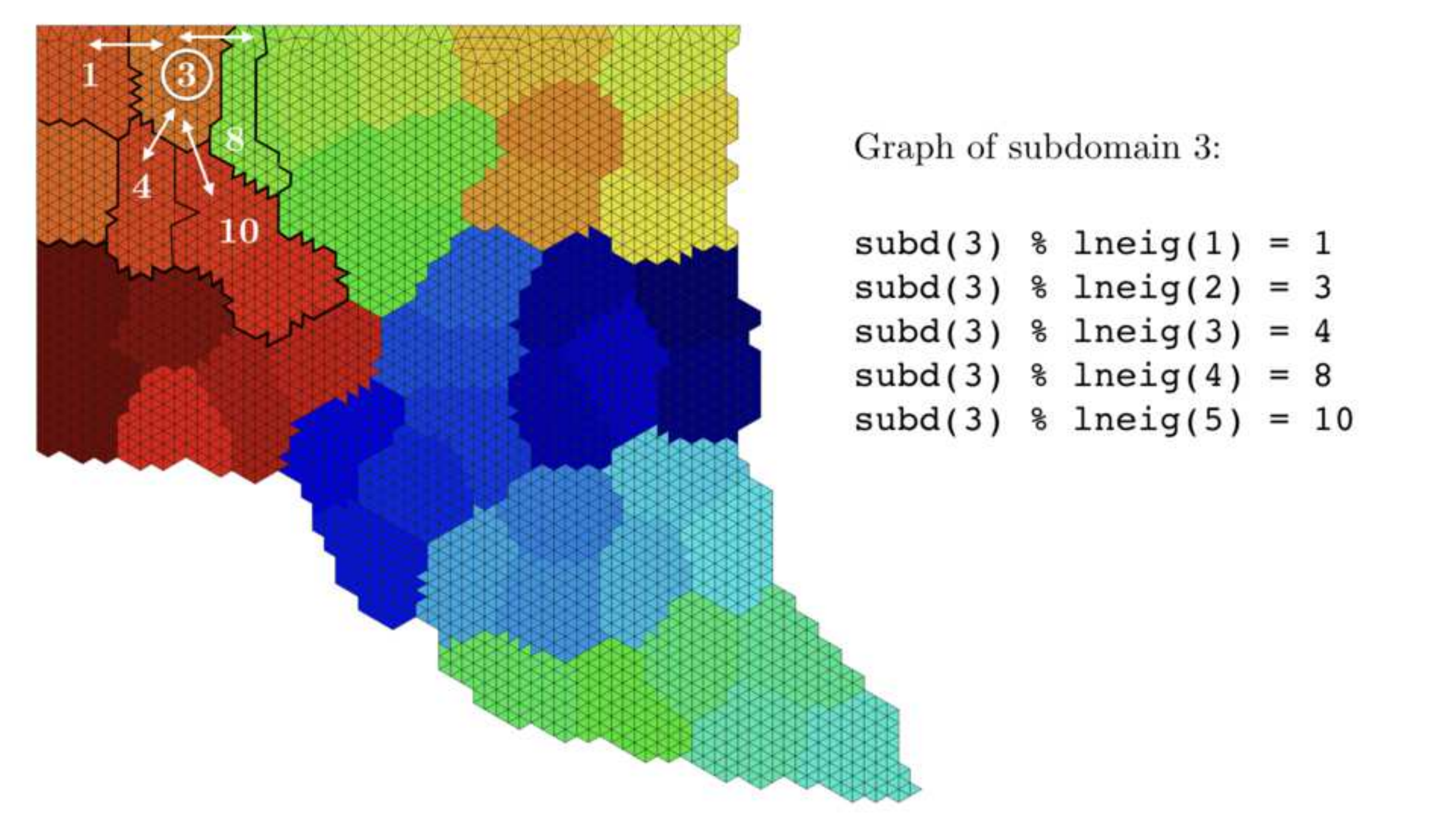}
    \caption{Task parallelism using multidependences: neighbors of subdomain 3 (including itself).}
  \label{fig:ompss}
\end{figure}

In this way, the tasking parallelization is possible by first computing the
adjacency list of each subdomain. Figure~\ref{fig:ompss} depicts this idea,
where the subdomain 3 has 5 neighbors (including itself).  Given that
adjacency list, it is then possible to use a \texttt{COMMUTATIVE} multidependence on
the neighbors. This causes the runtime to run as many subdomains as possible in
parallel that do not share any node.

As subdomains are processed, less of them will remain and the parallelism
available will decrease.  It is possible to get higher concurrency levels if,
of all the non-neighboring subdomains, we first process those with a bigger
number of neighbors: intuitively this potentially can free more subdomains
that are not neighbors. To this end, we prioritize subdomains with a higher
neighbor count.  We can achieve this using the \texttt{PRIORITY} clause.

Algorithm~\ref{alg:multideps} shows the final task parallelization. For each
subdomain: we create a task (step \ref{ite:task}); then we declare a commutative dependency with
all its neighbors (step \ref{ite:commutative}); we prioritize tasks with a higher number of
neighbors (step \ref{ite:priority}). These tasks are created by a single thread inside of a
parallel region (not shown in the listing) and the thread will not proceed
until all of them have been run by itself or other threads of the team (step \ref{ite:task_wait}).

 \begin{algorithm}[h!tbp]
  \begin{algorithmic}[1]
    \State Partition local mesh in $nsubd$ subdomains of size {\tt Chunk\_Size} 
    \State Store connectivity graph of subdomains in structure $subd$
    \For   {$isubd = 1,nsubd$}    
    \State $nneig = SIZE(subd(isubd)\%lneig) $
    \State \verb+!$OMP TASK & +                                            \label{ite:task}
    \State \verb+!$OMP COMMUTATIVE +
    \State \verb+([subd(subd%lneig(i)), i=1,nneig]) & +  \label{ite:commutative}
    \State \verb+!$OMP PRIORITY    (nneig) & +                             \label{ite:priority}
    \State \verb+!$OMP PRIVATE     (...) & +
    \State \verb+!$OMP SHARED      (subd, ...) +
    \For   {Elements $e$ in $isubd$}
    \State Compute element matrix and RHS: $\matra^e$, $\vectb^e$
    \State Assemble matrix $\matra^e$ into $\matra$
    \State Assemble RHS    $\vectb^e$ into $\vectb$
    \EndFor
    \State \verb+!$OMP END TASK +
    \EndFor
    \State \verb+!$OMP TASKWAIT +                                       \label{ite:task_wait}
  \end{algorithmic}
 \caption{Matrix assembly with commutative multidependences in each MPI partition}
 \label{alg:multideps} 
 \end{algorithm} 
 
%
%


\section{Dynamic Load Balancing}
\label{sec:dlb}
\subsection{MPI Load Imbalance}
\label{sec:mpi_lb}

As we explained in Section~\ref{sec:par_mpi}, using an MPI parallelization implies partitioning the
original mesh into $n$ subdomains. The essential characteristic of MPI is that it works on
distributed memory, thus, each MPI process will work on the data in its subdomain. This fact makes
the mesh partitioning crucial, as it will determine the load balance of the execution. Although
there are techniques to redistribute or repartition the mesh during the execution, these
are expensive as they require to move data between processes and to modify the code the code
to repartition when necessary.

The mesh partitioning software provides, in general, load balancing features which necessarily are based on optimizing a metric. As we discussed in the introduction, the main computational tasks of a CFD and
structure mechanics codes are the algebraic system assembly and the iterative solver.

We want to obtain a partition that ensures load balance in the matrix assembly. But in hybrid meshes the
number of elements might not be a good metric to measure the load balance, as their relative
weights in constructing the matrix may be different.

In this paper we focus on the assembly phase, for this reason we want to obtain a partition that
achieves a well balanced distribution of elements among the different MPI processes. In a hybrid
mesh, the computational loads of different elements are not the same. As a intuitive guess,
we will assign as a weight to each element the number of Gauss points used in the matrix assembly.

We define Load Balance as the percentage of time that the computational resources are doing useful
computation:
\begin{EQ}
  \mbox{Load Balance} = \frac{\mbox{Useful CPU time}}{\mbox{Total CPU time}} =\\
  = \frac{\displaystyle\sum_{i=1}^{n}t_i}{n \cdot \max_{i=1}^{n} t_i}
\end{EQ}

Let us define $w_e$ the weight of element $e$ and $n^{i}_e$ the number of elements of 
partition $i$. We define two theoretical load balance (LB) measures: the \textit{non-weighted load balance} which is the
ratio of the average number of elements to the maximum number of elements, as well as the
\textit{weighted load balance}, which represents the same including the weights given to METIS
(that is what METIS load balances). We also introduce the \textit{measured load balance} obtained by measuring 
the elapsed time (${\rm time}_i$) of each MPI task in the assembly or subgrid scale loop and dividing the average by the maximum of 
the elapsed times. 

\begin{EQ}[llll]
  &\mbox{Theo. weighted LB} &=&
  \frac{(\sum_i^{n_{\rm MPI}} \sum_e^{n^i_e} w_e ) / n_{\rm MPI} }{\mbox{max}_i ( \sum_e^{n^i_e} w_e)}, \\
  &\mbox{Theo. non-weighted LB} &=&
  \frac{( \sum_i^{n_{\rm MPI}} n^i_e) / n_{\rm MPI} }{\mbox{max}_i n^i_e}, \\
  &\mbox{Measured LB} &=&
  \frac{( \sum_i^{n_{\rm MPI}} {time}_i) / n_{\rm MPI} }{\mbox{max}_i ({time}_i)}.
\end{EQ}



\begin{figure}[h!tb]
 \centering
     \includegraphics[width=0.45\textwidth]{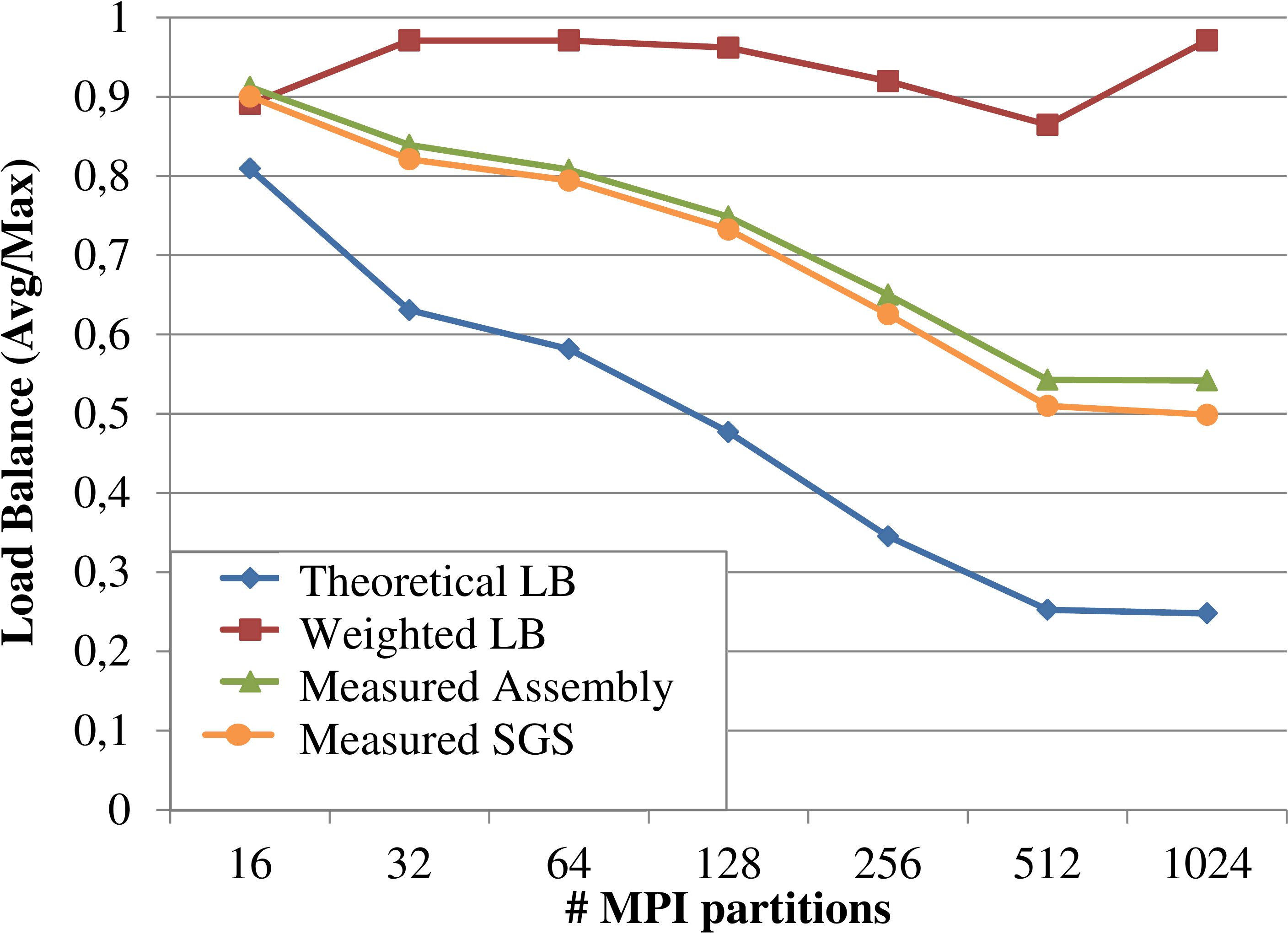}  
 \caption{Load Balance for the Respiratory system.} 
 \label{fig:lb_resp}
\end{figure}

\begin{figure}[h!tbp]
 \centering
     \includegraphics[width=0.45\textwidth]{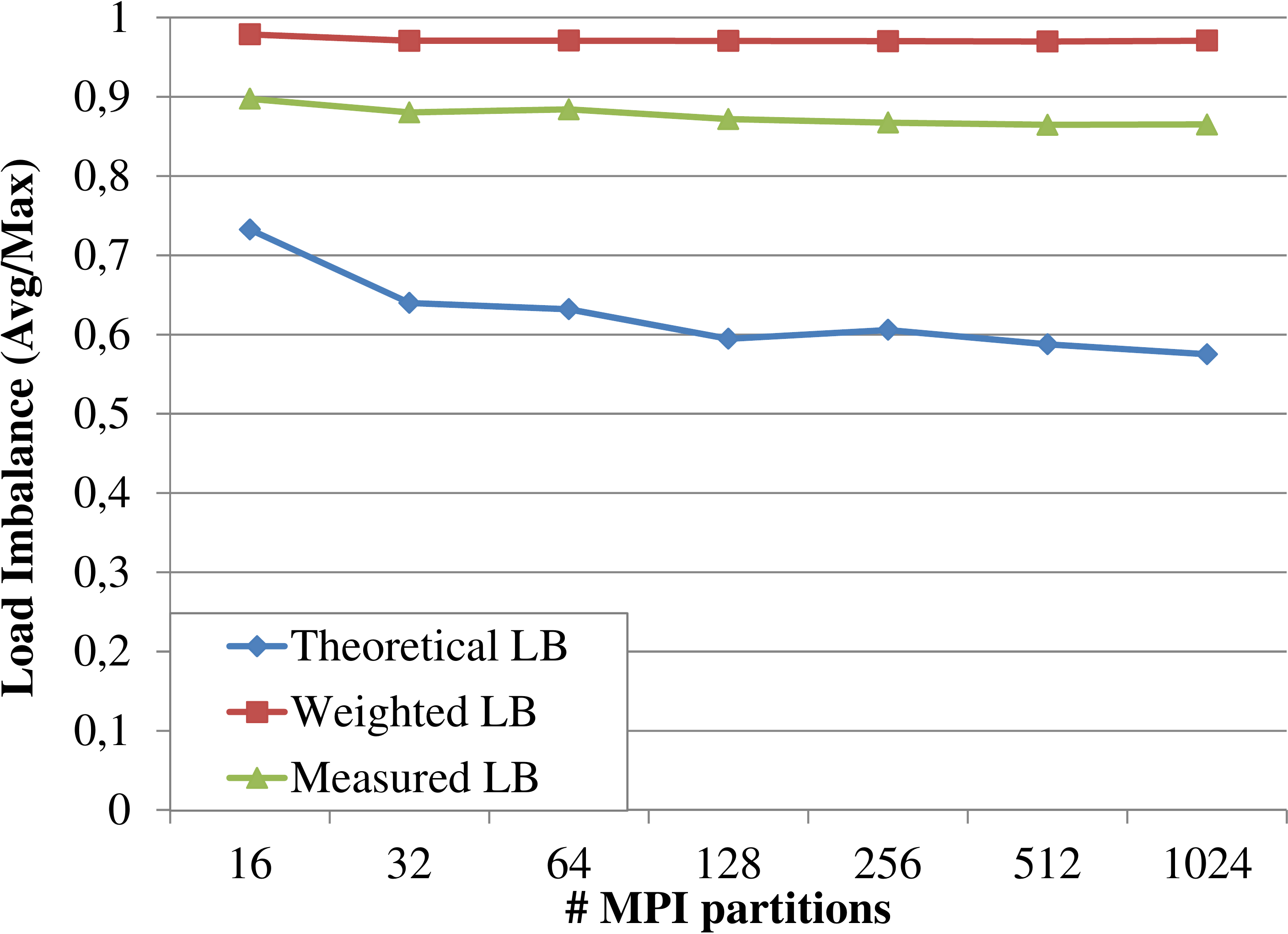}  
 \caption{Load Balance for the Iter simulation.} 
 \label{fig:lb_iter}
\end{figure}

Figures~\ref{fig:lb_resp} and \ref{fig:lb_iter} show the load balance measured for the two use cases presented in
the previous section. The X axis represents the number of MPI partitions used for each simulation. The
different values have been computed using the formulas presented above and the \textit{Measured LB}
has been measured from an execution of the simulation and averaged over 10 time steps.

We first observe that METIS provides a fairly good load balance based on the heuristic provided for
both meshes (weighted LB), even though an imbalance up to $14\%$ is observed for 512 partitions in the respiratory
system. We can say that the theoretical load balance achieved by METIS depends on the mesh structure
and the number of partitions, more partitions tend to higher load imbalances, we can observe this
specially in the case of the respiratory system.

But if we compare this theoretical result with the measured one, we can see a significant
difference. In practice the load imbalance increases with the number of partitions, specially for
the Respiratory system, for both the assembly and for the subgrid scale element loops (which are
quite similar). We conclude that the number of Gauss points is absolutely not a good measure of the
work per element. Indeed, the measured load balance follows the non-weighted load balance.
On the contrary, the load balance of the Iter simulation seems to follow the weighted load balance.
This means that the same heuristic cannot be used to obtain a well balanced partition of different
problems.
\\

\begin{figure}[h!tb]
  \centering
  \subfigure[Partitioning.]{
    \includegraphics[width=0.47\textwidth]{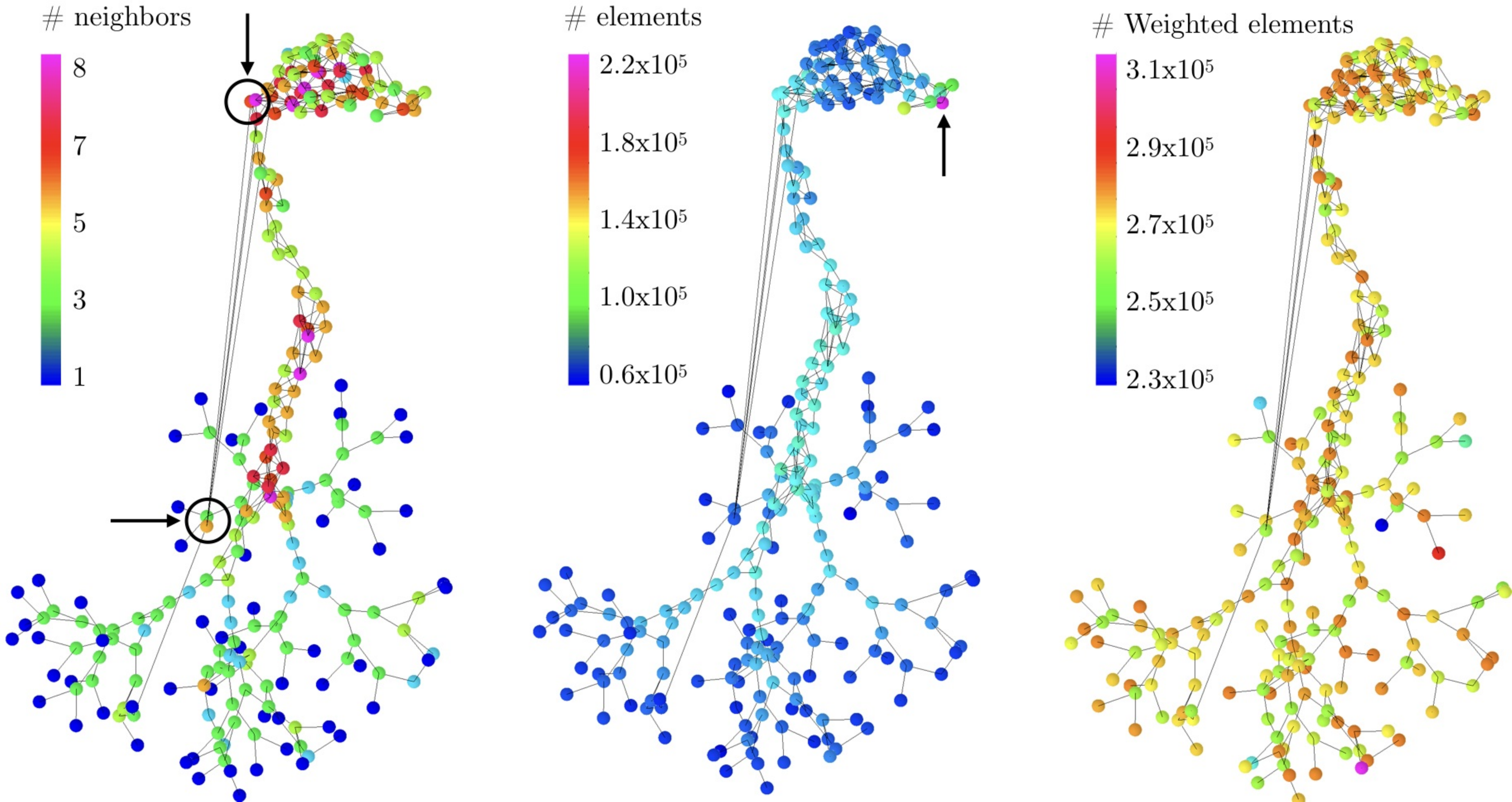} 
    \label{fig:gid_resp1}
  }\\
  \subfigure[Trace.]{
    \includegraphics[width=0.47\textwidth]{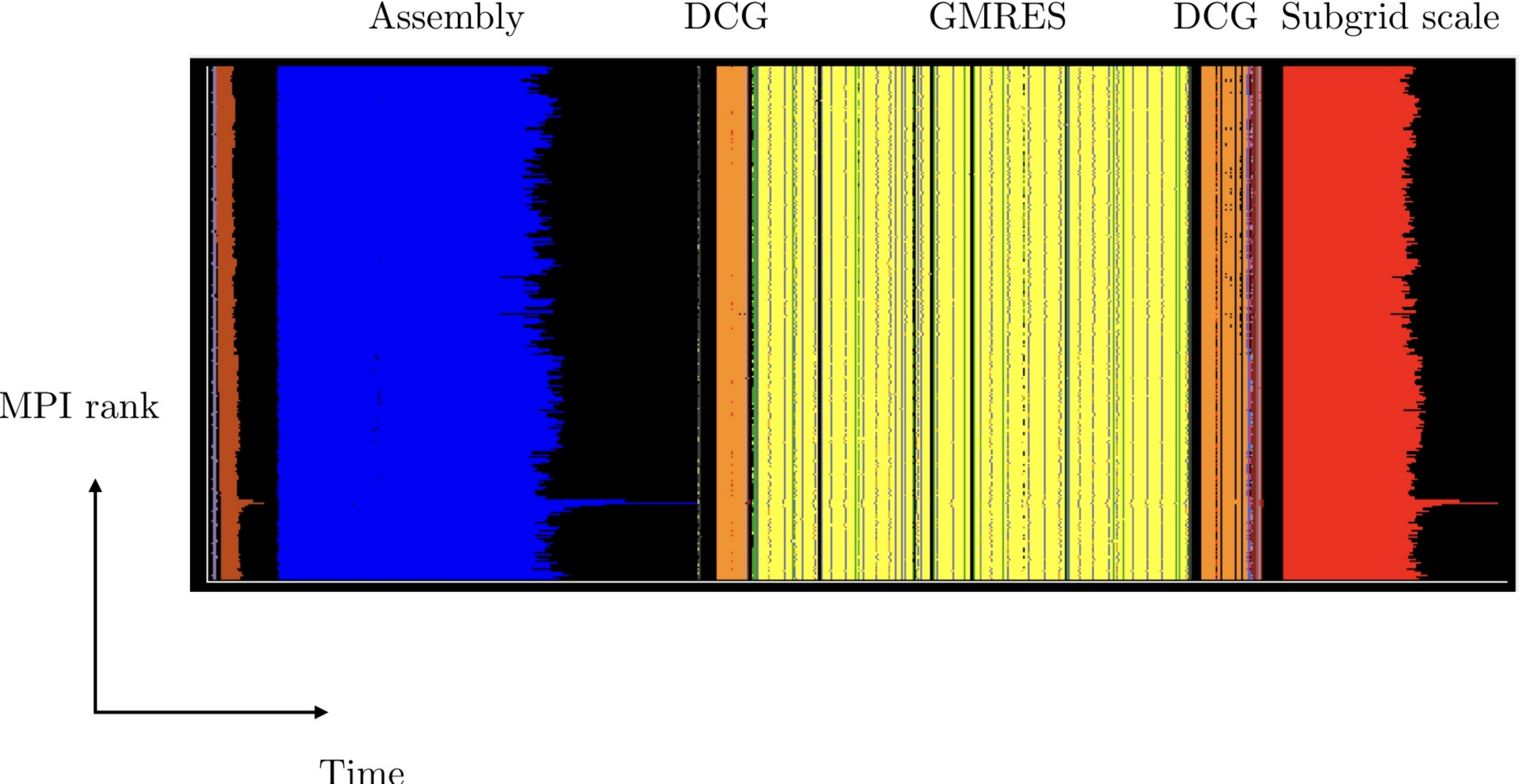}
    \label{fig:gid_resp2}
  }
  \caption{Respiratory system simulaiton on 256 CPUs.}
\end{figure}

\begin{figure}[h!tb]
  \centering
  \subfigure[Partitioning.]{
    \includegraphics[width=0.47\textwidth]{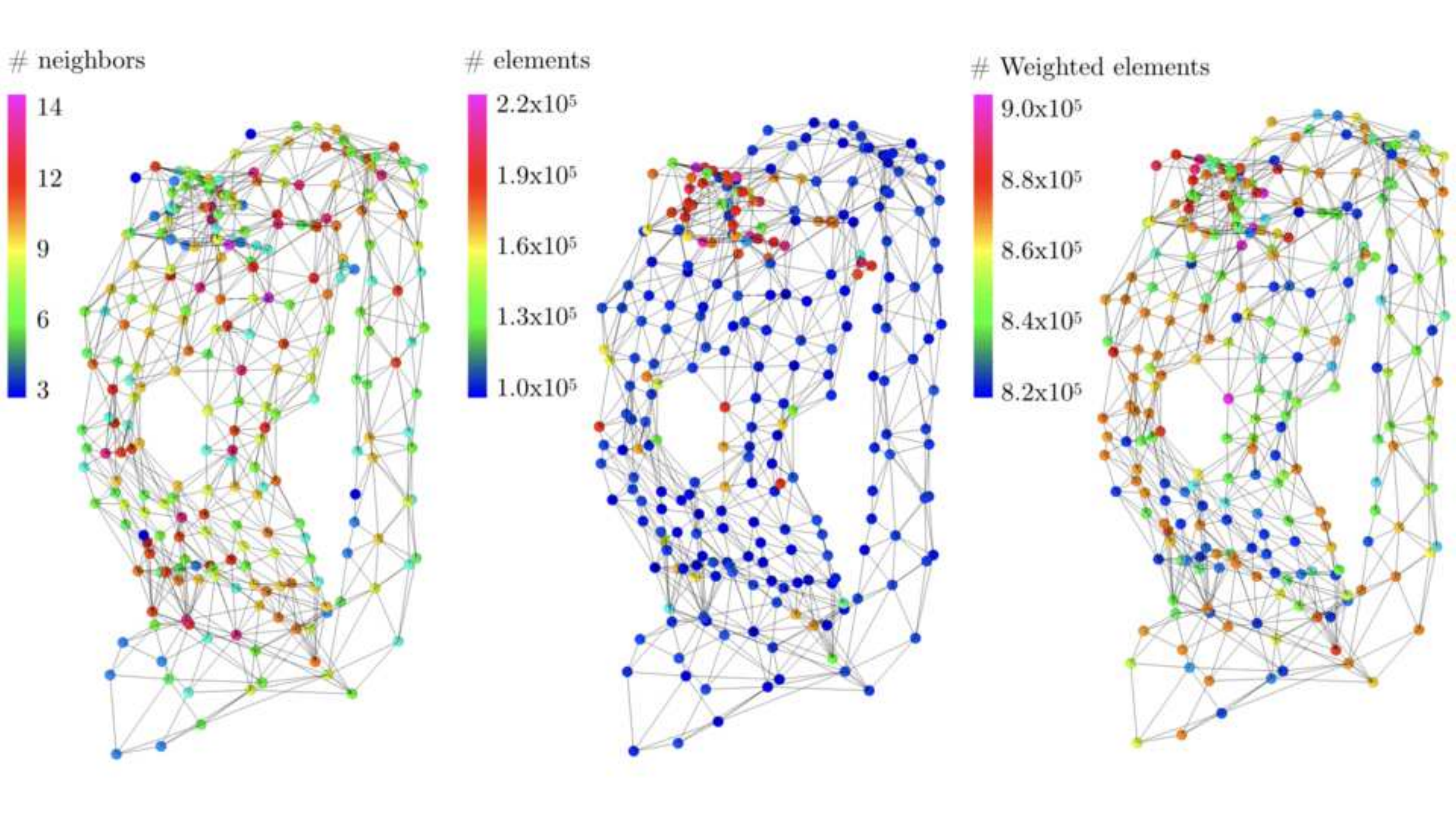} 
    \label{fig:gid_iter1}
  }\\
  \subfigure[Trace.]{
    \includegraphics[width=0.47\textwidth]{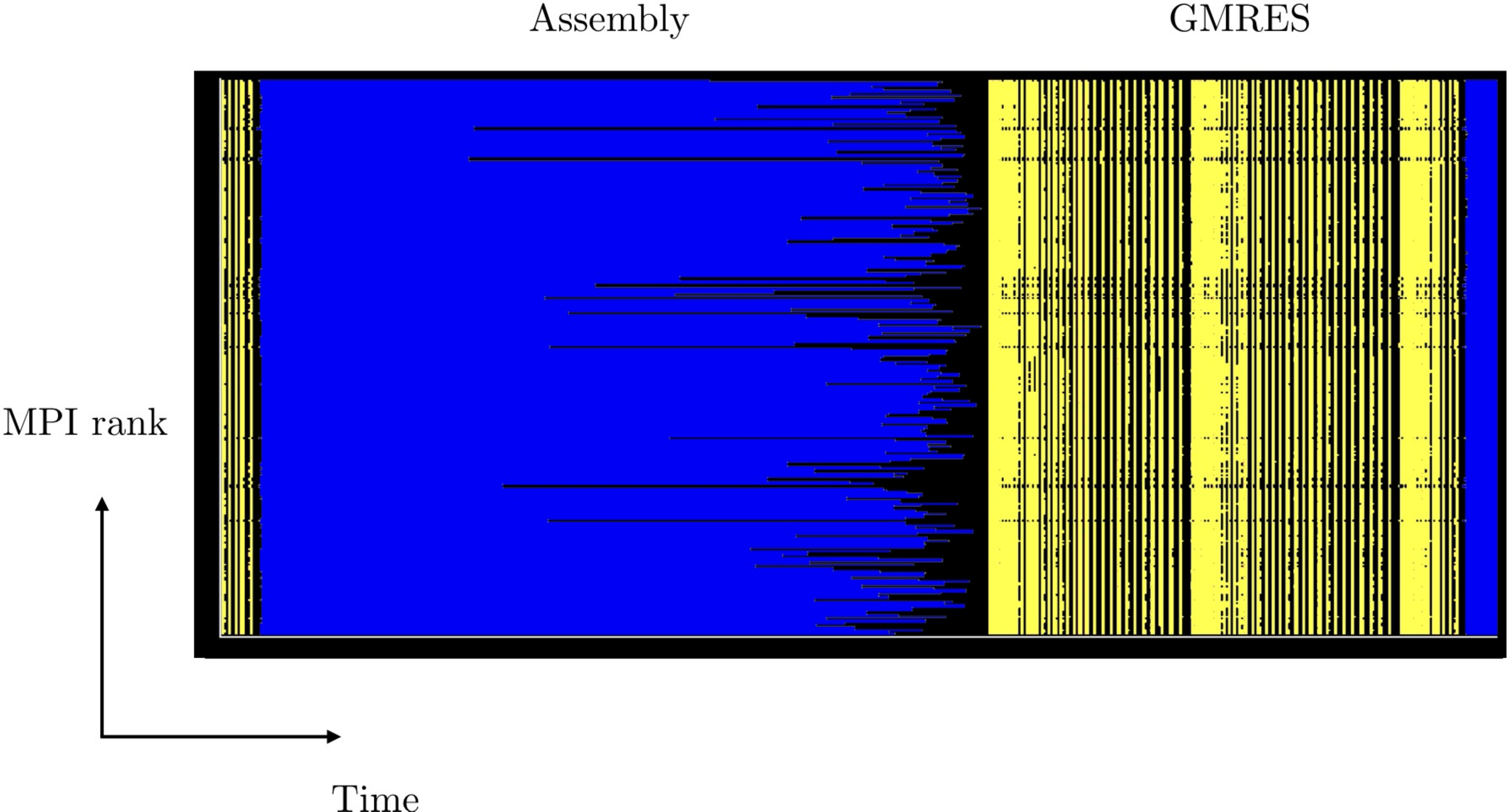}
    \label{fig:gid_iter2}
  }
  \caption{Iter simulation on 256 CPUs.}
\end{figure}

Finally, let us take a look at typical partitions and traces.
Figures \ref{fig:gid_resp1} and \ref{fig:gid_iter1} shows some statistics of the partitions
for the fluid and solid problems using 256 CPUs. The nodes are placed at the centers of gravity of the MPI
subdomains, while the edges represent neighboring relations. We observe that in the case of the
respiratory system, METIS happens partitions some MPI subdomains into non connected parts (identified
by the arrow). We also observe that one subdomain has much more elements than the others (identified
with an arrow in the middle figure). This is the subdomain located at front of the
face (see Figure \ref{fig:resp_mesh}), which is exclusively composed of tetrahedra.
Tetrahedra have less Gauss points and thus METIS admits more elements than in average.

The associated traces are shown in Figures \ref{fig:gid_resp2} and \ref{fig:gid_iter2}.
In the case of the Respirartory system, we can easily identify the subdomains responsible
for the load imbalance, near the bottom of the trace. These are the subdomains 
mentioned previously, in the front face region. This is because the element weight based on
the Gauss points given to METIS is not a good metric.

In the case of the Iter simulation, we can observe that we have some subdomains with much less
work than others. Once more, this indicates that the weights based on Gauss points is not
a good heuristic for load, although it affects much less the load balance than for the fluid simulation.

\subsection{DLB library}

Load imbalance is a concern that has been targeted since the beginning of parallel programming. In the literature, 
we can see that it has been attacked from very different points of view (data partition, data redistribution, 
resource migration, etc.). 

In this case we are using METIS to partition the mesh and obtain a balanced distribution among MPI tasks. But as we 
have seen in the previous section, the actual load balance obtained is far from optimal. There are several reasons for this, 
the geometry of the mesh and the weight of the elements given to METIS. 

Additionally, the algorithm or the physics (or both together) could produce very strong work imbalance by increasing
the computing needs locally (i.e. particle concentrations \cite{houzeaux16}, solid mechanics fracture, shock in compressible flows, etc.) 
For these reasons, we opt for a dynamic approach applied at runtime, with no need for an a-priori imbalance analysis. 

In this work we will use DLB~\cite{LeWI} (Dynamic Load Balancing Library). The DLB library aims at balancing MPI applications using a second level of parallelism (i.e. Hybrid parallelization MPI+OpenMP). 
Currently, the implemented modules balance hybrid MPI + OpenMP and MPI + OmpSs applications, where MPI is the outer level of parallelism and OpenMP or OmpSs are the inner ones. 

An important feature of the DLB library is that a runtime interposition technique is used to intercept MPI calls. 
With this technique we do not need to modify the application, the DLB library is loaded dynamically when running the 
application to load balance the execution.

The DLB library will reassign the computational resources (i.e. cores) of an MPI process waiting in an MPI blocking call, to another MPI process running on the same node that it is still doing computation.

\begin{figure}[h!bt]
\centering
\includegraphics[width=0.95\columnwidth,keepaspectratio]{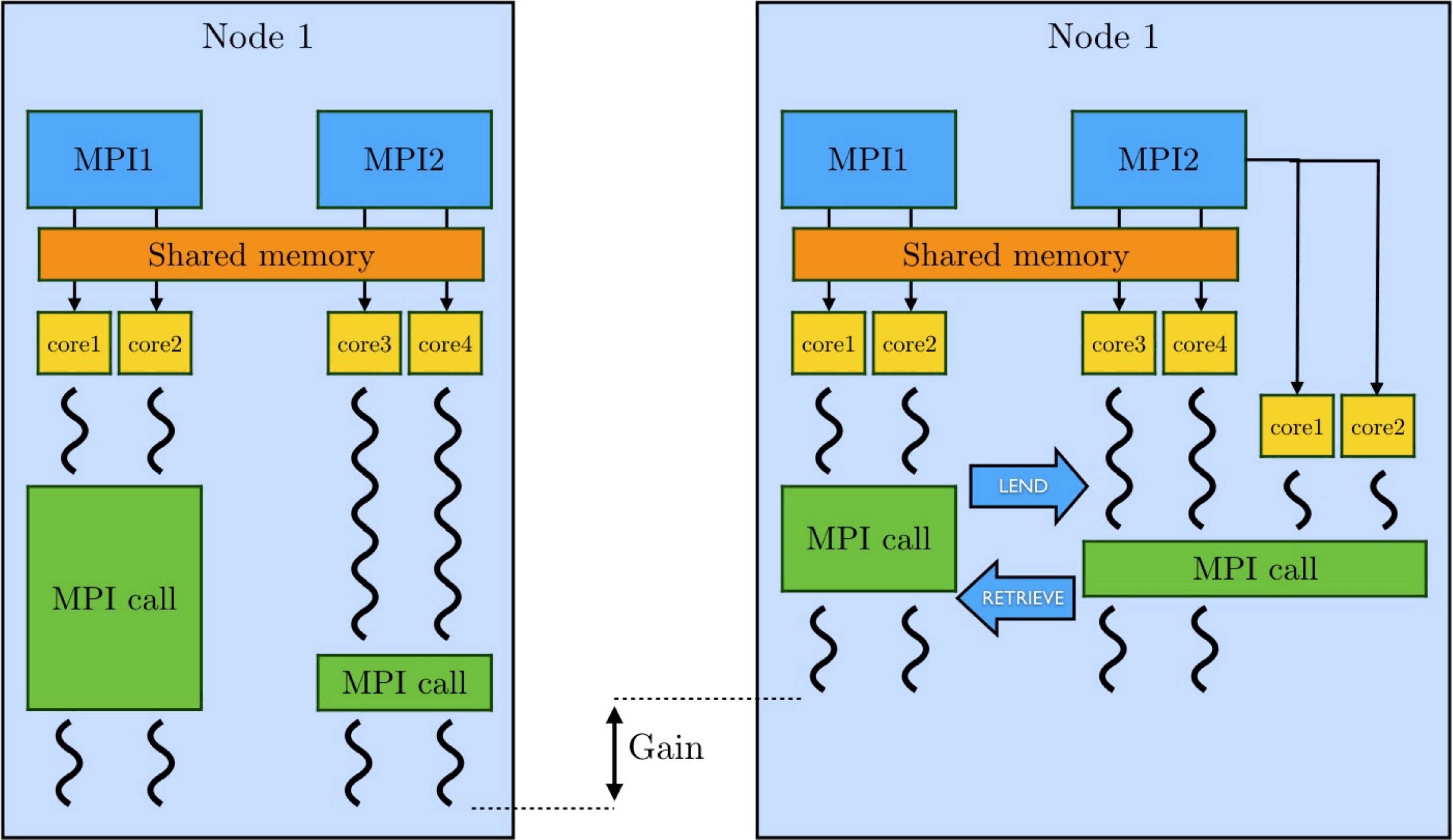}
\caption{(Left)  Unbalanced MPI+OpenMP application.
         (Right) Unbalanced MPI+OpenMP application balanced with DLB.}
\label{fig:lewi}
\end{figure} 

Figure~\ref{fig:lewi} illustrates the load balancing process. In the example, the application is running on a node with 
4 cores. Two MPI tasks are started on the same node, and each MPI task
spawns 2 OpenMP threads (represented by the wavy lines). Eventually, an
MPI blocking operation (in green) synchronizes the
execution. Regarding the assembly process, the wavy lines represent
the element loops and the MPI call represents the first MPI call in
the iterative solvers (namely the initial residual of the algebraic system).

On the one hand,
Figure~\ref{fig:lewi} (Left) shows the behavior of an unbalanced application where the excessive work of the threads running on core3 and core4
delays the execution of the MPI call. On the other side,
Figure~\ref{fig:lewi} (Right) shows the execution of the same application with the DLB library. We can see that 
when the MPI task 1 gets into the blocking call it will lend its two cores to the MPI task 2. 
The second MPI task will use the newly acquired cores and will be able to run with 4 threads. This will allow to finish the remaining computation faster. When the MPI 
task 1 gets out of the blocking call it retrieves its cores from the MPI task 2 and the execution continues 
with a core equipartition, until another blocking call is met.

The fact that DLB relies on the shared memory to load balance the different MPI tasks means that it needs to run more than one MPI task per node. In current many-cores architectures this is the normal trend

The dynamic load balancing algorithm illustrated previously relies on the OpenMP parallelization.
One important characteristic of this strategy is that {\it not} the whole application needs to be
fully parallelized, as the second level of parallelism can be introduced only for load balancing
purposes, in the main imbalanced loops of the code.

\section{Performance Evaluation}
\label{sec:results}

\subsection{Environment and Methodology}

All the experiments presented in Section \ref{sec:fluid} have been executed on MareNostrum 3
supercomputer. Each node of MareNostrum 3 is composed of two Intel Xeon processors (E5-2670), 
each of this two sockets includes 8 cores and 16 GB of main memory. In total each compute node has
16 cores with 32GB of main memory.

We have used the Intel MPI library version 4.1.3 and as the underlying Fortran compiler Intel
13.0.1. For OpenMP we have used Nanos 0.12 \cite{ompss-web}\cite{ompss} with the source to source
compiler Mercurium 2.0~\cite{Mercurium}. For the dynamic load balancing we have used the DLB library
1.3.

We have executed the experiments on 16, 32 and 64 nodes of MareNostrum 3 that correspond to 256, 512
and 1024 cores, respectively. For each experiment we will consider 5 different configurations of MPI
processes and threads inside the node:
\begin{itemize}
\item \textbf{1x16:} 1 MPI process with 16 threads, this is the pure hybrid approach, where MPI is
  used across nodes and OpenMP/OmpSs inside the shared memory node. In this configuration, DLB
  cannot load balance, but we show it for completeness.
\item \textbf{2x8:} 2 MPI processes with 8 threads each. This is another typical configuration
  when running in nodes with two sockets, each MPI process is mapped to one socket, and 8 threads
  are spawn on each socket.
\item \textbf{4x4:} 4 MPI processes with 4 threads each.
\item \textbf{8x2:} 8 MPI processes with 2 threads each.
\item \textbf{16x1:} 16 MPI processes with 1 thread each. In this case, the shared memory level is
  only used for load balancing. This configuration is useful when the application is not fully
  parallelized with OpenMP/OmpSs and it is launched as a pure MPI application.    
\item \textbf{Pure MPI:} As a reference we show the performance of the pure MPI version of the
  application, in this case 16 MPI processes are launched on each node.
\end{itemize}

For each experiment we will execute different versions, in Table~\ref{tab:methods} we present
detailed summary of each data series that we will see in the charts. 
\begin{table*}[htb]
  \centering
  \caption{Different methodologies tested.}
  \begin{tabular}{llllll} 
    \hline
    Hybrid          & Programming   & Shared mem.             & Dynamic         & Algorithm                          & Race condition      \\
    method          & model         & parallelism             & load balance    &                                    & treatment           \\
    \hline           
    No coloring     & MPI + OpenMP  & Loop                    & No              & Alg. \ref{alg:no_Coloring}         & ATOMIC              \\
    Coloring        & MPI + OpenMP  & Loop                    & No              & Alg. \ref{alg:Coloring}            & Coloring technique  \\
    Multideps       & MPI + OpenMP  & Task                    & No              & Alg. \ref{alg:multideps}           & Multidependencies   \\
    DLB no coloring & MPI + OpenMP  & Loop                    & DLB             & Alg. \ref{alg:no_Coloring}         & ATOMIC              \\
    DLB coloring    & MPI + OpenMP  & Loop                    & DLB             & Alg. \ref{alg:Coloring}            & Coloring technique  \\
    DLB multideps   & MPI + OpenMP  & Task                    & DLB             & Alg. \ref{alg:multideps}           & Multidependencies   \\
    Pure MPI        & MPI           & None                    & No              & Alg. \ref{alg:sequential_assembly} & Not applicable      \\
    \hline
  \end{tabular}
  \label{tab:methods}
\end{table*}

We have divided the evaluation into four parts:
\begin{itemize}
\item \textbf{Chunk size:} In this section we will study the impact of the chunk size on the
  performance of the different parallelizations and when using DLB. From this evaluation, we will
  try to find the optimum chunk size, and use it for the following experiments.
  
\item \textbf{Execution Time:} In this evaluation we will show the performance obtained by the
  three shared memory parallelization alternatives: No Coloring, Coloring, and Multidependencies,
  in terms of elapsed execution time. We will also see the impact in performance of using the
  dynamic load balancing mechanism.
  
\item \textbf{Hardware Counters:} In this section we demonstrate our hypothesis in the performance
  of each parallelization, based in the different hardware counters collected during the execution.
  
\item \textbf{Scalability:} Finally we will present some scalability tests of the
  \textit{Respiratory system} simulations using up to 16K cores of MareNostrum 3.  
\end{itemize}


\subsection{Chunk Size Study}

In this section, we want to evaluate the impact of the chunk size on the performance of the
different parallelization alternatives and DLB. In OpenMP the chunk size determines the number of iterations (in our test case 
one iteration corresponds to the computation of one element) that are executed sequentially by one thread. When using a parallel loop
approach (both coloring and no coloring) the chunk size is defined using the schedule
clause \texttt{SHEDULE(DYNAMIC, <chunk size>)}. In the multidependencies version the chunk is defined by the number of elements that 
are assigned to each subdomain. In Table \ref{tab:num_tasks} we summarize the number of chunks that
are created in the different scenarios that we are executing, when partitioning the problem in 256,
512 and 1024 MPI partitions. As we have seen in the previous section, for a given partition the
number of elements assigned to each MPI process may be different, for this reason in the table we
show 3 values, the maximum number of chunks, the minimum and the average of all the MPI processes.
\begin{table*}[htb]
  \centering
  \begin{center}
    \begin{tabular}{r|rrr|rrr|rrr}
      &\multicolumn{3}{c|}{256 Partitions}          & \multicolumn{3}{c|}{512 Partitions}  & \multicolumn{3}{c}{1024 Partitions} \\  
      Chunk size &	Max & 	Min & 	Average & 	Max & 	Min & 	Average & 	Max & 	Min & 	Average\\
      \hline
      100	   &	2004&	522 &	691,3   &	1369&	255 &	345,4   &	697 &	101 &	172,5 \\
      500	   &	400 &	104 &	137,8	&	273 &	51  &	68,7    &	139 &	20  &	34,1 \\
      1000	   &	200 &	52  &	68,7	&	136 &	25  &	34,1	&	69  &	10  &	16,8 \\
      2000	   &	100 &	26  &	34,1	&	68  &	12  &	16,8	&	34  &	5   &	8,2\\
      \hline
    \end{tabular}
  \end{center}
  \caption{Respiratory system: number of chunks created for the multidependences version.}
  \label{tab:num_tasks}
\end{table*}

All the experiments have been executed with 16 MPI
processes per node and 1 thread per process. With this configuration, we want to evaluate the impact
of the chunk duration, the amount of chunks in the queue and the impact in the malleability when
using DLB. Using only one thread will avoid seeing the overhead of several threads accessing the
shared structures or invalidating the memory.

\paragraph{Respiratory system simulation}

\begin{figure}[h!tbp]
 \centering
  \subfigure[16 Nodes (256 cores)]{
  \includegraphics[width=0.47\textwidth]{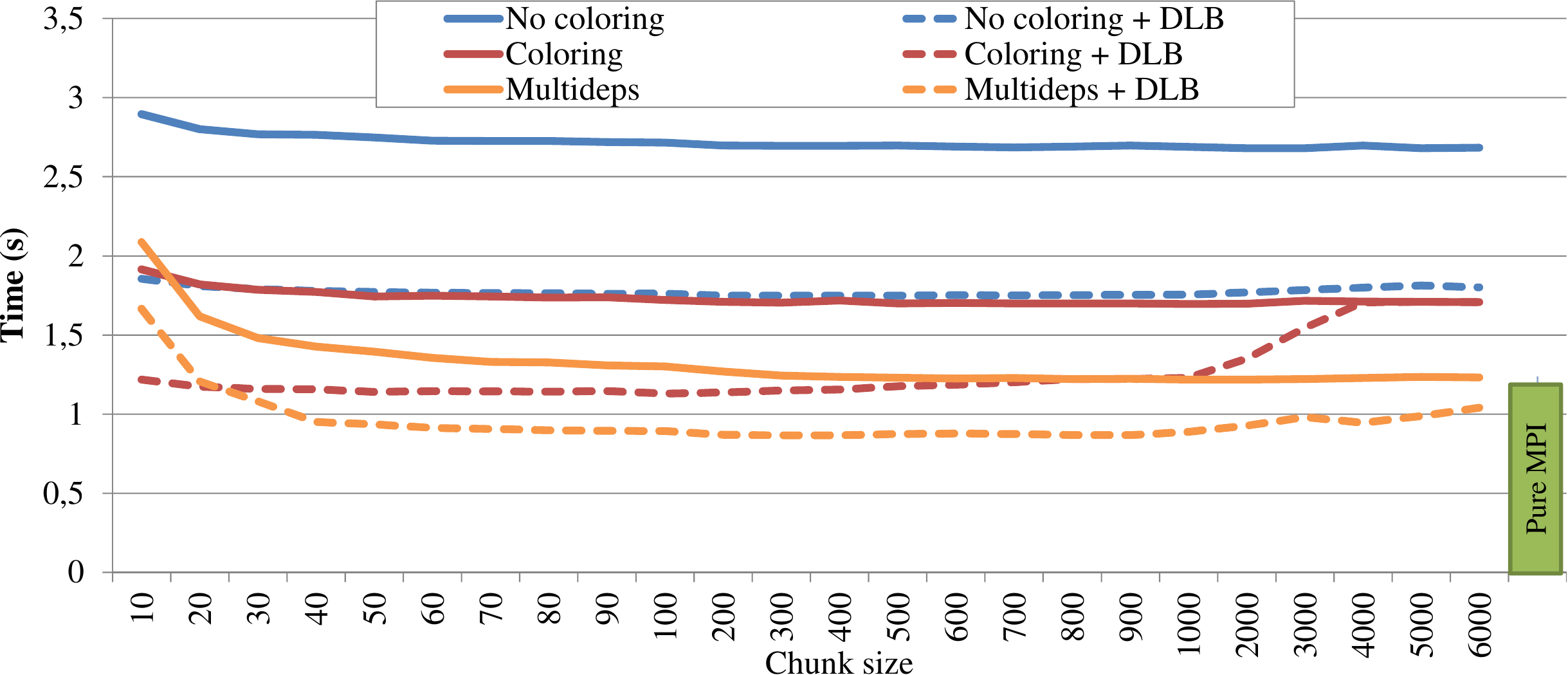}
  \label{fig:resp_16_ass_chunk}
  }
  \subfigure[32 Nodes (512 cores)]{
  \includegraphics[width=0.47\textwidth]{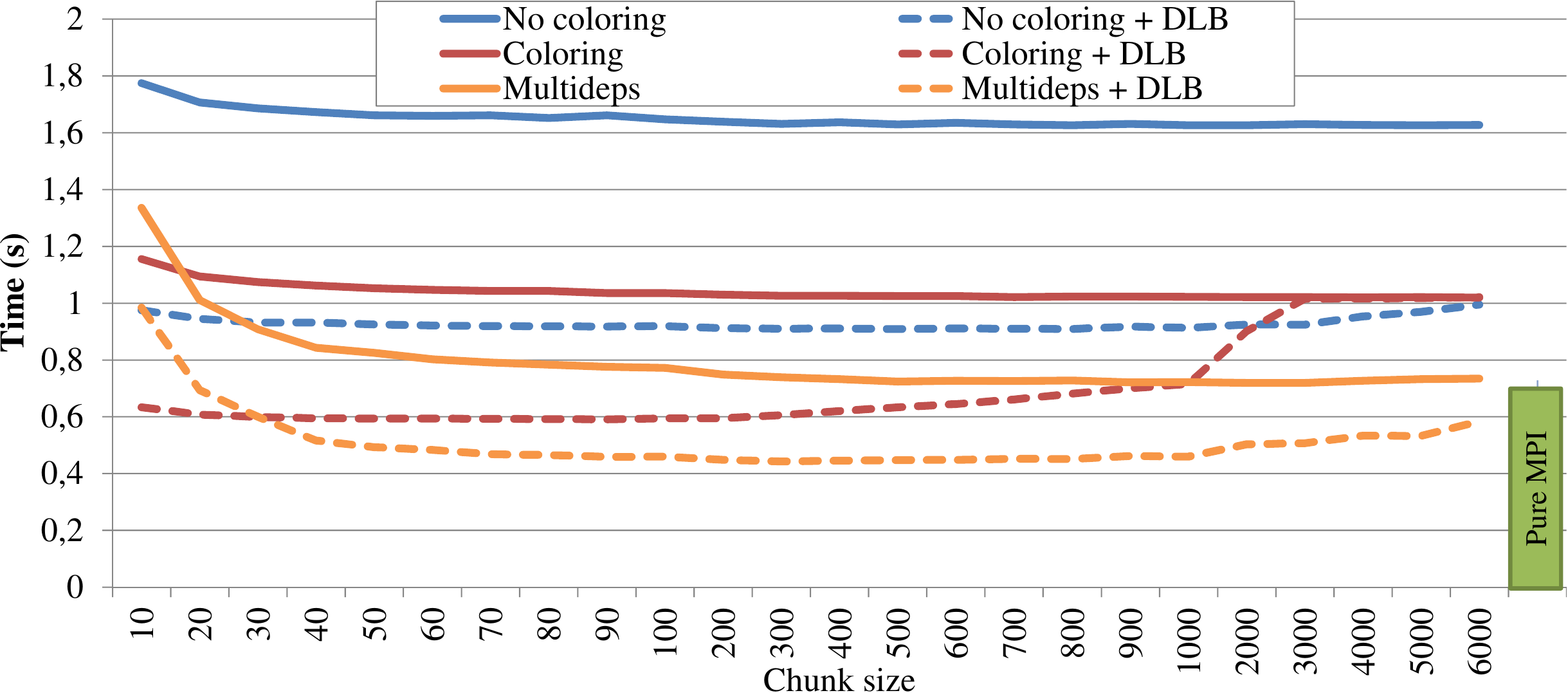}
  \label{fig:resp_32_ass_chunk}
  }
 \subfigure[64 Nodes (1024 cores)]{
  \includegraphics[width=0.47\textwidth]{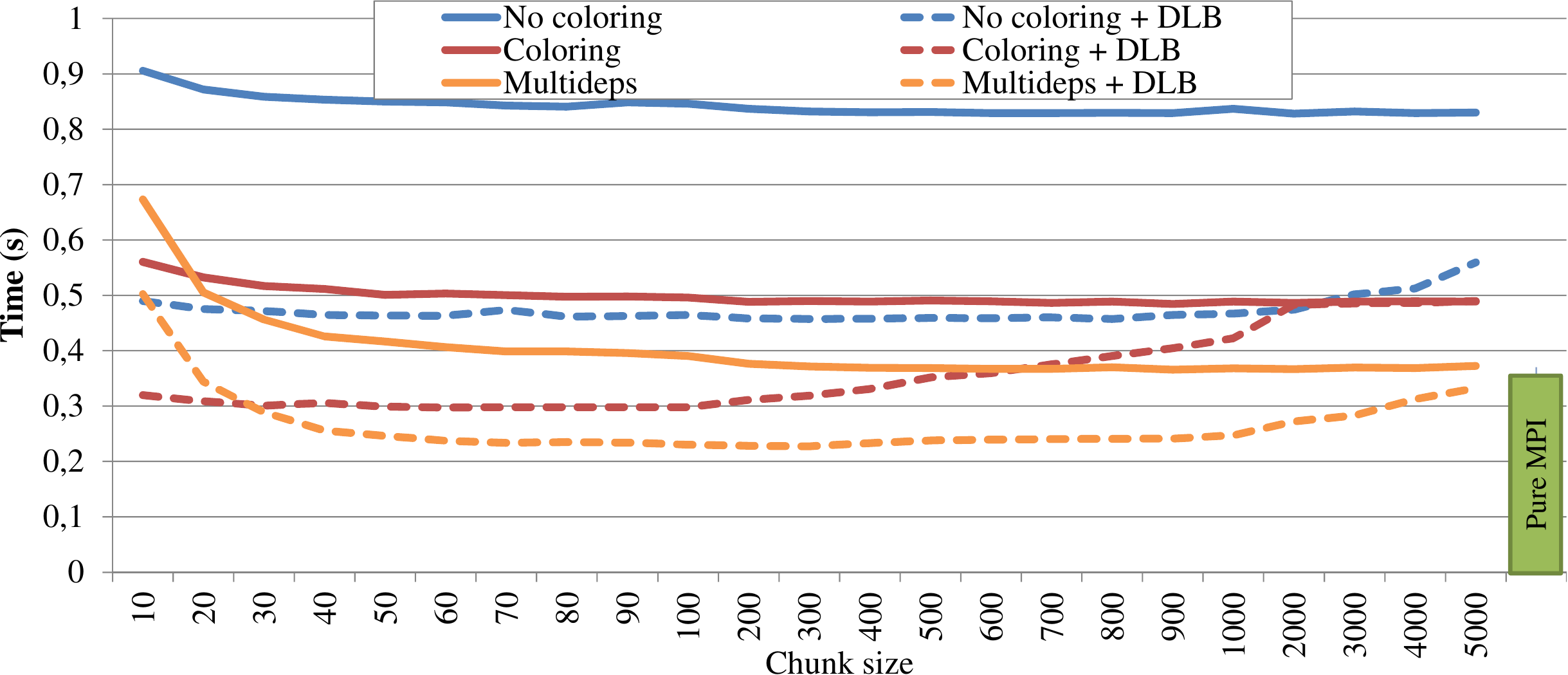}
 \label{fig:resp_64_ass_chunk}
 }
 \caption{Respiratory system, matrix assembly: chunk size impact on execution time.}
 \label{fig:ass_respiratory_chunk}
\end{figure}

In Figure~\ref{fig:ass_respiratory_chunk} we can see the execution time of the matrix assembly in
the Respiratory system simulation. The X axis represents the different chunk sizes
used. Figures \ref{fig:resp_16_ass_chunk}, \ref{fig:resp_32_ass_chunk} and
\ref{fig:resp_64_ass_chunk} correspond to executions on 16, 32 and 64 nodes of MareNostrum 3,
respectively.

In the case of the execution without DLB, the No Coloring and Coloring versions are not affected by
the chunk size, until reaching very small values, chunk sizes of 10 or 20. On the other hand, the
Multidependencies version starts to degrade its performance at chunk sizes between 30 and 60
depending on the number of partitions. This is related to the number of chunks that are created and
their \textit{commutative} relationship. When using bigger chunk sizes the execution time is not
affected for none of the three parallelizations.

The impact of the chunk size when using DLB comes from the fact that bigger chunks imply less
malleability because threads can not leave a chunk until it is finished. The Coloring version with
DLB is the most affected one by big chunks, after chunks of size 200 the performance slowly degrades
until it reaches the same performance as the Coloring version with DLB. This is because the
parallelism of the Coloring version opens and closes for each color, and bigger chunks means fewer
chunks to be distributed among the threads, and therefore, less parallelism. If there is not enough
parallelism when DLB tries to spawn more threads to load balance, there is not enough work for all
of them. In the case of Coloring and Multidependencies using big chunk sizes can limit the
performance obtained by DLB when the number of chunks created in each MPI process is not enough to balance the load.\\

\begin{figure}[h!tb]
 \centering
 \subfigure[16 Nodes (256 cores)]{
  \includegraphics[width=0.47\textwidth]{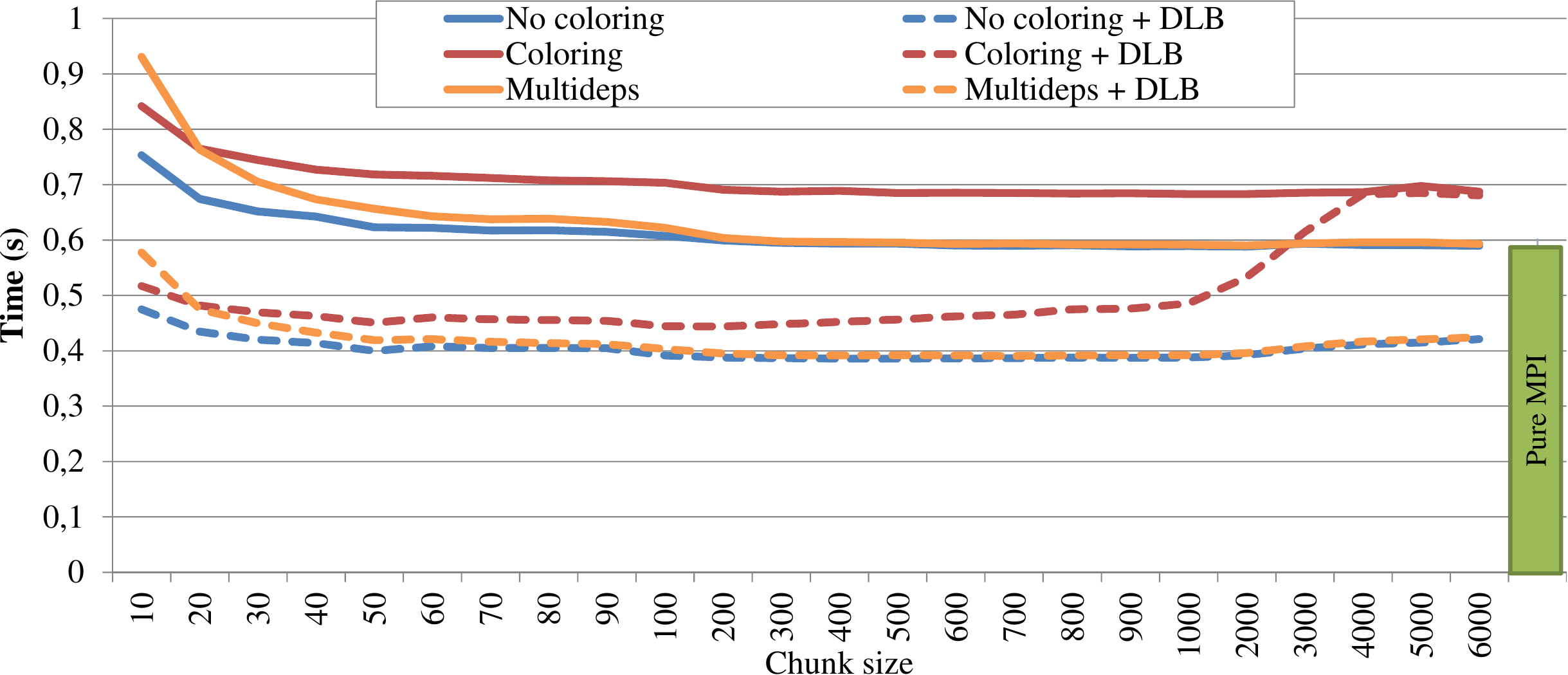}
 \label{fig:resp_16_sgs_chunk}
 }
 \subfigure[32 Nodes (512 cores)]{
  \includegraphics[width=0.47\textwidth]{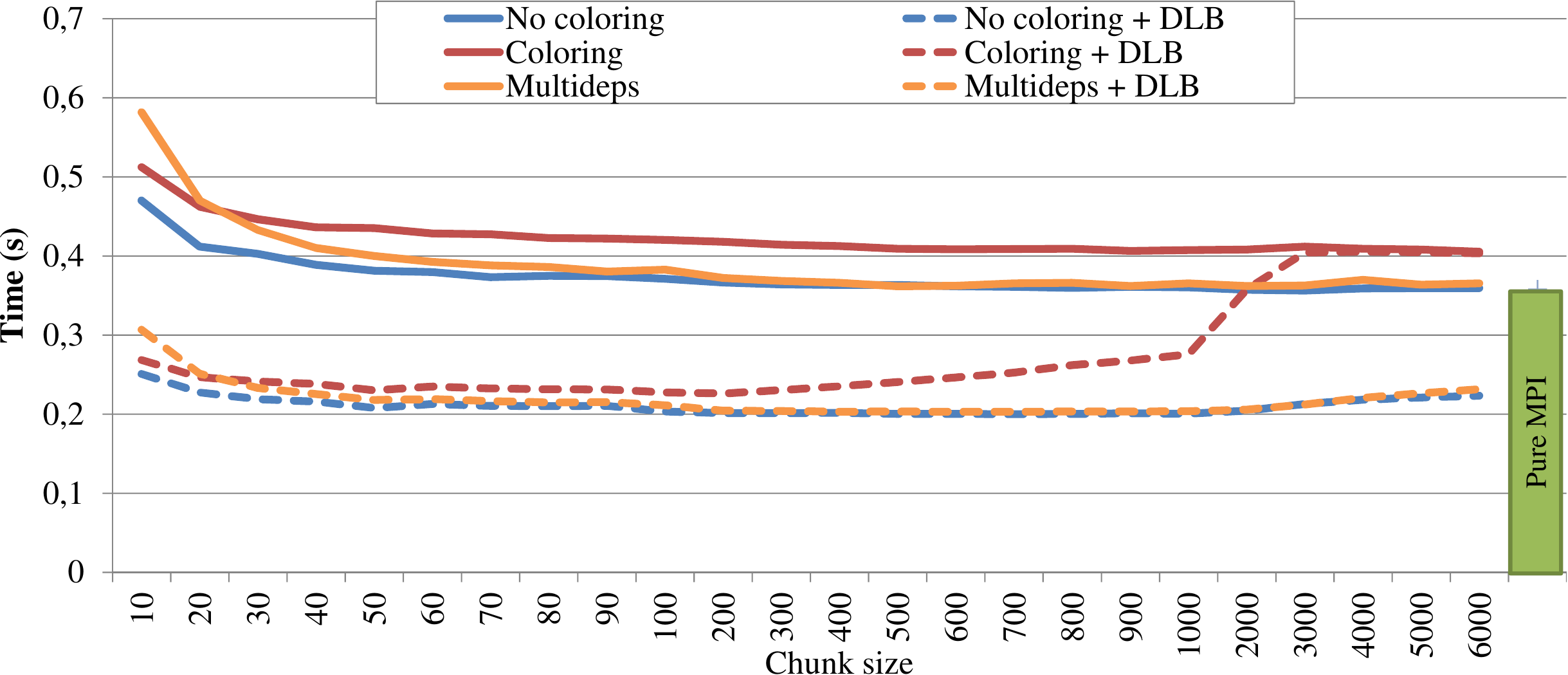}
 \label{fig:resp_32_sgs_chunk}
 }
 \subfigure[64 Nodes (1024 cores)]{
  \includegraphics[width=0.47\textwidth]{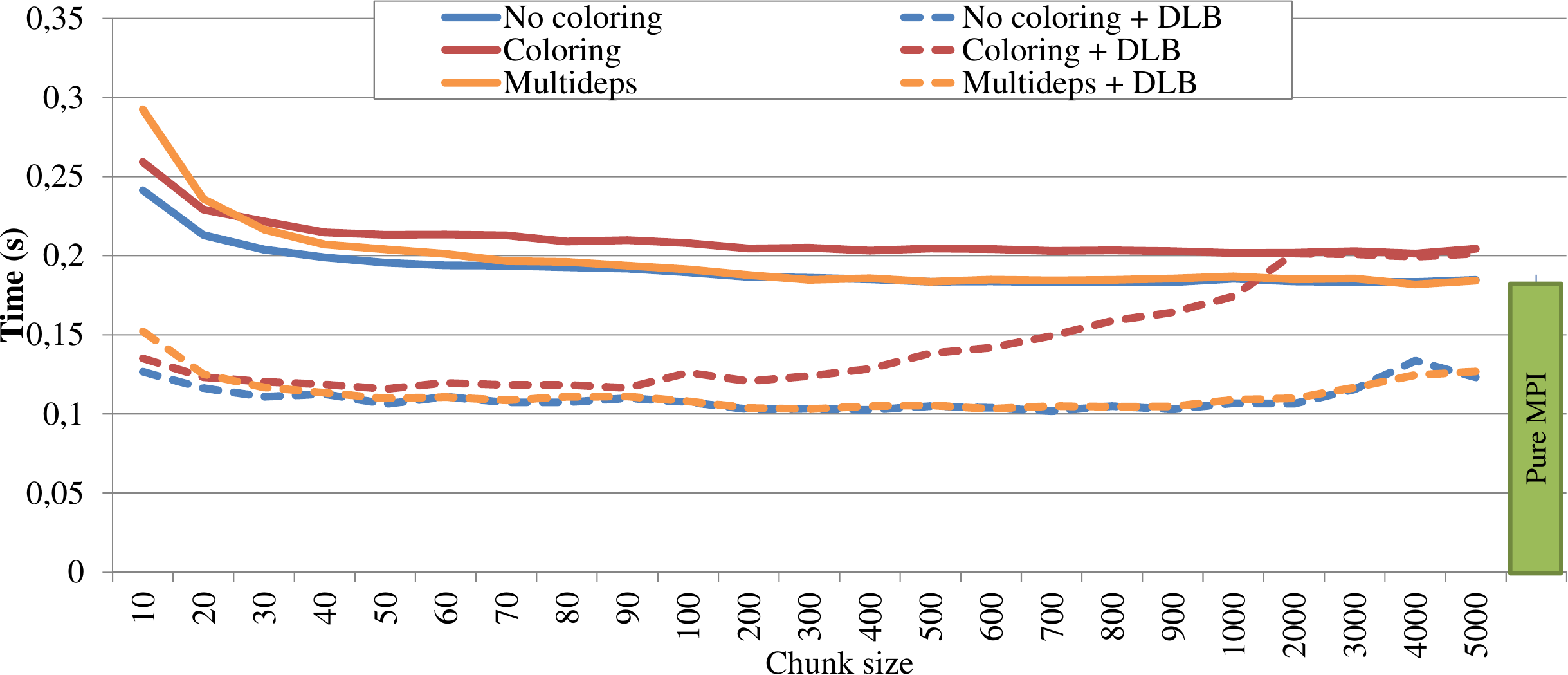}
 \label{fig:resp_64_sgs_chunk}
 }
 \caption{Respiratory system, subgrid scale: chunk size impact on execution time.}
 \label{fig:sgs_respiratory_chunk}
\end{figure}

Figure~\ref{fig:sgs_respiratory_chunk} shows the execution time of the subgrid scale computation
for the Respiratory system simulation, when varying the chunk size.
The conclusions for this experiments are very similar to the previous ones. The limit in the chunk size is around 20, from this size the
performance of the Coloring and No coloring versions degrade. For very small chunks sizes the
overhead of creating the tasks and scheduling them is too high. In this case, the limit chunk size
is bigger because the durations of the tasks are smaller if we compare the total execution time and
considering that the number of elements that are processed in the parallel loop is the same.

Big chunk sizes, like before, affect only when running with DLB. And the worst performance with big
chunks is obtained by the Coloring parallelization. We consider that a chunk size of 200 is the
optimum value for all the versions and configurations. This will be the chunk size used in the
experiments executed in the following sections for the respiratory simulation.

\paragraph{Iter simulation}

\begin{figure}[h!tb]
 \centering
 \subfigure[16 Nodes (256 cores)]{
    \includegraphics[width=0.47\textwidth]{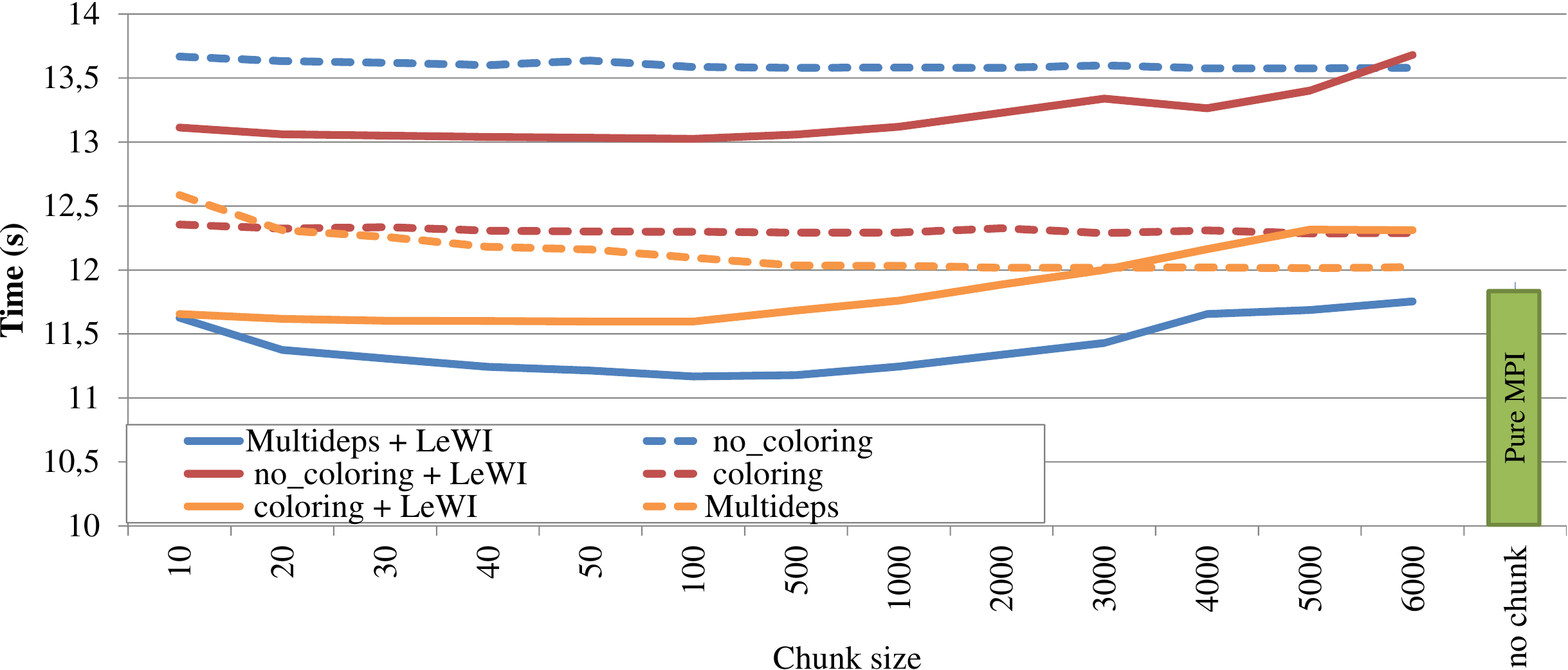}
 \label{fig:F4E_16_chunk}
 }
 \subfigure[32 Nodes (512 cores)]{
  \includegraphics[width=0.47\textwidth]{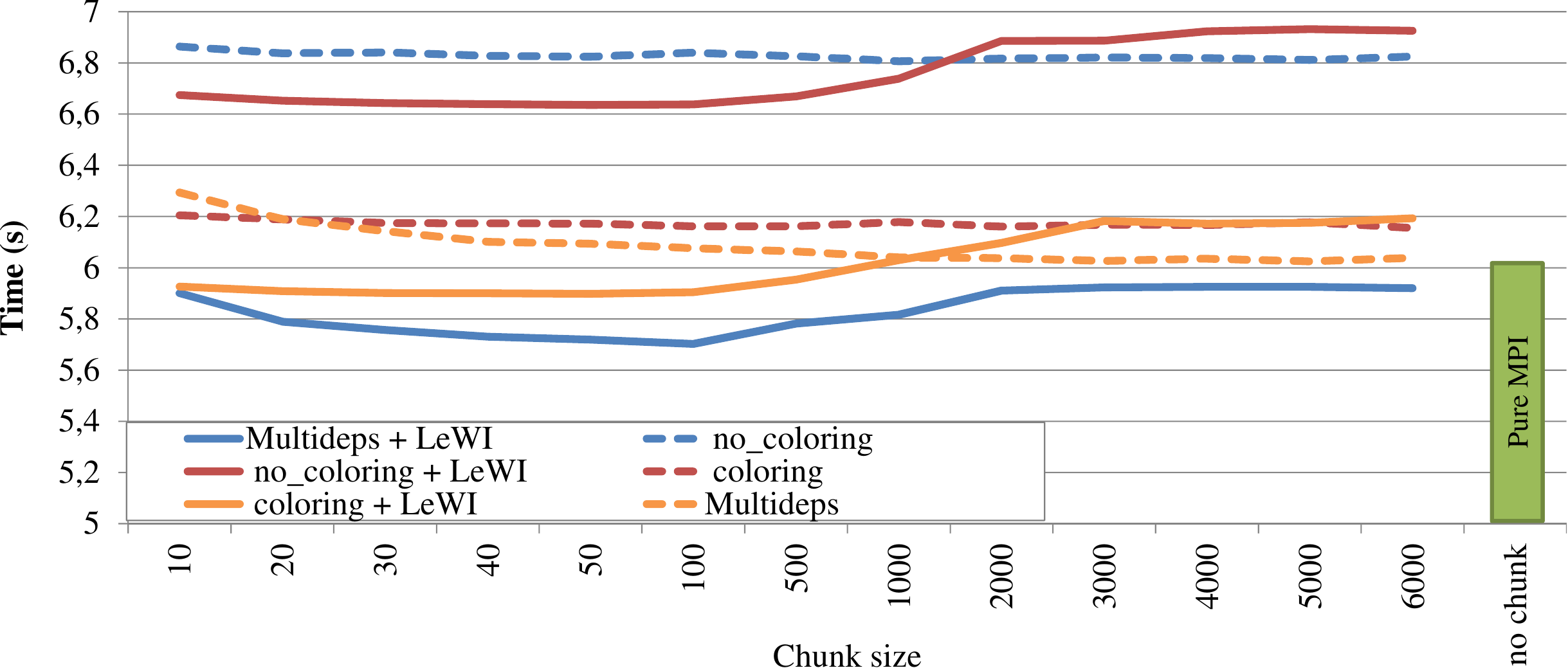}
 \label{fig:F4E_32_chunk}
 }
 \subfigure[64 Nodes (1024 cores)]{
  \includegraphics[width=0.47\textwidth]{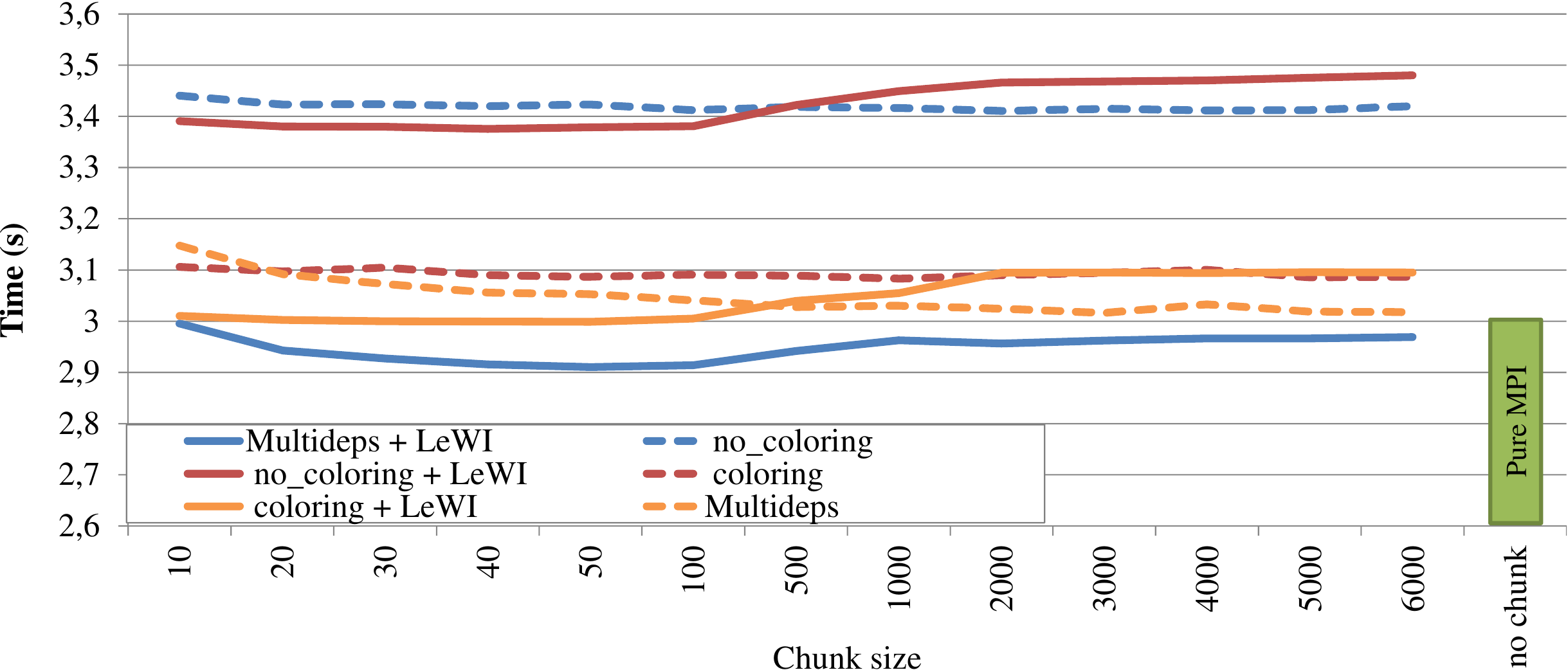}
 \label{fig:F4E_64_chunk}
 }
 \caption{Iter, matrix assembly: chunk size impact on execution time.}
 \label{fig:F4E_chunk}
\end{figure}

Figure~\ref{fig:F4E_chunk} shows the execution time of the matrix assembly for the Iter simulation.
In these charts, we have adjusted the scale of the Y axis to see some
differences between the different series. Each chart corresponds to a number of nodes (16, 32 and
64), while the X axis represents the different chunk sizes. In this case, the Coloring and No Coloring
versions are not affected by the chunk size. This is because the duration of these chunks in this
simulation is much higher than in the Respiratory system case. The matrix assembly in solid
mechanics requires a costly calculation of the constitutive model at the Gauss points, involved
in the Jacobian matrix.

On the other hand, the Multidependencies version is still limited by
small chunk sizes, because its limitation does not only come from the duration of the chunks but
also from the amount of chunks in the queue and with a commutative relationship.

When using DLB the use of big chunk sizes implies a loss of performance. The Coloring version is the
most affected one, because of the synchronizations between the loops computing the different
colors. The no coloring and Multidependencies are affected in a similar way.

For this simulation the optimal chunk size in all the series is around 100. This the chunk size that
we will use for this simulation in the following experiments.

\subsection{Execution Time}

In this section we will compare the execution times of the matrix assembly and subgrid scale
computations for the Respiratory system and matrix assembly for the Iter simulations, with and without
DLB, for each parallelization technique listed in Table \ref{tab:methods}.

\paragraph{Respiratory system simulation}

\begin{figure}[h!tb]
 \centering
  \subfigure[16 Nodes (256 cores)]{
  \includegraphics[width=0.47\textwidth]{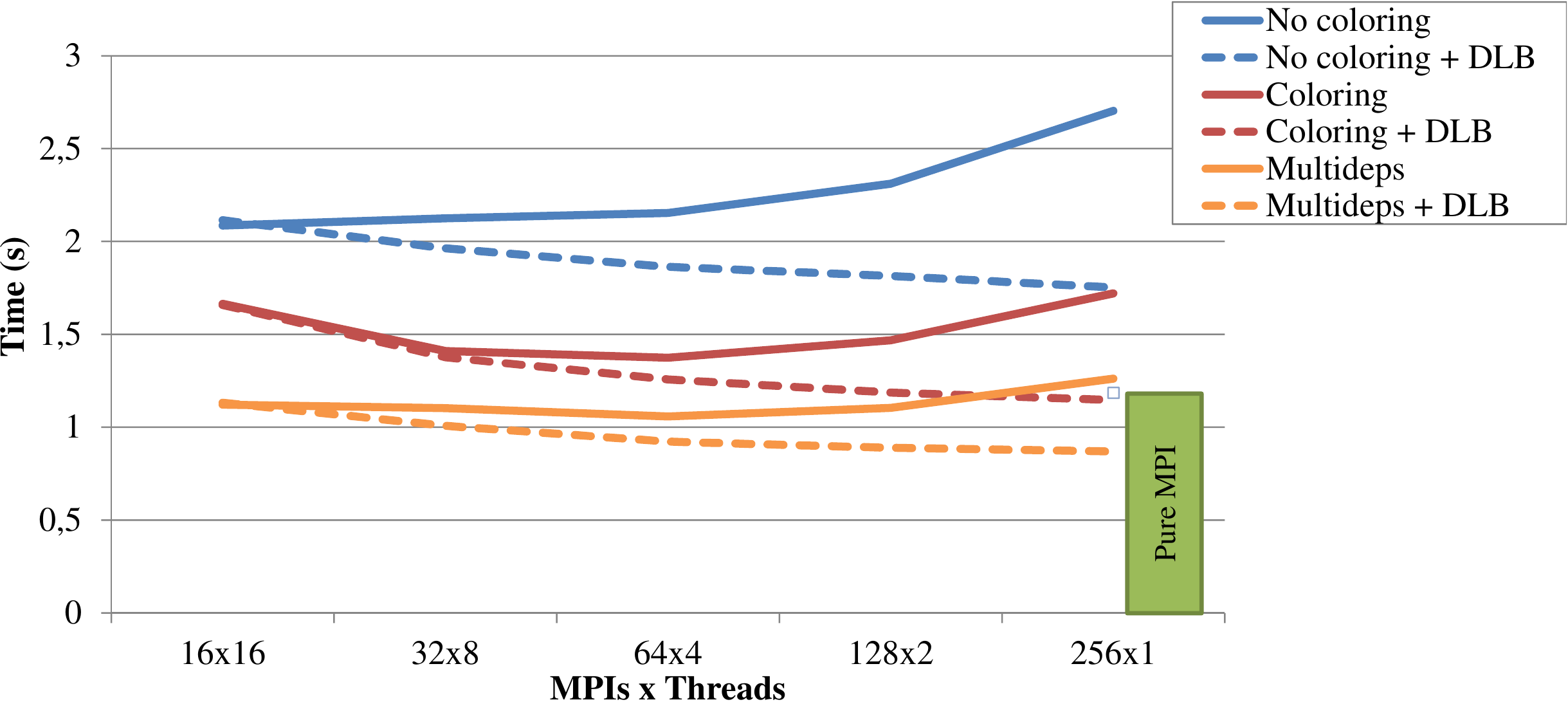}
  \label{fig:resp_16_ass}
  }
  \subfigure[32 Nodes (512 cores)]{
  \includegraphics[width=0.47\textwidth]{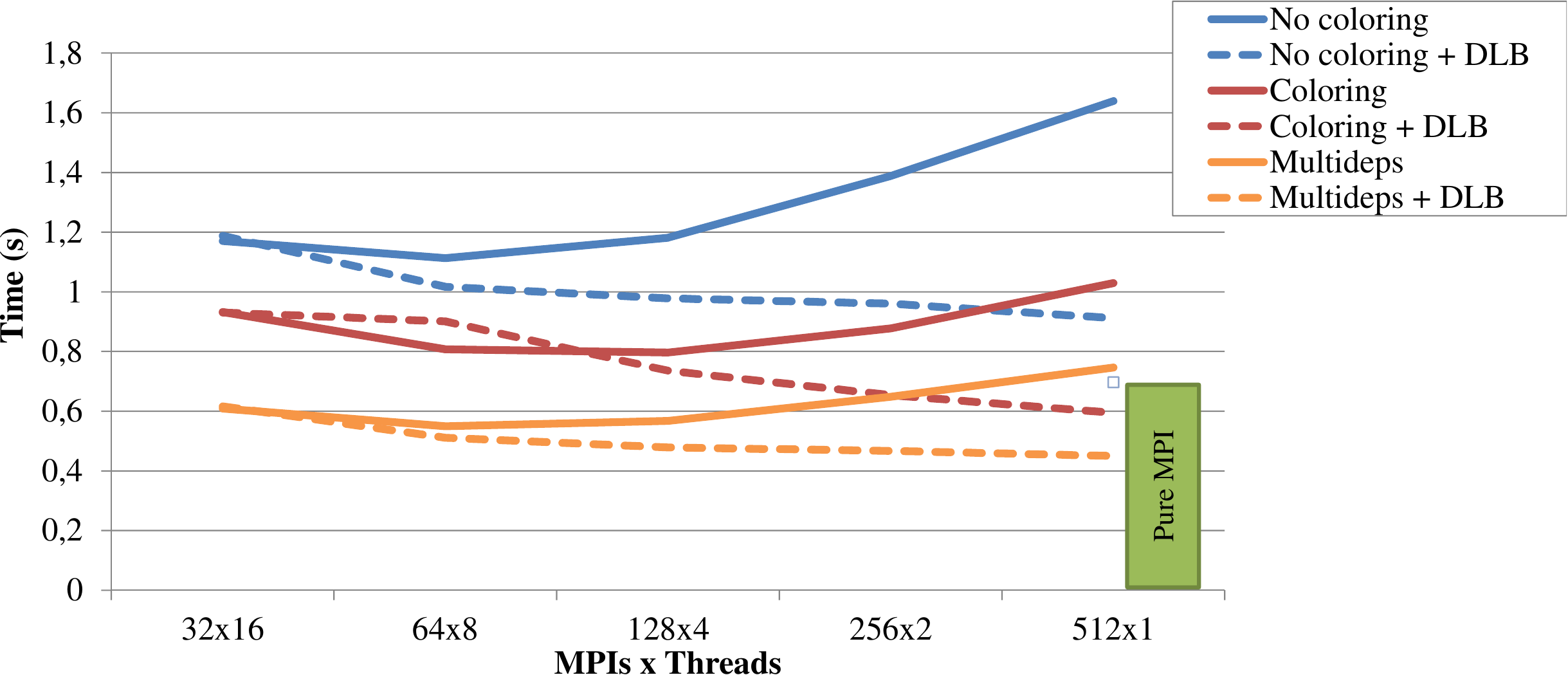}
  \label{fig:resp_32_ass}
  }
  \subfigure[64 Nodes (1024 cores)]{
  \includegraphics[width=0.47\textwidth]{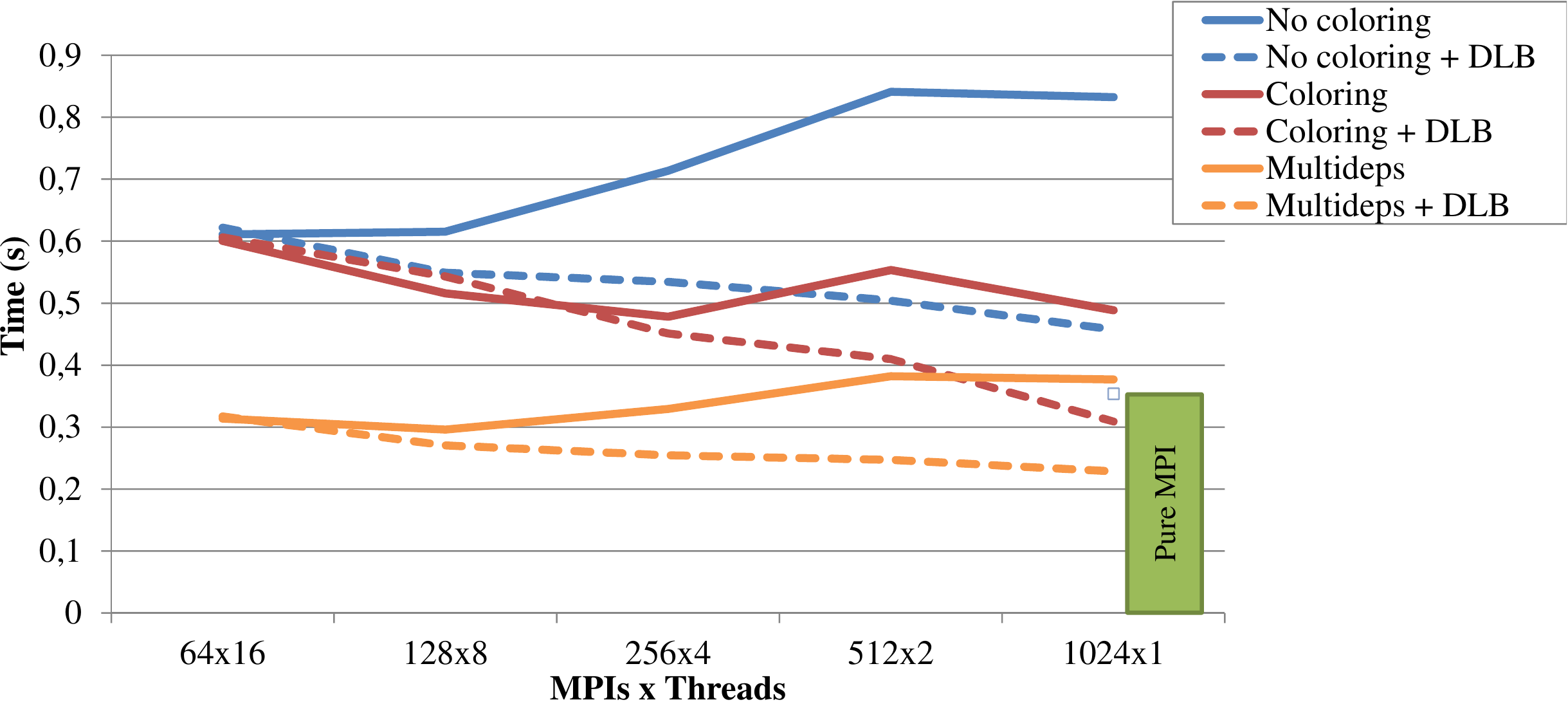}
  \label{fig:resp_64_ass}
  }
 \caption{Respiratory system, matrix assembly: execution time.}
 \label{fig:assembly_respiratory}
\end{figure}

Figure~\ref{fig:assembly_respiratory} shows the execution time of the matrix assembly of
the Respiratory system simulation.
Figures~\ref{fig:resp_16_ass},\ref{fig:resp_32_ass} and \ref{fig:resp_64_ass} correspond to
executions on 16, 32 and 64 nodes respectively. The X axis represents the different configurations
of MPI processes and threads used in each case, while the Y axis represents the average execution
time of the assembly over ten time steps.

When comparing the three different implementations of the parallelization without DLB, we can see
the difference in performance that they obtain. Being the No Coloring version the worst one,
performing worse than the pure MPI version in all the configurations. As we already mentioned in
previous sections, the problem of the No coloring technique is the use of ATOMICS to avoid the race
condition.

The Coloring parallelization yields a better execution time than the No coloring one, but still far
from using the pure MPI version, due to the worst data locality. Finally, the Multidependencies
implementation achieves the best performance. When using a configuration filling the nodes with MPI
processes and only one thread per MPI process (i.e. configurations 256x1, 512x1 and 1024x1) the
execution time obtained is very close to the pure MPI version. In this case the OpenMP level is not
used and we are just measuring the overhead introduced. For the other configurations the
Multidependencies implementation achieves a better performance than the MPI pure version. In general
the best configuration is to spawn one MPI process per socket (2 per node) and use the OpenMP
parallelism within the socket with 8 threads (i.e. configurations 32x8, 64x8 and 128x8).

All the versions present a worse imbalance when increasing the number of MPI processes per node (and
decreasing the number of threads), because the load imbalance increases with the number of
MPI processes, see Figure~\ref{fig:lb_resp} (Left). Except in the case of just one MPI process per
node and 16 threads, where the threads may access memory belonging to the other socket of the node
and these data accesses are slower. When using 8, 4 or 2 MPI processes per node, each MPI process is
pinned to one of the sockets of the node. Therefore, all the data accesses will be to the local
memory of the socket.

When looking at the executions with DLB we observe that in all the cases DLB improves the
performance of the analogous execution without DLB. The only situation where DB can not be applied
is when running one MPI process per node and 16 threads per MPI process because DLB needs more than
one MPI process in each node to load balance. Nevertheless, in this situation DLB does not add any
overhead.

Although the performance of the parallelization affects DLB, in some cases, the load balance can
overcome the overheads of the parallelization and obtain a better performance than the pure MPI
version. For example, in Figure~\ref{fig:resp_32_ass}, when running in 64 nodes (512 cores) and 512
MPI processes with one thread per process, the performance of the Coloring version is 47\% slower
than the pure MPI version, but when using DLB the performance of the Coloring execution is 14\%
faster than the pure MPI.

It is interesting to see how the performance of DLB improves with the number of MPI processes on the
node, being the best configuration to fill the nodes with MPI processes and only one thread for
OpenMP. This can be explained because having more MPI processes gives DLB more flexibility to load
balance. i.e. if we use 2 MPI processes per node configuration, the load balance can only be applied
to the two MPI processes running on the same node.

In all the cases the best situation is to use the commutative Multidependencies with DLB, which can
represent a 37\% faster execution than the pure MPI version.
\\

\begin{figure}[h!tb]
 \centering
 \subfigure[16 Nodes (256 cores)]{
  \includegraphics[width=0.47\textwidth]{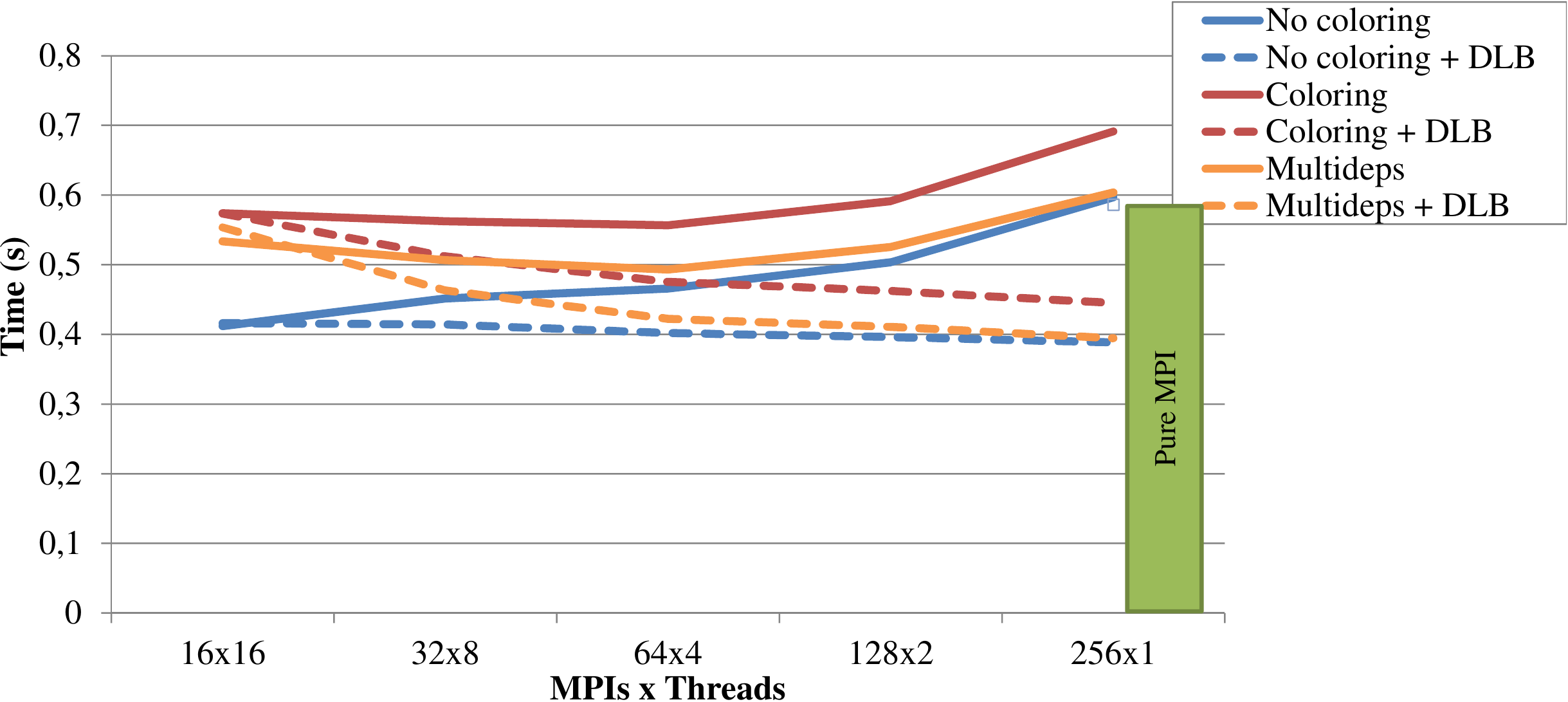}
 \label{fig:resp_16_sgs}
 }
 \subfigure[32 Nodes (512 cores)]{
  \includegraphics[width=0.47\textwidth]{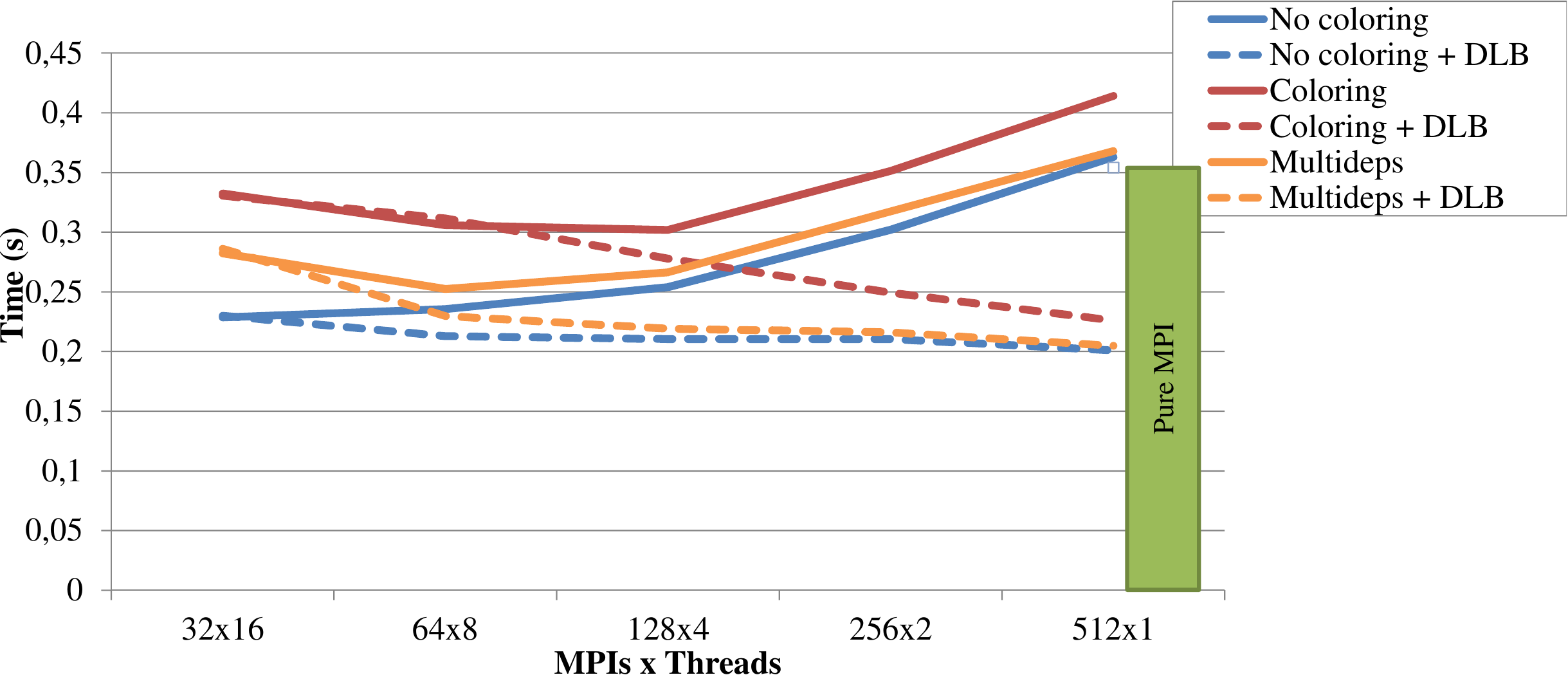}
 \label{fig:resp_32_sgs}
 }
 \subfigure[64 Nodes (1024 cores)]{
  \includegraphics[width=0.47\textwidth]{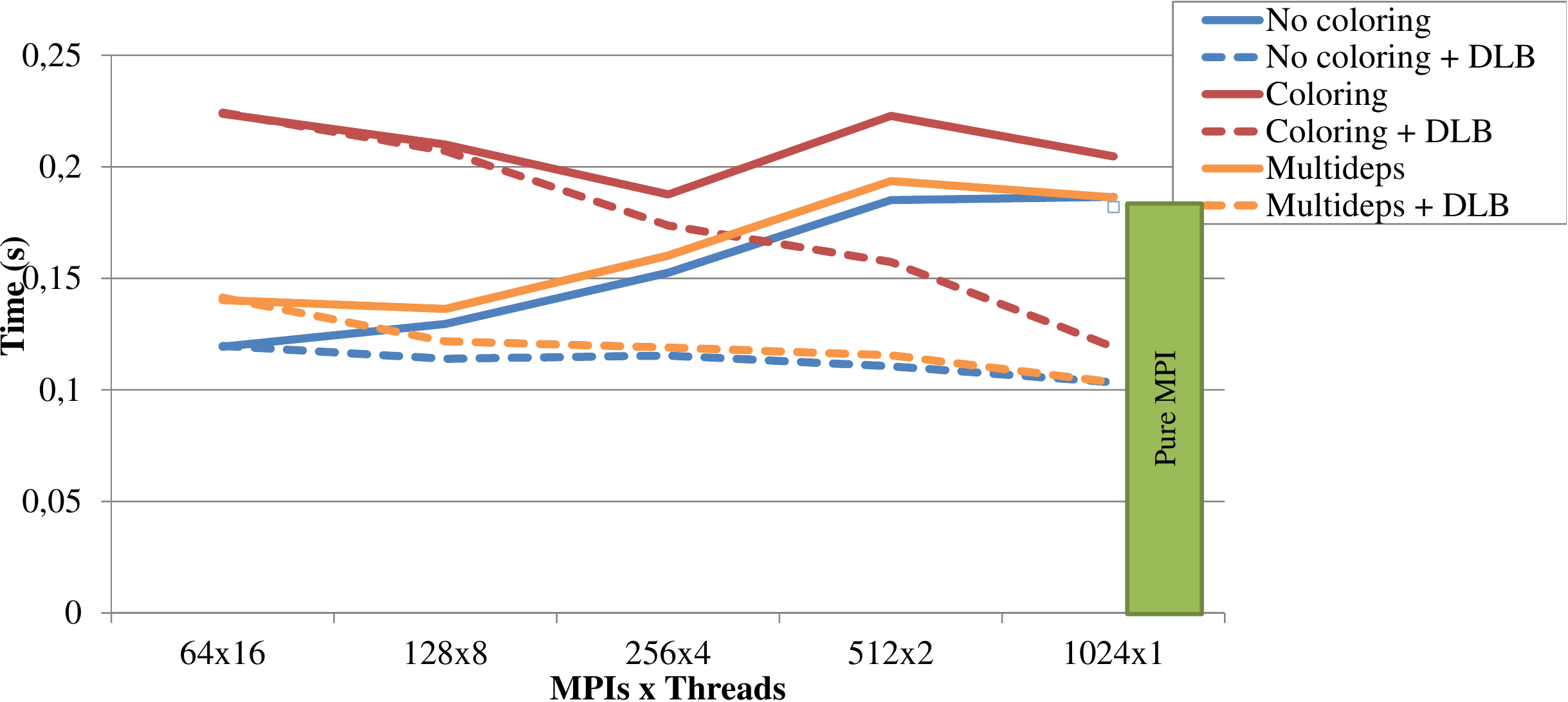}
 \label{fig:resp_64_sgs}
 }
 \caption{Respiratory system, subgrid scale: execution time.}
 \label{fig:sgs_respiratory}
\end{figure}

Figure~\ref{fig:sgs_respiratory} shows the execution time of the subgrid scale calculation, for the same
experiments.
In this case the No Coloring version obtains a performance close to the pure MPI
execution when using 1 thread per MPI process. In the subgrid scale computation, the \texttt{ATOMIC} clause is
not necessary, as it is obtained element-wise. For this reason, the performance compared the to the pure MPI version is much better
than the one obtained in the matrix assembly. Moreover, in the subgrid scale using a hybrid method
with the No coloring version is the best configuration.

The coloring version performs worse than the no coloring because it still presents the bad locality
issue. The Multidependencies parallelization has a performance close the that of the No coloring one for
a low number of threads. When increasing the number of threads to 16, the execution time is higher
because all the threads accessing the shared queue of commutative tasks in the OpenMP
runtime. However, it still improves the performance of the pure MPI version.

When using DLB, the performance is improved in all the cases. The best configuration with DLB is to
use 16 MPI processes per node and one thread per process independently of the parallelization
strategy used. When using the no coloring or Multidependencies version the performance with DLB is
almost constant independently of the configuration of MPI processes and threads used. This means
that DLB is able to solve all the load imbalance within the node and that it is independent of the
configuration decided by the user. When running in 64 nodes, the version of Multidependencies with
DLB and one thread per MPI process is 44\% faster than the pure MPI version.

\paragraph{Iter simulation}

\begin{figure}[h!tb]
 \centering
 \subfigure[16 Nodes (256 cores)]{
  \includegraphics[width=0.47\textwidth]{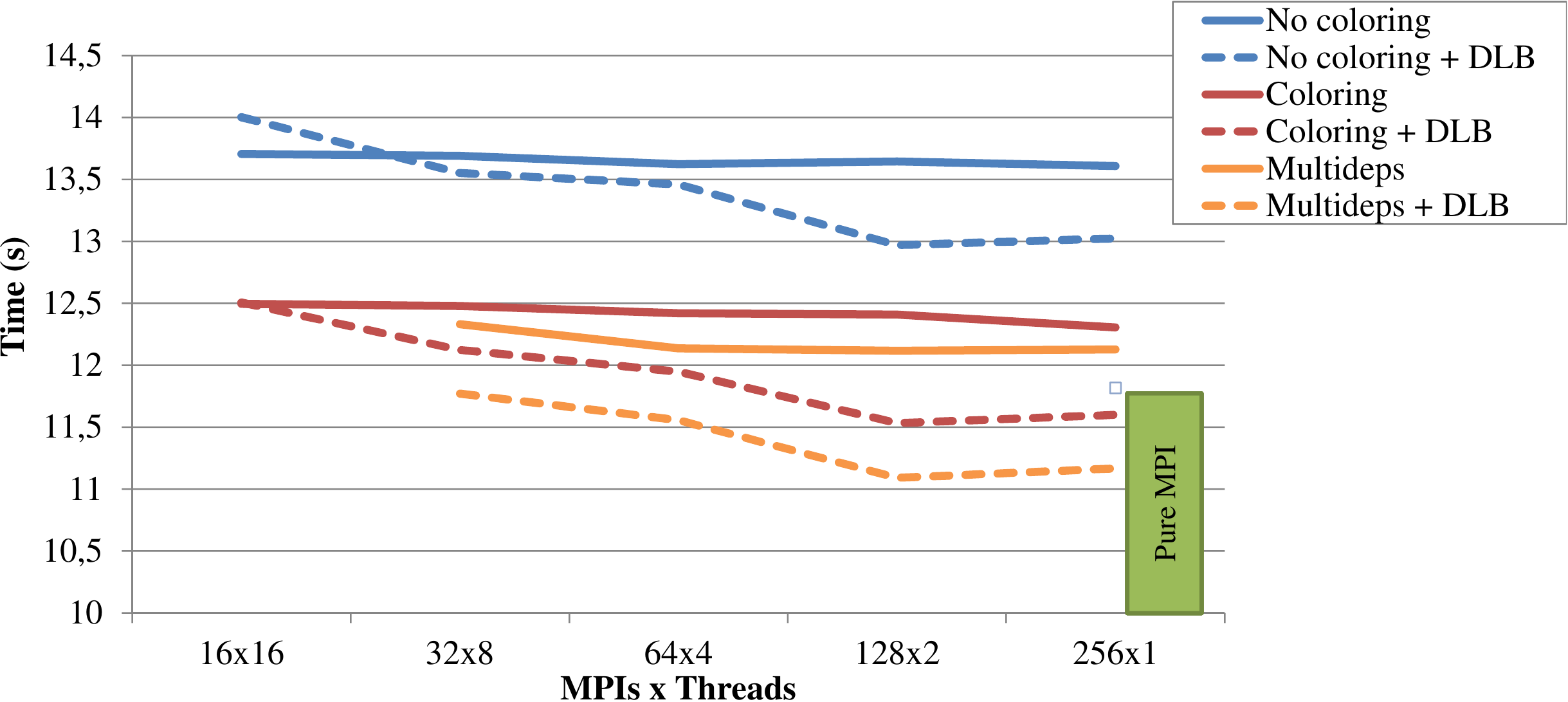}
 \label{fig:F4E__16}
 }
 \subfigure[32 Nodes (512 cores)]{
  \includegraphics[width=0.47\textwidth]{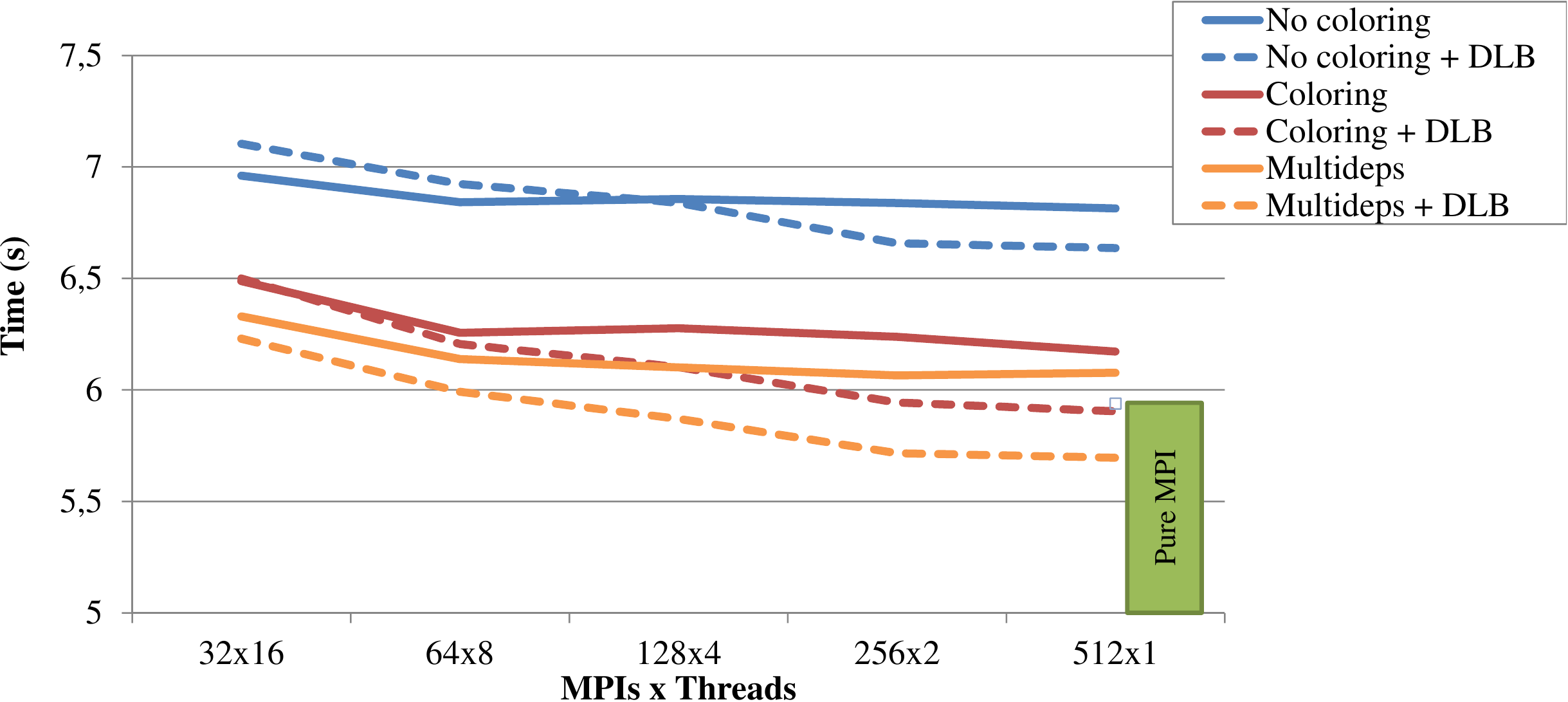}
 \label{fig:F4E_32}
 }
 \subfigure[64 Nodes (1024 cores)]{
  \includegraphics[width=0.47\textwidth]{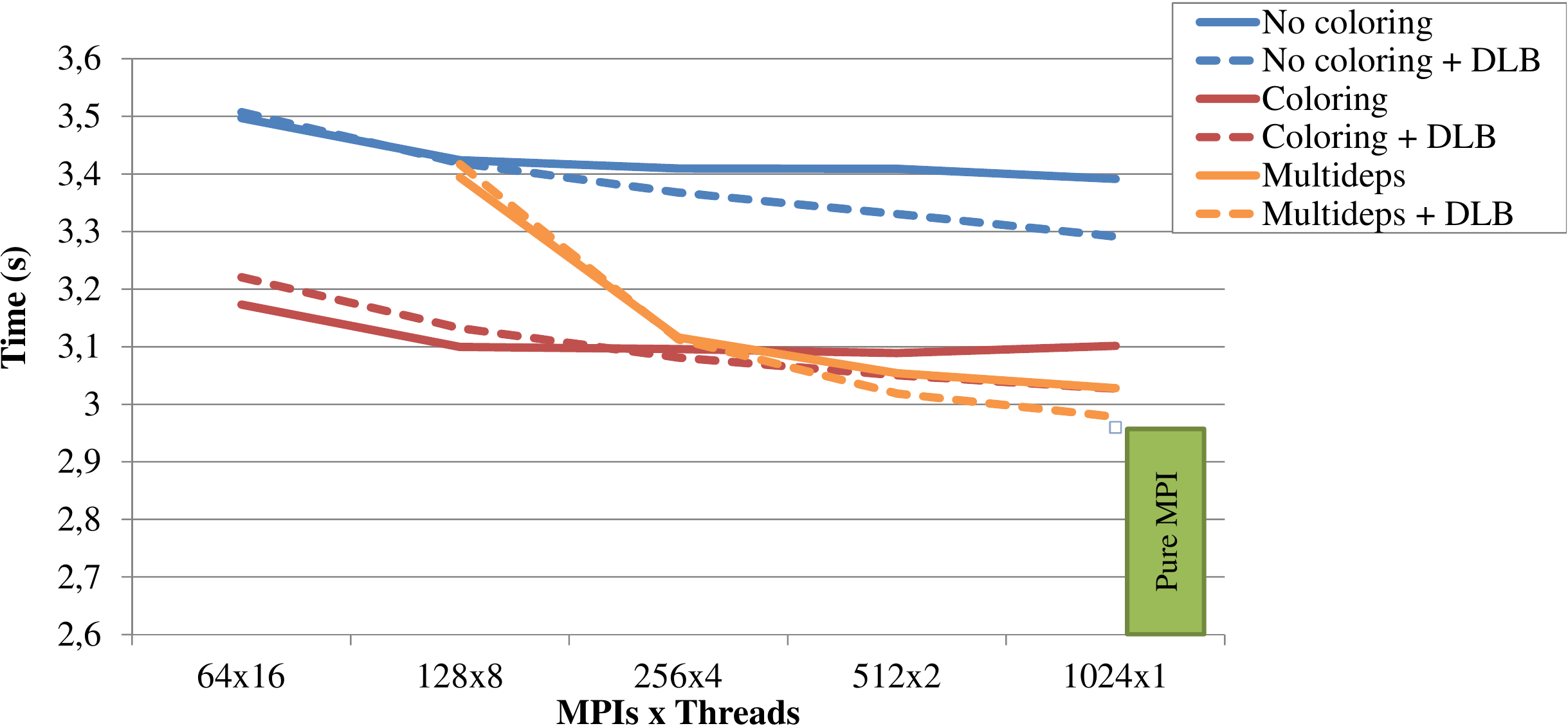}
 \label{fig:F4E_64}
 }
 \caption{Iter, matrix assembly: execution time.}
 \label{fig:F4E}
\end{figure}

Figure~\ref{fig:F4E} shows the execution time of the matrix assembly phase for the
\textit{Iter} simulation.
The different charts correspond to executions on 16, 32 and 64 nodes. The X axis, we can see the
different configurations of MPI processes and OpenMP threads. As already noticed in
Figure~\ref{fig:lb_resp} (Right), the imbalance exhibited by this simulation is not very high. Therefore, the
performance improvement that we can expect with DLB will not be as high as the ones obtained in the
Respiratory system simulation. Note that in this case; we have increased the scale of the X axis
in order to discern better the different values.

When comparing the performance of the different parallelizations, we can see that the No Coloring
version is slower than Coloring and Multidependencies because of the \texttt{ATOMIC} overhead. In
this case, the difference between the Coloring and Multidependencies is not very significant because
in this simulation the amount of computation per element is higher than when solving the Respiratory
system. This means that the data locality has less impact on the overall performance.

The simulations executed with DLB are always faster than their analogous ones without DLB. As in the
case of the fluid, the best performance is obtained when using 16 MPI processes per node and one thread per MPI process. When
using DLB with Multidependencies and 16 MPI processes per node, the performance is better that the
pure MPI version (around 5\% improvement).

\subsection{Hardware Counters Study}

In this subsection, we are going to evaluate the different parallelizations (No Coloring, Coloring,
and Multidependencies) based in different performance counters in order to support some of the
performance explanations we have used in the previous sections. The data shown in the following
charts have been obtained with Paraver from an Extrae trace of a real execution using PAPI
5.4.1~\cite{PAPI-web} \cite{PAPI}.

We are going to see the results for both simulations the respiratory system and the iter simulation. In all the
cases we have launched 256 MPI processes with one OpenMP thread per MPI process (16 MPI processes
per node). This configuration is used to see the impact of the parallelization in the performance of
the code but to avoid seeing the contention between the different threads, as the optimum
distribution of threads per MPI process has already been discussed in the previous section.

\paragraph{Respiratory system simulation}

\begin{figure}[h!tb]
  \centering
  \includegraphics[width=0.47\textwidth]{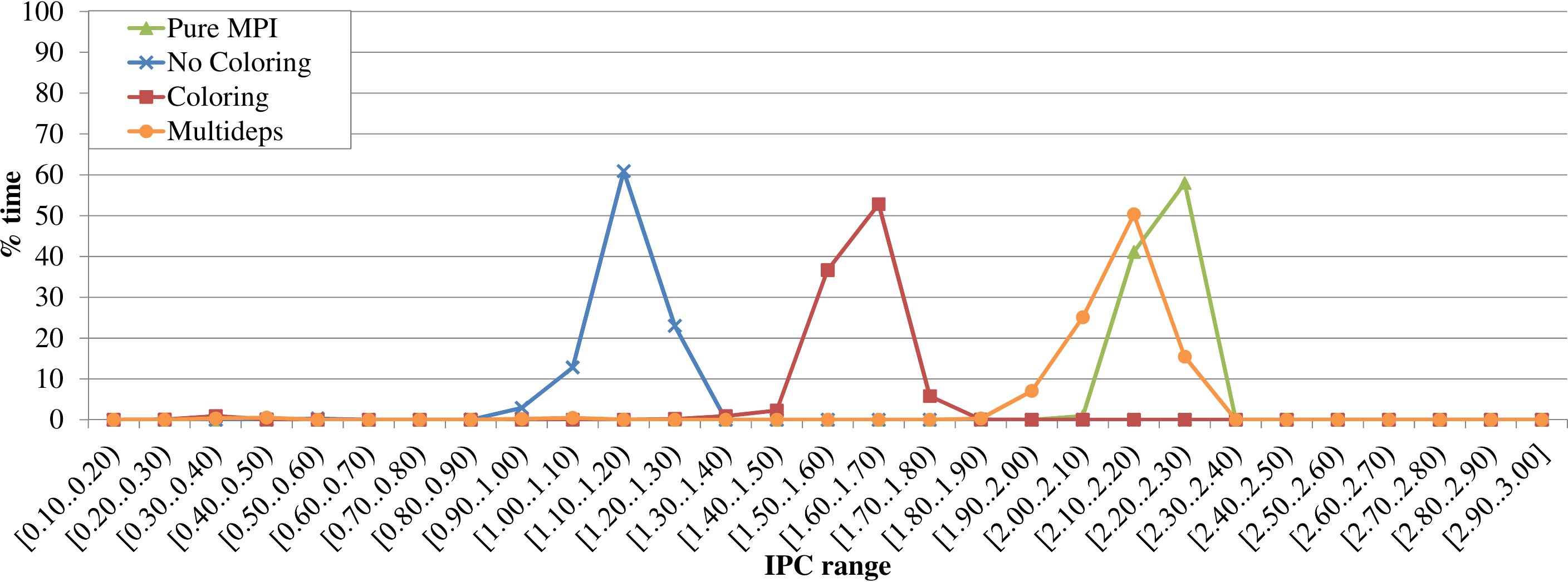}
  \caption{Respiratory system, matrix assembly: IPC.}
  \label{fig:ipc_assembly}
\end{figure}

Figure~\ref{fig:ipc_assembly} is a normalized histogram of the IPC obtained during the matrix assembly.
The X axis represents the different intervals of IPC measurements, while the Y axis gives the
percentage of time of each IPC range, with respect to the total CPU time spent in the matrix
assembly. We observe that the pure MPI version had an IPC between 2.1
and 2.3. When using the No Coloring version of the parallelization the IPC went down to 1.1. This
matches the previous performance results were the No Coloring version was two times slower than the
pure MPI.

The IPC of the Coloring version is between 1.5 and 1.7, better than the No Coloring but still far
from the IPC obtained by the pure MPI version. When using the Multidependencies parallelization, the
IPC is between 2 and 2.2 almost the same as the one achieved by the pure MPI version.
\\

\begin{figure}[h!tb]
   \centering
   \includegraphics[width=0.47\textwidth]{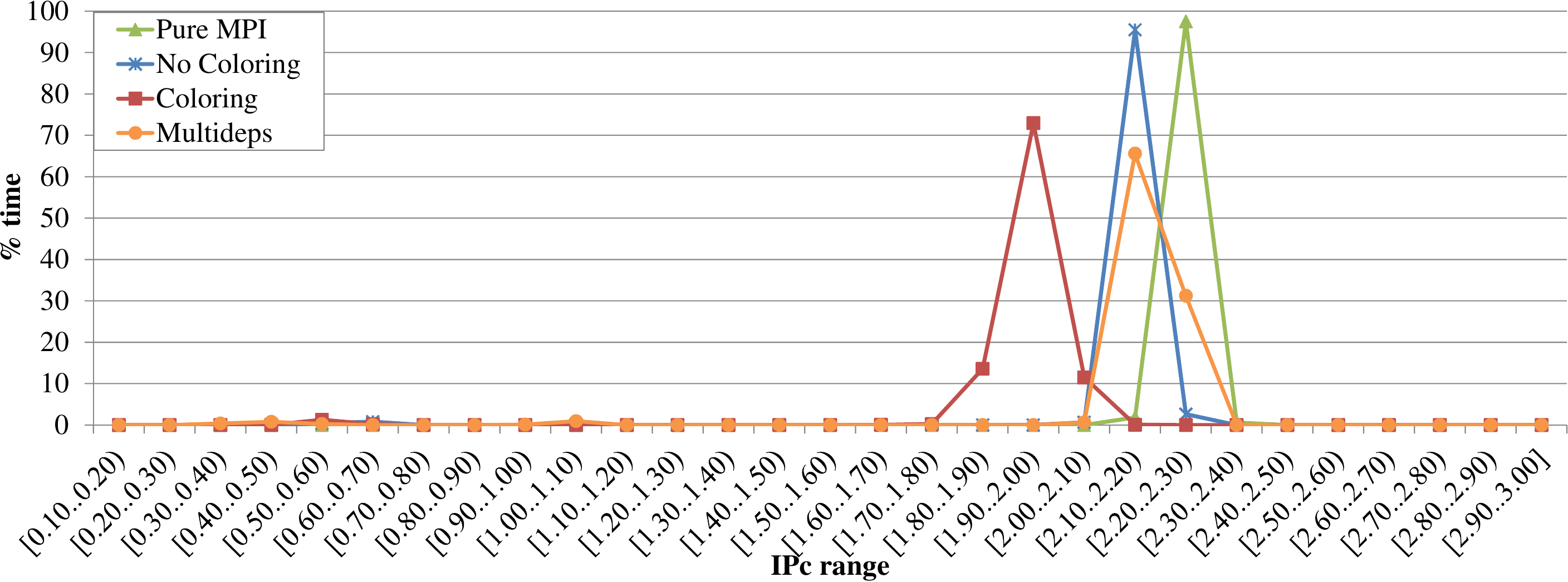}
   \caption{Respiratory system, subgrid scale: IPC.}
  \label{fig:ipc_sgs}
\end{figure}

Figure~\ref{fig:ipc_sgs} gives the IPC for the subgrid scale computation phase.
The IPC for the pure MPI and
the Multidependencies versions are the same as in the matrix assembly. The No Coloring version in
this phase presents a much higher IPC because it does not need the \texttt{ATOMIC} clause.
An IPC equivalent to the one achieved with Multidependencies is obtained. On the other hand, the Coloring
parallelization has a worse IPC than the others because of the loss in data locality. But we can see
that the performance loss is not as important as in the matrix assembly phase, this is because in
the subgrid scale computation, the pressure over the memory is not as high as in the matrix assembly phase.
\\

\begin{figure}[h!tb]
  \centering
  \subfigure[Number of instructions]{
    \includegraphics[width=0.47\textwidth]{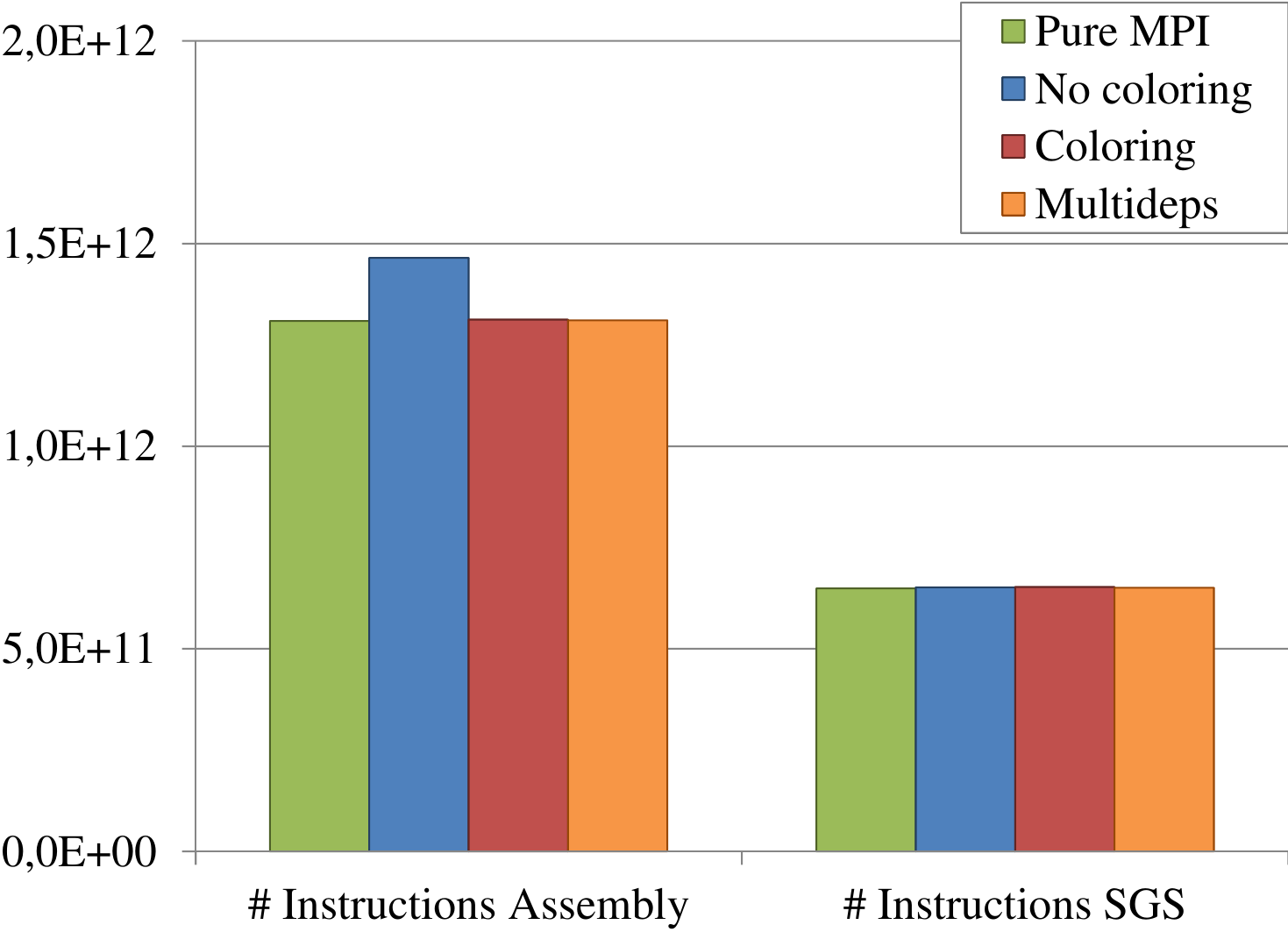}
    \label{fig:instructions_resp}
  }
  \subfigure[Number of L3 cache misses]{
    \includegraphics[width=0.47\textwidth]{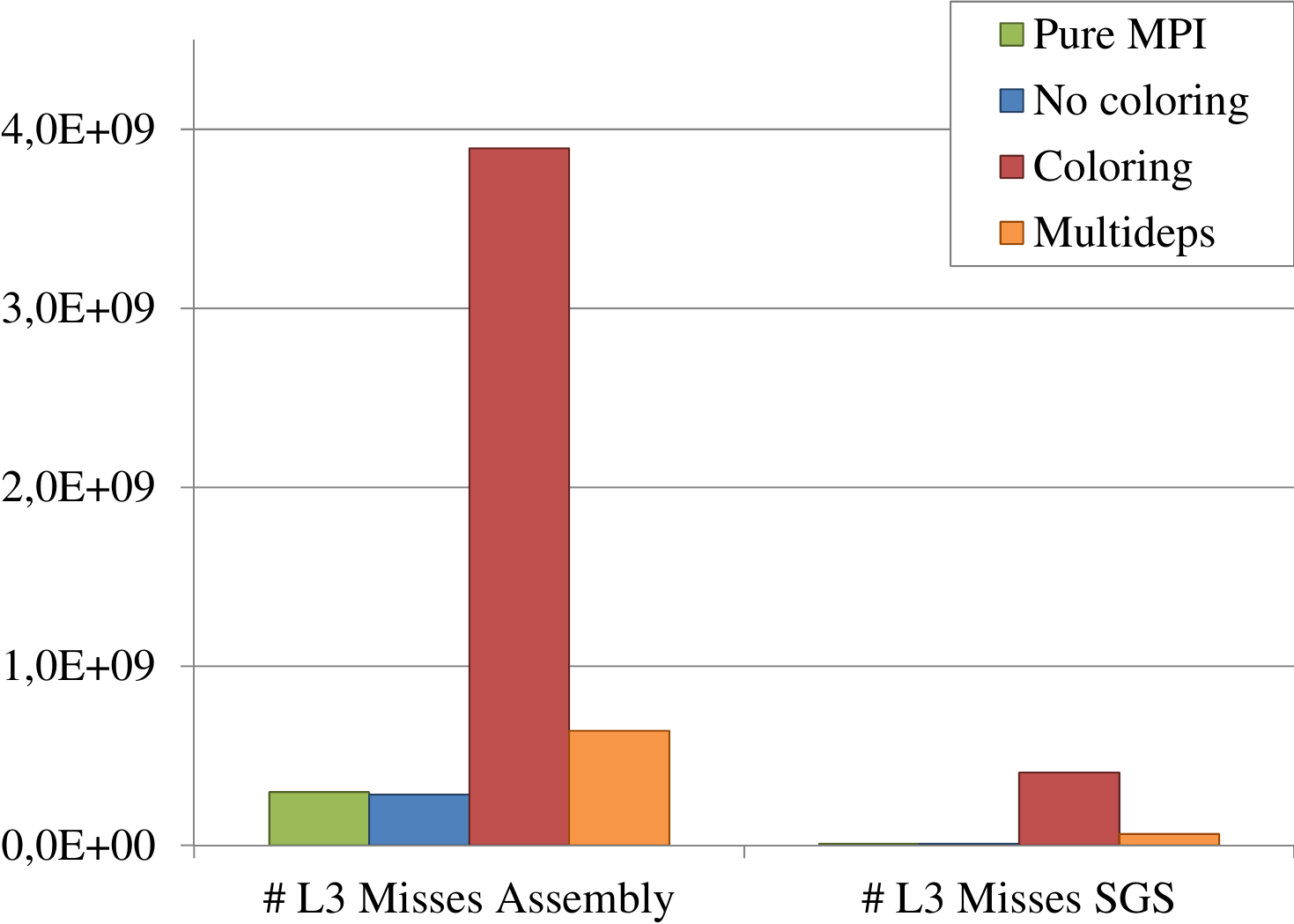}
    \label{fig:misses_resp}
  }
  \caption{Respiratory system: performance counters.}
\end{figure}

To back up our conclusions on the previous charts, we have measured the total number of instructions
executed during a matrix assembly and subgrid scale computation. The results obtained are shown 
in Figure~\ref{fig:instructions_resp}.
In these charts we can see that the number of instructions
executed in the subgrid Scale computation is much lower than the ones necessary to compute the
matrix assembly. When comparing the different parallelizations we can observe that the number of
executed instructions in the different parallelizations is the same, this means that the amount of
computation for the different parallelizations is the same. The difference in the performance
come from other sources, for example the cache misses.

Figure~\ref{fig:misses_resp} shows the cache misses in L3 during the execution of the matrix
assembly and the subgrid scale computation. As we already said, the pressure on the memory is much
higher in the matrix assembly than in the subgrid scale computation. When comparing the different
parallelizations, we observe that the No coloring version has the same number of cache misses than
the pure MPI version, as the execution order is the same. On the other hand, the number of cache
misses in the Coloring version is much higher due to loss of data locality when computing
elements that are not contiguous in memory. The Multideps version presents more cache misses than
the pure MPI version but far from the number of cache misses achieved by the Coloring version.

\paragraph{Iter simulation}

\begin{figure}[h!tb]
  \centering
  \includegraphics[width=0.47\textwidth]{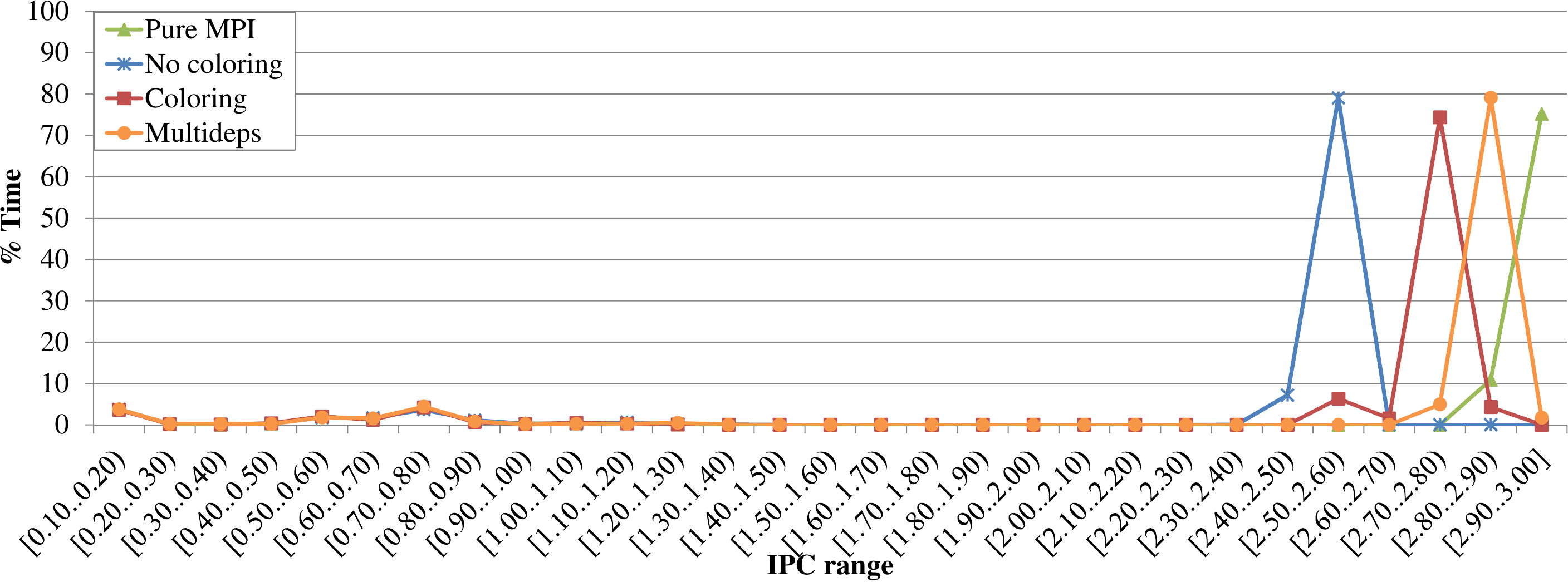}
  \caption{Iter, matrix assembly: IPC.}
  \label{fig:ipc_F4E}
\end{figure}

Figure~\ref{fig:ipc_F4E} shows the IPC obtained in the matrix assembly for the \textit{Iter}
simulation.
As we said before, the matrix assembly of this problem has a higher computational load per
element than that of the fluid problem, and this can be seen in the higher IPC in all the cases. The pure MPI version has an IPC
around 3 almost during the whole phase. The No Coloring version goes down to an IPC of 2.5 but still
far from dividing the IPC by two that we observed in the Respiratory system simulation. Again,
this confirms that there is more computation going on, and the impact of the \texttt{ATOMIC} clause
is not as high as in the other case.

The Coloring parallelization presents an IPC of 2.7 because of the worst data locality, and the
Multidependencies version obtains an IPC of 2.9 achieving almost the same performance as the pure
MPI version.

\begin{figure}[h!tb]
  \centering
  \subfigure[Number instructions]{
    \includegraphics[width=0.22\textwidth]{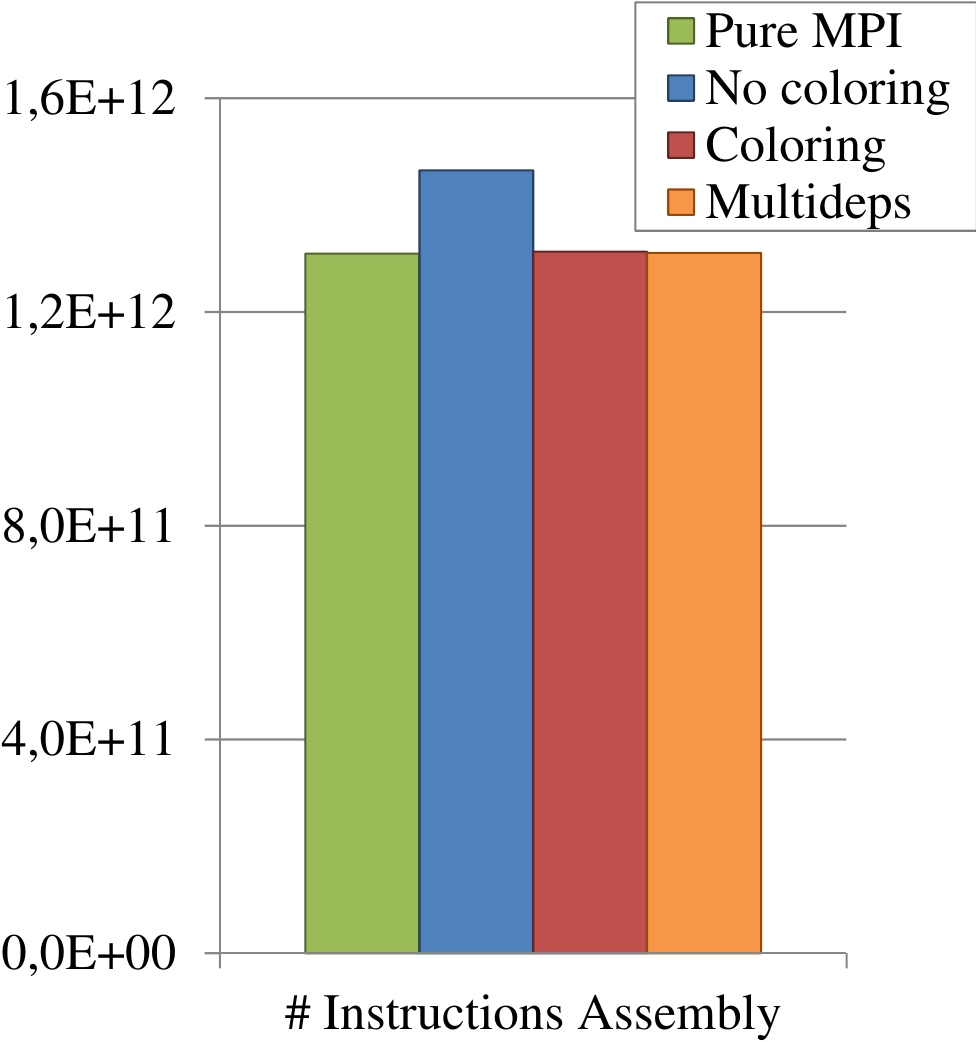}
    \label{fig:instructions_iter}
  }
  \subfigure[Number of L3 cache misses]{
    \includegraphics[width=0.22\textwidth]{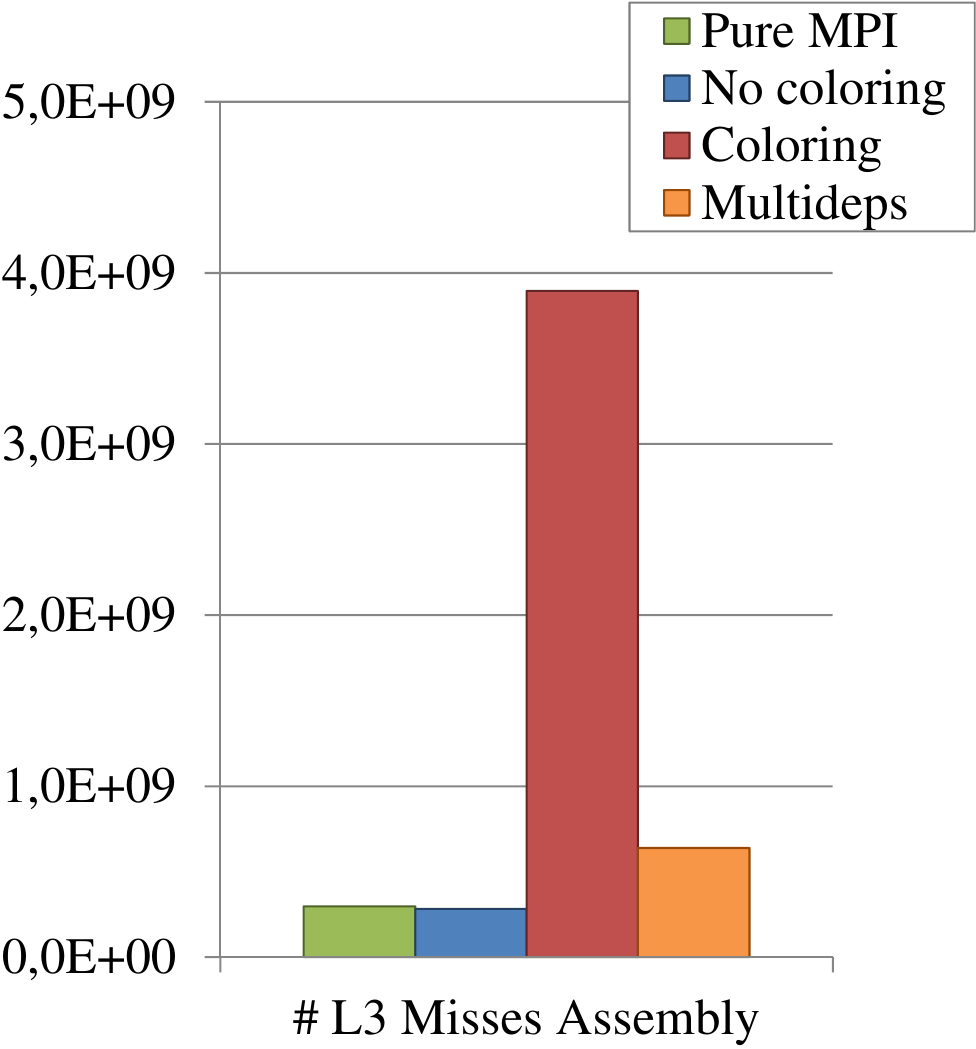}
    \label{fig:misses_iter}
  }
  \caption{Iter, matrix assembly: Performance counters.}
\end{figure}

Figure~\ref{fig:instructions_iter} shows that the number of instructions necessary to
compute the matrix assembly is the same for all the parallelizations.
By looking at the L3 cache misses for the different parallelizations (Figure~\ref{fig:misses_iter}, we observe the same
conclusions as in the Respiratory system simulation: the No coloring version has the same cache misses as
the pure MPI; the Coloring parallelization presents a higher number of misses and the
Multidependencies version has a worse data locality than the pure MPI but far from the number of
misses of the Coloring one.
\begin{figure}[h!tb]
   \centering
   \includegraphics[width=0.47\textwidth]{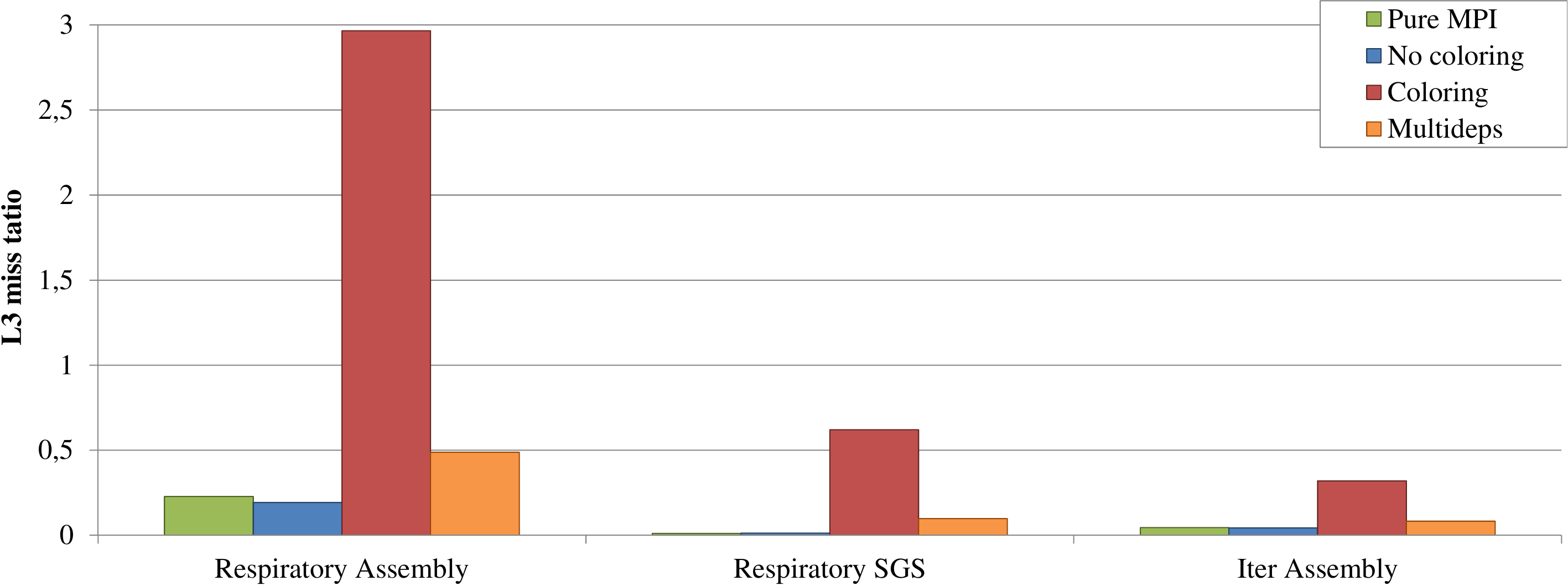}
   \caption{Iter, matrix assembly: L3 Miss ratio.}
  \label{fig:l3_missratio}
\end{figure}

We define the miss ratio as the number of cache misses issued for each 1000 instructions executed,
computed as follows: $L3MissRatio=\frac{\#MissesL3 * 1000}{\#Instructions}$. In
Figure~\ref{fig:l3_missratio} we can see the L3 miss ratio for the different simulations and
computation phases. Based in this chart we can asses that the pressure in the data access is much
higher in the matrix assembly of the respiratory simulation than in the other computations. In the
case of the matrix assembly for the iter simulation the miss ratio is lower than for the subgrid
scale computation of the respiratory simulation.

\subsection{Scalability}

During this evaluation, we had the opportunity to run some strong scalability tests on MareNostrum 3
with up to 16384 cores (1024 nodes). For this, we have used the mesh multiplication strategy
described in \cite{Houzeaux:2013fk} to obtain a mesh of 141 million elements from the original mesh shown in Figure \ref{fig:resp_mesh}.
In these experiments, we wanted to demonstrate that DLB can scale
up to using thousands of cores and also that even working at the node level the use of DLB can help
improving the performance significantly in this kind of executions. 

\begin{figure}[h!tb]
 \centering
 \subfigure[Matrix Assembly]{
  \includegraphics[width=0.45\textwidth]{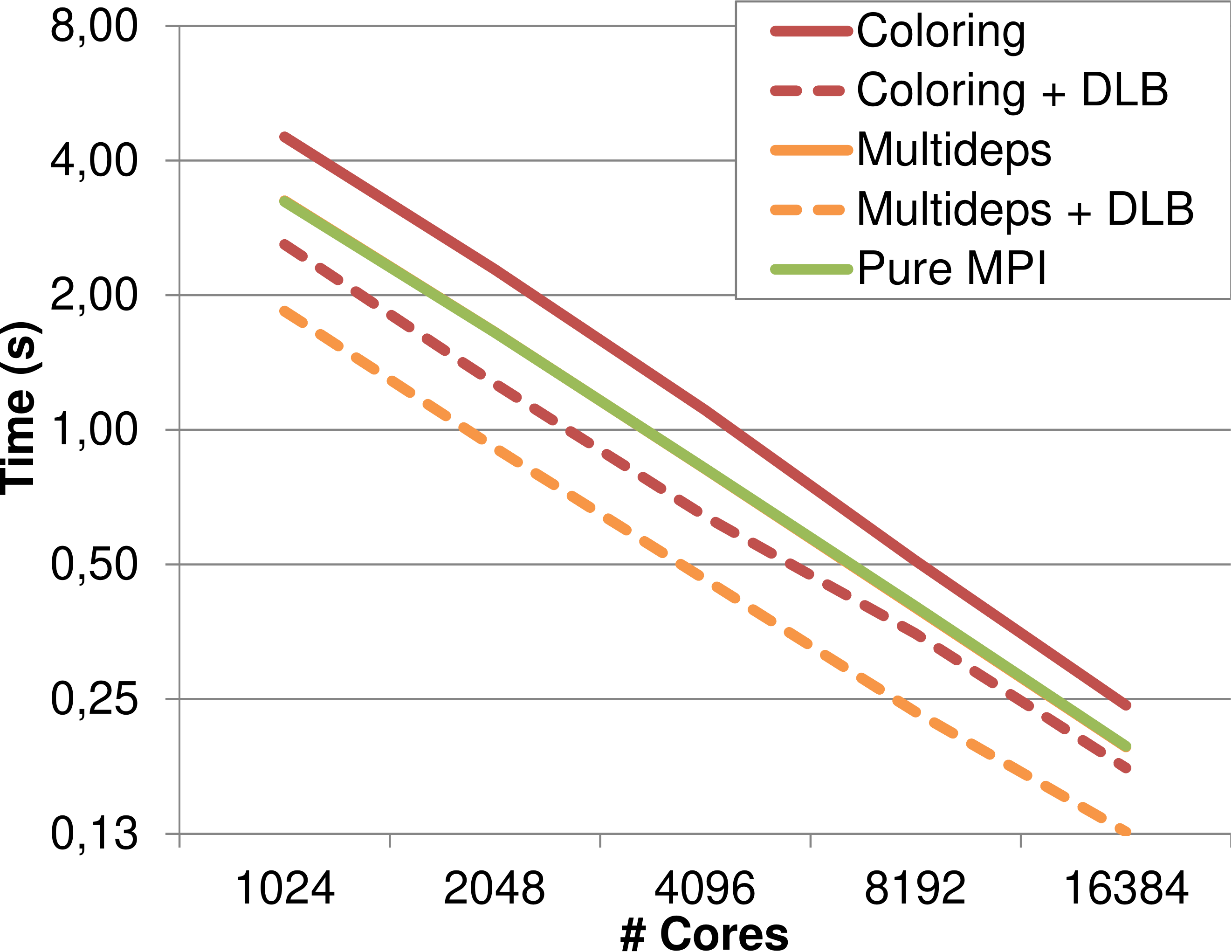}
 \label{fig:scalabilty_time2_ASS}
 }
  \subfigure[Subgrid Scale]{
  \includegraphics[width=0.45\textwidth]{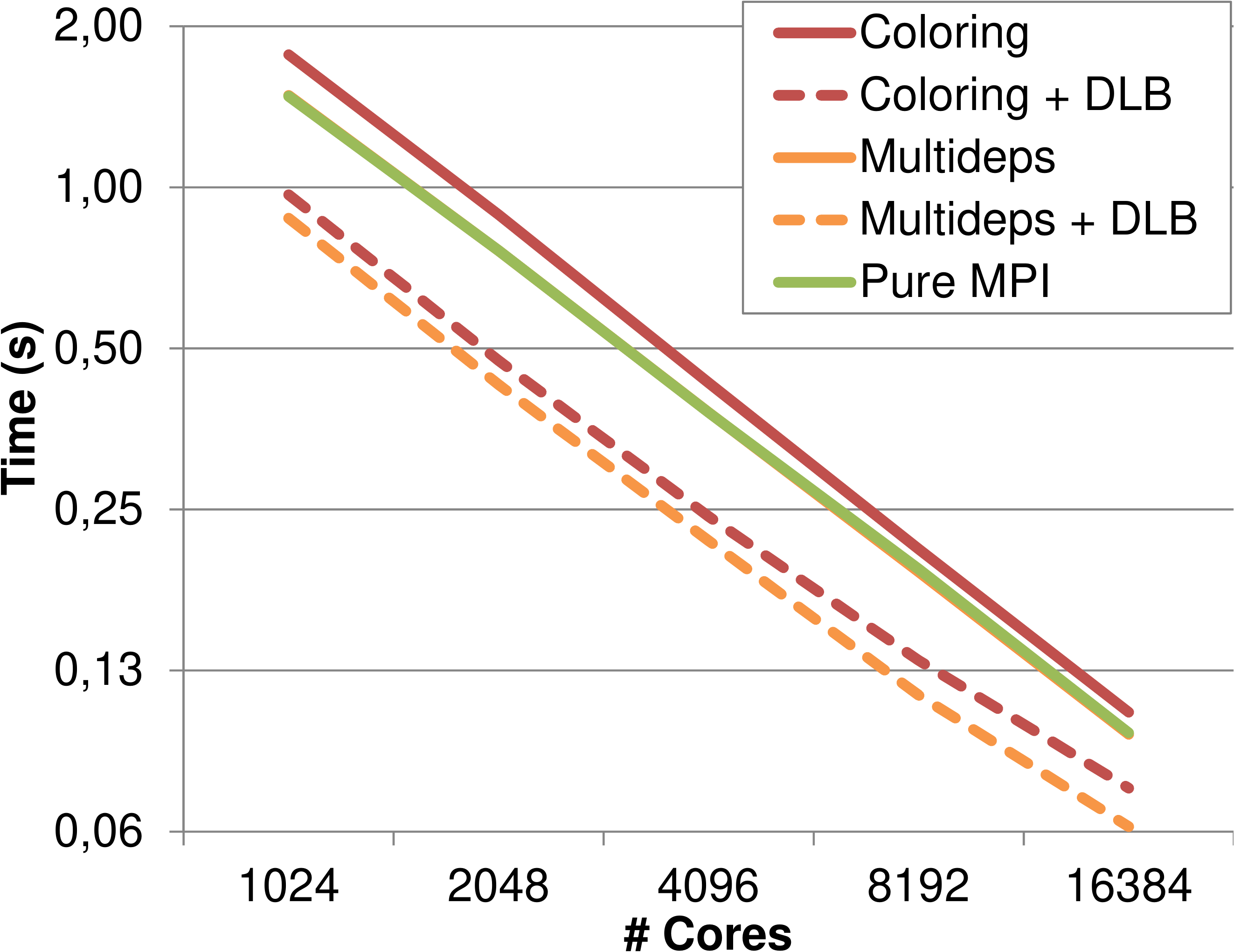}
 \label{fig:scalabilty_time2_SGS}
 }
 \caption{Respiratory system: execution time up to 16k cores.}
 \label{fig:scalability_time}
\end{figure}

In these executions we have simulated the Respiratory system using the best configurations
observed in the previous experiments, 16 MPI processes per node with 1 thread per process and chunks
of 200 elements. The values presented are the average execution time of 10 time steps for each phase
of the computation.
\\

Figure~\ref{fig:scalability_time} shows the execution time of the matrix assembly and the
subgrid scale. In the X axis, we can see the number of cores used to run and in the Y axis the
elapsed time in seconds in a logarithmic scale. As we have seen before, the performance of the
coloring version is worse than the pure MPI. On the other hand, the Multidependencies
parallelization obtains the same performance as the pure MPI version independently of the number of
cores used.

\begin{figure}[h!tb]
 \centering
 \subfigure[Matrix Assembly]{
  \includegraphics[width=0.45\textwidth]{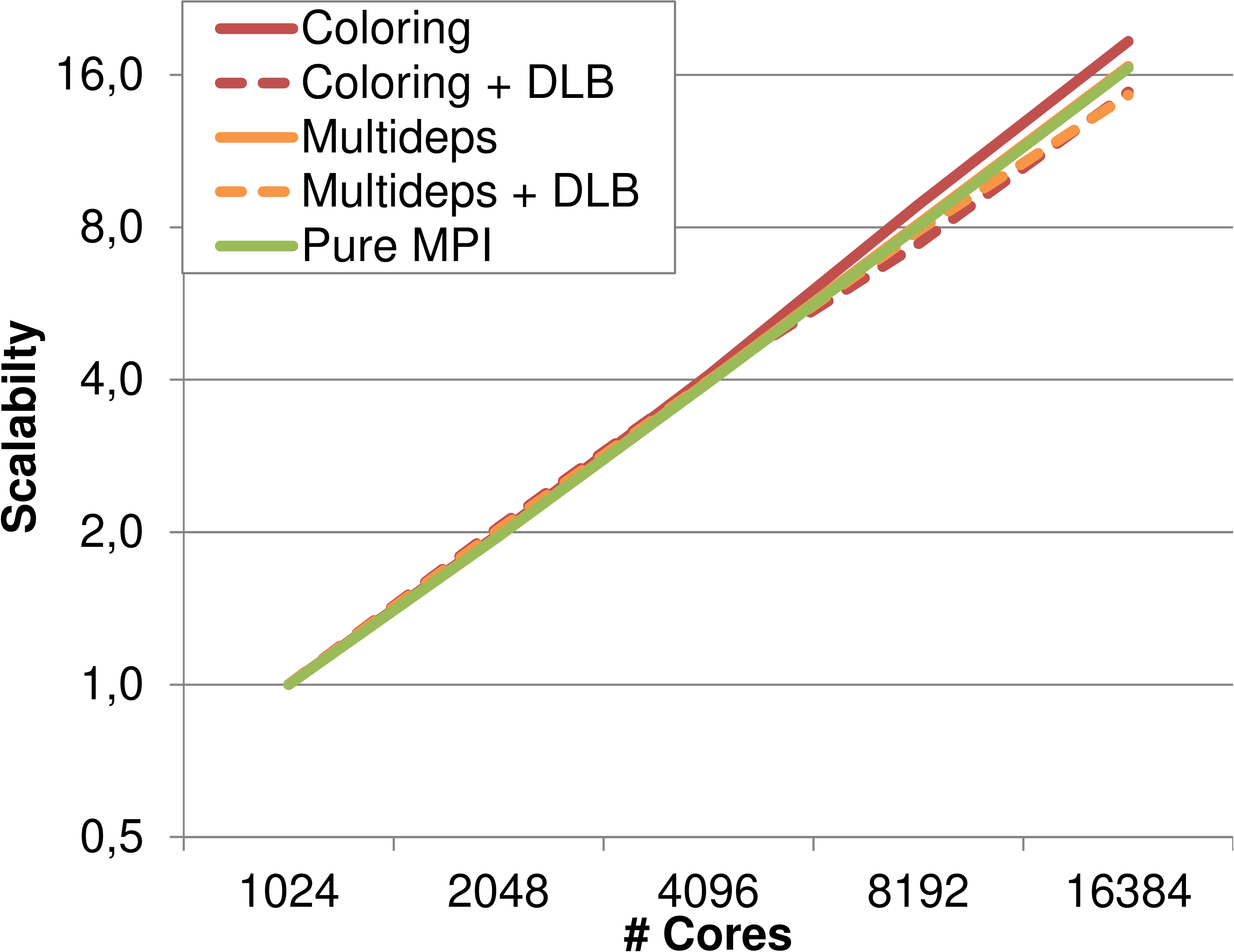}
 \label{fig:scalabilty_ASS}
 }
 \subfigure[Subgrid Scale]{
  \includegraphics[width=0.45\textwidth]{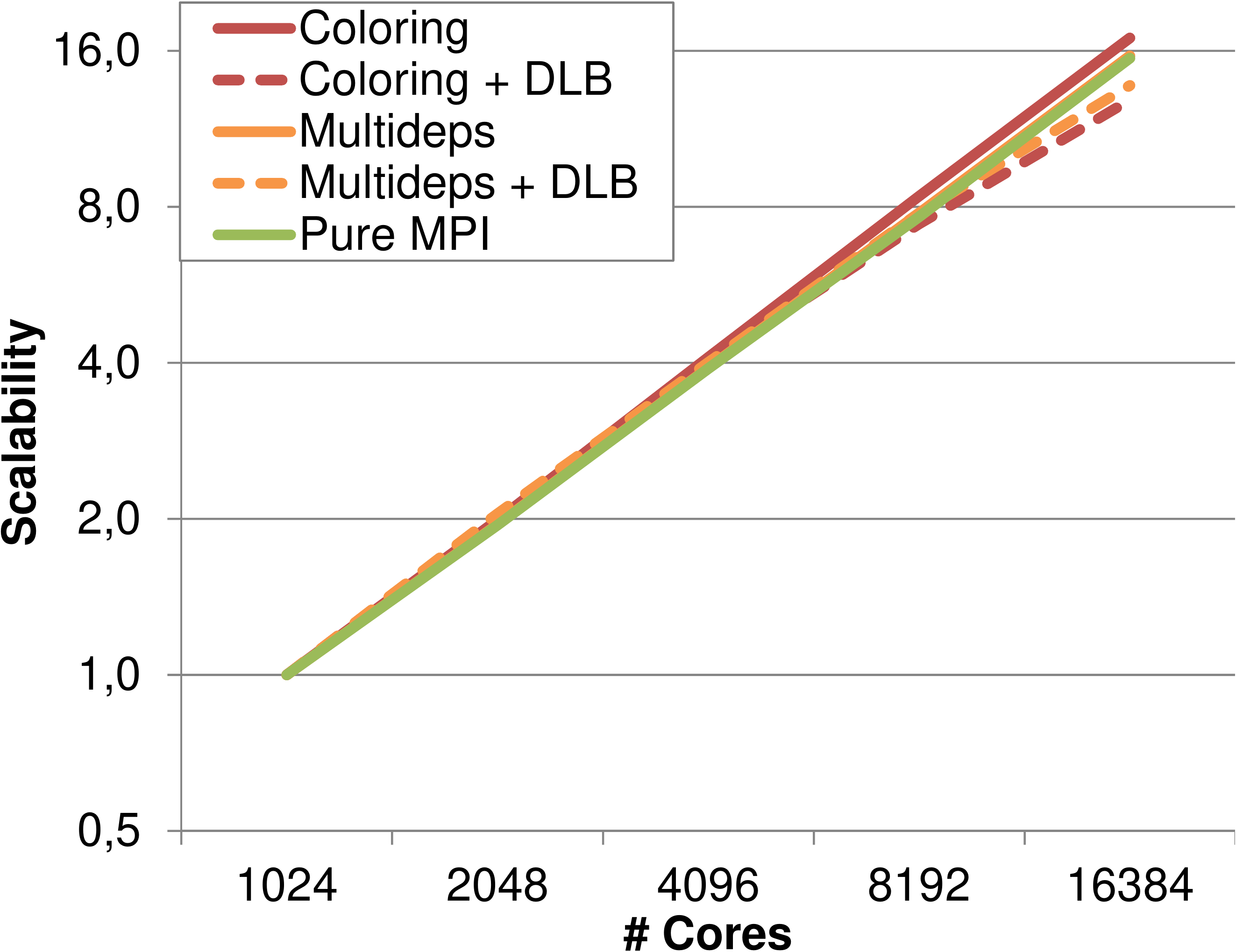}
 \label{fig:scalabilty_SGS}
 }
 \caption{Respiratory system: scalability up to 16k cores.}
 \label{fig:scalability}
\end{figure}

If we look at the results obtained using DLB, we can see that the execution time is reduced
significantly when running with the Coloring or the Multidependencies parallelizations, but
specially with the last one. The most interesting thing is to see how the gain when using DLB is
maintained independently of the number of nodes used. In particular, comparing with the pure MPI
version the gain goes from 1.55 to 1.75 using 1024 and 16384 cores, respectively,

Figure~\ref{fig:scalability} shows the scalability as it is usually presented by application
developers, using as base case the smallest number of resources used of the same
version:
\begin{EQ}
  \textrm{Scalability}_{x,y}=\frac{time_{1024,y}}{time_{x,y}}.
\end{EQ}

We want to show how misleading this metric can be. In this chart, the best scalability is obtained by
the Coloring version without DLB. But this version is the one that obtains the worst execution
time. On the other hand, the executions with DLB (both with Coloring and Multidependencies) has a
worse scalability curve, but a better execution time.

\begin{figure}[h!tb]
  \centering
  \subfigure[Matrix Assembly]{
    \includegraphics[width=0.45\textwidth]{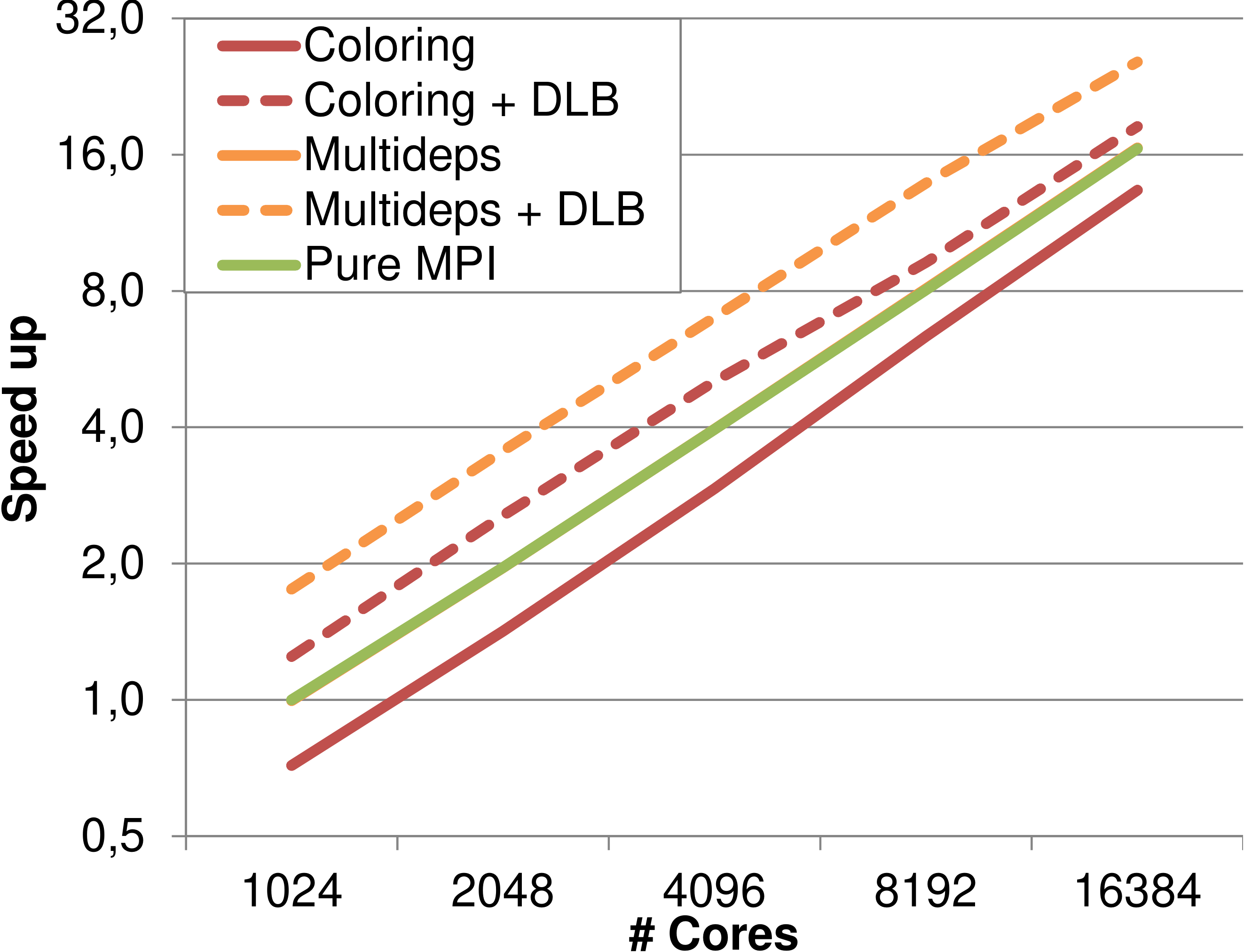}
    \label{fig:scalabilty_speedup_ASS}
  }
  \subfigure[Subgrid Scale]{
    \includegraphics[width=0.45\textwidth]{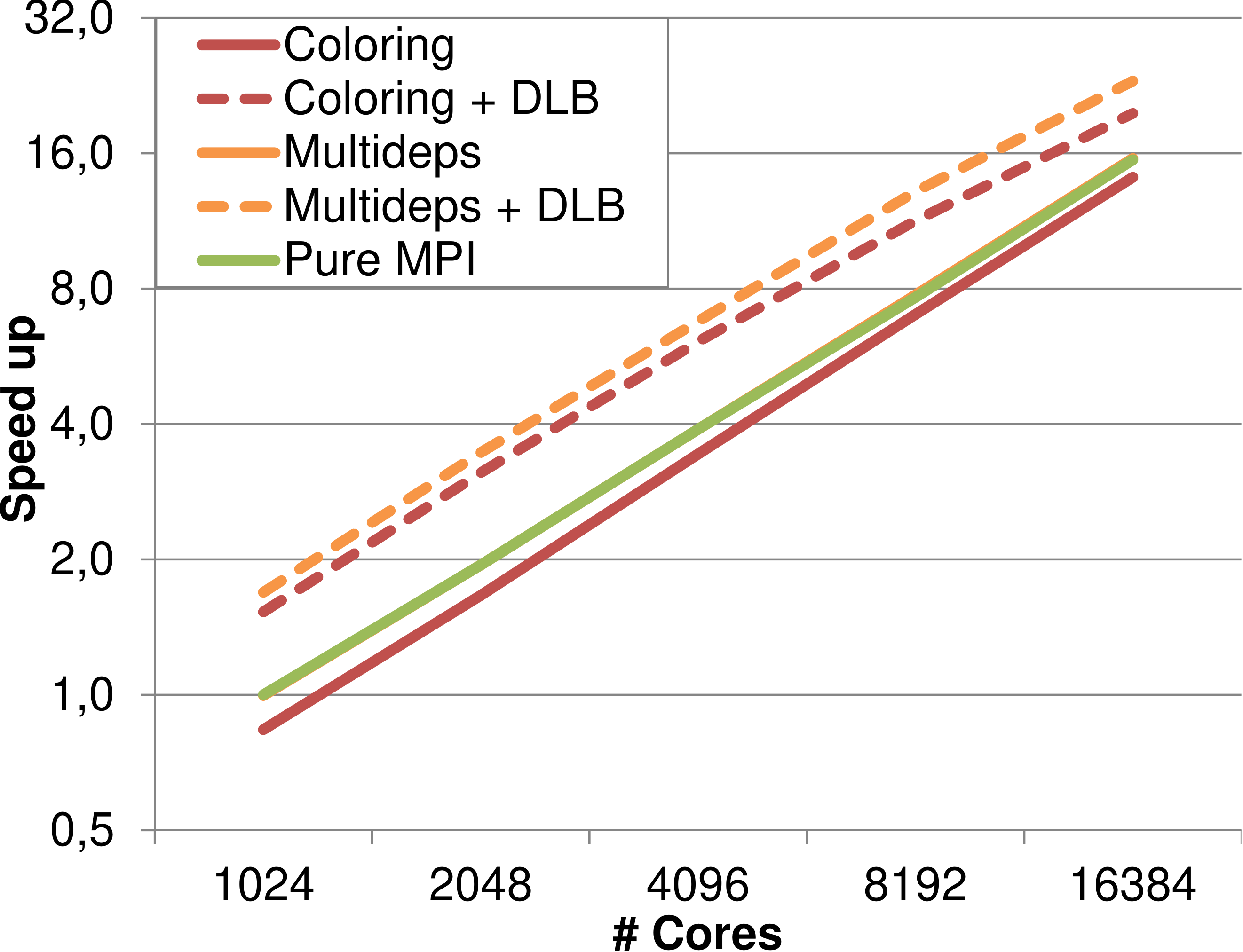}
    \label{fig:scalabilty_speedup_SGS}
  }
  \caption{Respiratory system: speedup up to 16k cores.}
  \label{fig:scalability_speedup}
\end{figure}

Figure~\ref{fig:scalability_speedup} shows the speed up with respect to the pure MPI version on 1024
cores. In this case, all the versions are computed versus the same baseline scenario and are thus
directly comparable:
\begin{EQ}
\textrm{Speedup}_{x,y}=\frac{ time_{ 1024,\textrm{Pure MPI}} }{time_{x,y}}.
\end{EQ}

In this chart, we can observe how the performance of the DLB versions is significantly better even when
running on a large number of cores, being able to obtain a speed up of 23 when using 16 times more
resources.

\section{Conclusions}

In this paper we have presented two runtime mechanisms to improve the performance of a computational
dynamics code. Both approaches can be used without important modifications in the source code and
are applied at runtime. We have tested both mechanisms in the solutions of two production computational mechanics
problems, involving a fluid and a solid. These two problems present different performance issues.

On one hand, we have presented the use of multidependences to avoid a race condition in the matrix
assembly, for which we have demonstrated that the performance can be improved up to a 60\% when using the
multidependences approach with respect to the use of ATOMICS. We have explained using a hardware counters
analysis this improvement, related to avoiding the use of ATOMICS and obtaining a better spatial locality.

On the other hand, we have used a dynamic load balancing library (DLB) to improve the load balance in
some phases. DLB can be used without modifying the source code and we have shown an improvement in
performance of up to 50\%. Moreover, we have seen that the use of DLB releases the user from
choosing the better configuration for a hybrid parallel programing (i.e. distribution of MPI
processes and threads).

DLB can be used also in MPI pure applications, just by adding OpenMP pragmas where necessary, in
this case the second level of parallelism is only used for load balancing purposes.

Finally we have shown that both mechanisms can scale up to 16384 cores obtaining the best results
with the multidependences and DLB versions.

\bibliography{biblio}
\bibliographystyle{acm}



\end{document}